\documentclass[fleqn,usenatbib]{mnras}

\usepackage{newtxtext,newtxmath}
\usepackage[T1]{fontenc}
\usepackage{ae,aecompl}

\usepackage{graphicx}	% Including figure files
\usepackage{amsmath}	% Advanced maths commands
\usepackage{multirow}
\usepackage{threeparttable}
\usepackage[export]{adjustbox}
\usepackage{fontawesome}\usepackage[dvipsnames]{xcolor}%voy por aqui%%%%%%%%%%%%%%%%%%%%%%%%%%
\usepackage{natbib}
\title[Circumnuclear ionising clusters: NGC~7742]{Physical properties of circumnuclear ionising clusters. I. NGC~7742}

\author[S. Zamora and A. I. Díaz]{
S. Zamora,$^{1, 2}$\thanks{E-mail: sandra.zamora@uam.es}\thanks{PhD fellow of Ministerio de Educación y Ciencia, Spain, BES-2017-080509, CEAL-AL/2017-02}
Ángeles I. Díaz,$^{1, 2}$
\\
$^{1}$Departamento de Física Teórica, Universidad Autónoma de Madrid, 28049 Madrid, Spain\\
$^{2}$CIAFF, Universidad Autónoma de Madrid, 28049 Madrid, Spain\\
}

\date{Accepted 2023 July 03. Received 2023 June 30; in original form 2022 December 22}

\pubyear{2023}

\begin{document}
\label{firstpage}
\pagerange{\pageref{firstpage}--\pageref{lastpage}}
\maketitle

\begin{abstract}
This work aims to derive the physical properties of the CNSFRs in the ring of the face-on spiral NGC 7742 using IFS observations. We have selected 88 individual ionising clusters that power HII regions populating the ring of the galaxy that may have originated in a minor merger event. For the HII regions the rate of Lyman continuum photon emission is between 0.025 and 1.5  10$^{51}$ which points to these regions being ionised by star clusters. Their electron density, ionisation parameter, filling factor and ionised hydrogen mass show values consistent with those found in other studies of similar regions and their metal abundances as traced by sulphur have been found to be between 0.25 and 2.4 times solar, with most regions showing values slightly below solar. 
The equivalent temperature of the  ionising clusters is relatively low, below 40000 K which is consistent with the high elemental abundances derived. The young stellar population of the clusters has contributions of ionising and non-ionising populations with ages around 5 Ma and 300 Ma respectively. The masses of ionising clusters once corrected for the contribution of underlying non-ionising populations were found to have a mean value of 3.5 $\times$ 10$^4$ M$_{\odot}$, comparable to the mass of ionised gas and about 20 \% of the corrected photometric mass. 

\end{abstract}

\begin{keywords}
galaxies: abundances -- galaxies: ISM  -- galaxies: star clusters: general -- galaxies: starburst -- ISM: abundances -- Nebulae, (ISM:) H II regions -- Nebulae 
\end{keywords}

%%%%%%%%%%%%%%%%%%%%%%%%%%%%%%%%%%%%%%%%%%%%%%%%%%
\section{Introduction}
\label{sec:introduction}

Careful determinations of abundance distributions over galaxies at the early stages of their evolution could provide important pieces of information about their formation processes since gas accretion or gas ejection episodes will leave an imprint on these abundance distributions. However, several important caveats exist: (a) the assumption that the ionisation of the gas in inner regions of galaxies is due only to star formation processes; (b) that the star formation modes dominating at high redshifts are similar to those encountered in the local universe. Regarding the first, it is nowadays generally accepted that some connection exists between star formation and activity in galactic nuclei, and young stars appear as one component of the unified model of AGN giving rise to the blue featureless continuum which is observed in Seyfert 2 galaxies where the broad line region is obscured. But regarding the second, that is that the star formation modes dominating at high redshifts are similar to those encountered in the local universe, this might not be the case. Recently, large and massive clumps of star formation have been detected in more than half of the resolved z > 1 galaxies in the Hubble UDF \citep[see][]{2005ApJ...627..632E}. These star-forming entities are in galaxies at all distances covered by the ACS (0.07 < z < 5). They have sizes of about 2 kpc, estimated ages of ~10 Ma and masses often larger than 10$^8$ M$_\odot$. They are so luminous that they dominate the appearance of their host galaxies. Massive clumps like these are found in galaxies with a variety of morphologies, from somewhat normal ellipticals, spirals, and irregulars, to types not observed locally, including chain galaxies and their face-on counterparts, clump-cluster galaxies.

Interestingly enough these star-forming entities which seem to constitute the star formation mode in galaxies at high redshifts resemble the well known circumnuclear star-forming regions (CNSFRs), a common mode of star formation found close to galactic nuclei. These regions, many of them a few hundred pc in size and showing integrated H$\alpha$  luminosities which overlap with those of HII galaxies (typically higher than 10$^{39}$ erg s$^{-1}$), seem to be composed of several HII regions ionised by luminous compact stellar clusters whose sizes, as measured from high spatial resolution Hubble Space Telescope (HST) images, are seen to be of only a few pc. These regions are young (age < 10 Ma and massive (up to 2 $\times$ 10$^8$ M$_\odot$) \citep{Hagele2007,Hagele2013}. In the UV-B wavebands, they contribute substantially to the emission of the entire nuclear region, even in the presence of an active nucleus \citep[see e.g.][]{2002ApJ...579..545C}. In a galaxy like NGC~3310 the starburst “ring” is the strongest organized source of far-UV (FUV) emission and 30\% of the total observed FUV emission is produced within a radius of 10”. At redshifts of z ~ 2–3, this structure would be confined to a region 0.2” in diameter for $\Omega$ = 1 and would appear point-like in low-resolution observations. Consequently, in the absence of diagnostic spectroscopy, a high-redshift NGC~3310–like object  could be mistaken for an active galactic nucleus (AGN). 

CNSFRs in nearby galaxies, being close to the galactic nuclei, are expected to be of high metal abundance. However, detailed long-slit spectroscopic analyses show that most of them have abundances consistent with solar values \citep{2007MNRAS.382..251D}. Also, their ionisation structure as mapped by suitable emission line ratios is more similar to that of HII galaxies than to galactic disc GEHR, pointing to relatively hard ionising sources, not expected at high metallicities. As mentioned above, similar effects have been found for a considerable sample of star-forming galaxies at 1.0<z<1.5 \citep[see e.g.][]{2008ApJ...678..758L}. The answer might be related to the influence of a hidden low luminosity AGN, the presence of shocks in zones of high specific star formation rates, or harder ionizing continuum sources among other possibilities. 

The above mentioned work of \citet{2007MNRAS.382..251D} was based on long-slit spectroscopy, which is very time consuming, and involved a few CNSFRs in three selected galaxies with the total number of regions studied amounting to a dozen. Obviously, the best strategy to study the complex star forming regions in circumnuclear rings is the use of Integral Field Spectroscopy (IFS).  The Multi-Unit Spectroscopic Explorer (MUSE) available at VLT offers the opportunity to carry out this detailed study program. Typical circumnuclear rings have sizes of less than 1kpc (20 arcsec at a distance of 10 Mpc), hence are easily accommodated in the large field of view of MUSE (1 arcmin$^2$) which also provides the necessary combination of high spatial (0.3 - 0.4 arcsec) and spectral resolution (R $\simeq$  2000 - 4000). The use of this technique greatly reduces the observing time and can increase the number of analysed clusters by an order of magnitude. On the other hand, the usually high abundances of the objects involved and their low excitation produce very weak [OIII] lines, difficult to measure with confidence, precluding the use of these lines for the analysis. The extended range of wavelength to the red provided by MUSE allows the use of Sulphur as an alternative abundance and excitation tracer for the characterisation of the HII regions and ionising clusters \citep[see ]{2022MNRAS.511.4377D}.

In this first paper we present the study of the physical properties of the CNSFRs in the ring of the face-on galaxy NGC~7742 using MUSE observations publicly available and the full spectral region observed, from 4800 to 9300 Å. NGC~7742 is classified as an SA(r)b galaxy. It has a weakly active nucleus classified as T2/L2 in \citet{1997ApJS..112..315H}, that corresponds to a transition object whose spectrum is dominated by emission lines characteristic of both LINER and HII regions. Its morphology is dominated by a nuclear ring which is easily identified by prominent bumps on the luminosity profiles in different photometric bands at galactocentric distance between 9 and 11 arcsec which corresponds to around 1 kpc at the assumed distance of 22 Mpc \citep{1988cng..book.....T}. The feature shows up most importantly in the U band thus pointing to a star formation origin \citep{1996ASPC...91...83W}. Apart from the ring, the surface brightness profile can be represented by the combination of two exponential discs and one central bulge \citep[][]{2006AJ....131.1336S}. The galaxy shows a high degree of circular symmetry at different spatial levels: core, ring, and main body, and hence constitutes a very good case for the study of formation mechanisms of nuclear rings in non-barred galaxies. It is also one of the approximately 10\% spirals showing a gaseous counter-rotating disc \citep{2004A&A...424..447P}, that was first reported by \citet{2002MNRAS.329..513D}. 

The kinematics of gas and stars in the central part of this galaxy has been studied in detail by \citet{Contrarrotante}, also using MUSE data. In their work they have mapped the ring counter-rotation and have found evidence for two distinct stellar populations: the older of them counter-rotates with the gas while the younger one, concentrated to the ring, co-rotates with the gas. They conclude that the ring has been originated in a minor merger event that took place probably 2-3 Ga ago.

The present work is centred in the study of the individual ionising clusters that power the HII regions populating the ring of NGC~7742. The observations on which the work is based are presented in section 2 together with the description of the data reduction; section 3 is devoted to the description of the measurement methods and the data analysis; section 4 presents the results; the discussion is given in section 5 and our final conclusions are in Section 6.

\section{Observations and data reduction}
\label{observations}

In this work we analyse the almost face-on galaxy NGC~7742 that shows a prominent circumnuclear star-forming ring using publicly available observations obtained by the IFS MUSE.  Some characteristics of this galaxy are given in Table \ref{tab:galaxy characteristics}.

%%%%%%%%%%%%%%%%%%%%%%
\begin{table}
\centering
\caption{NGC~7742 global properties.}
\label{tab:galaxy characteristics}
\begin{tabular}{cc}
\hline
Galaxy & \href{https://ned.ipac.caltech.edu/byname?objname=NGC7742&hconst=67.8&omegam=0.308&omegav=0.692&wmap=4&corr_z=1}{NGC~7742} \\ \hline
RA J2000 (deg)$^a$ & 356.065542\\ 		
Dec J2000 (deg)$^a$ & 10.767083\\ 
Morphological type & SA(r)b\\ 
Luminosity Class & LC II\\ 	
Nuclear type & LINER/HII \\
z & 0.00555\\ 
Distance (Mpc)$^b$ & 22.2 \\
Scale (pc/arcsec)$^c$ & 92\\ 	
\hline
\end{tabular}
\begin{tablenotes}
\item $^a$ \citet{2006AJ....131.1163S}.\\
\item $^b$ \citet{1988cng..book.....T}.\\
\item $^c$ Cosmology corrected scale.
\end{tablenotes}
\end{table}
%%%%%%%%%%%%%%%%%%%%%

The Multi-Unit Spectroscopic Explorer, MUSE \citep{MUSE} is an integral-field spectrograph (IFS) located at the Nasmyth focus of the Very Large Telescope (VLT) on the Unit Telescope 4 (UT4), of the European Southern Observatory (ESO) at Cerro Paranal, Chile. It operates in the visible wavelength range, covering from 4800 \AA\ to 9300 \AA\ with a nominal dispersion of 1.25 \AA /pixel with a spectral resolving power from 1770 (at 4800 \AA ) to 3590 (at 9300 \AA ) in the blue and red arms respectively. It is composed of 24 integral field units (IFUs) which, in the Wide Field Mode (WFM), provides a field of view (FoV) of 60 arcsec$^2$ with a spatial sampling of 0.2 arcsec$^2$. 

NGC~7742 was observed as part of the first MUSE Science Verification run on 2014 June 22 under ESO Programme \href{http://archive.eso.org/wdb/wdb/eso/sched_rep_arc/query?progid=60.A-9301(A)}{60.A-9301(A)} (PI: M. Sarzi). The observing time was split in two exposures of 1800s with an offset of 1 arcsec in declination and a rotation of 90º between observations and with a median seeing of 0.63 arcsec. Offset sky observations were taken before or after the target observations for adequate sky subtraction. 

We have used the ESO Phase 3 Data Release. The reduction of the data was performed by the Quality Control Group at ESO in an automated process applying version 0.18.5 of the MUSE pipeline \citep{MUSEpipeline}. They used calibration images taken as part of the standard MUSE calibration plan using different pointings of the object. The corrections applied to each exposure were: subtraction of the master-bias, division by a master-flat-field and illumination correction between all slices of the IFU. Corrections for twilight and differential atmospheric refraction were also made. Data were wavelength calibrated, corrected for telluric absorption and sky subtracted using a dedicated offset exposure of 300s. Finally, the data were flux calibrated. The astrometric solution was provided and the data were resampled into a datacube. Finally, the produced individual datacubes were weighted by their respective exposure time and resampled into a single combined one. 

We have also used additional data from Hubble Space Telescope (HST) that were acquired on 1995 July 9 with the Wide Field and Planetary Camera 2 (WFPC2) as a part of the program \href{ https://www.stsci.edu/cgi-bin/get-proposal-info?observatory=HST&id=6276}{GTO/wfc 6276} (IP: J. Westphal) providing high-resolution images with a spatial resolution  of $\simeq$ 0.1 arcsec
pixel$^{-1}$ and FoV of 150 arcsec$^2$. The data have been obtained from the Hubble Legacy Archive and are organised in 3 exposures of 700 s each in the U broad band obtained with the F336W filter.  The reduction of these data has been performed by the Space Telescope Science Institute (STScI) using available calibration files taken for this observation and keeping in mind different dithering positions. The pipeline provides standard calibrations like correction for permanent camera defects, the temperature-dependence of the WF4 detector gain, bias, dark current and flat field corrections, the position-dependent exposure time and the absolute detector efficiency. Additionally, STScI has reprocessed all WFPC2 data including improvements to the time-dependent UV contamination, the variation in the bias level in WF4 and other subtle details.

\section{Results and analysis}
\label{sec:results}

\subsection{Emission line and continuum maps}
%%%%%%%%%%%%%%%%%%%%%%%%
\begin{figure*}
\includegraphics[width=\textwidth]{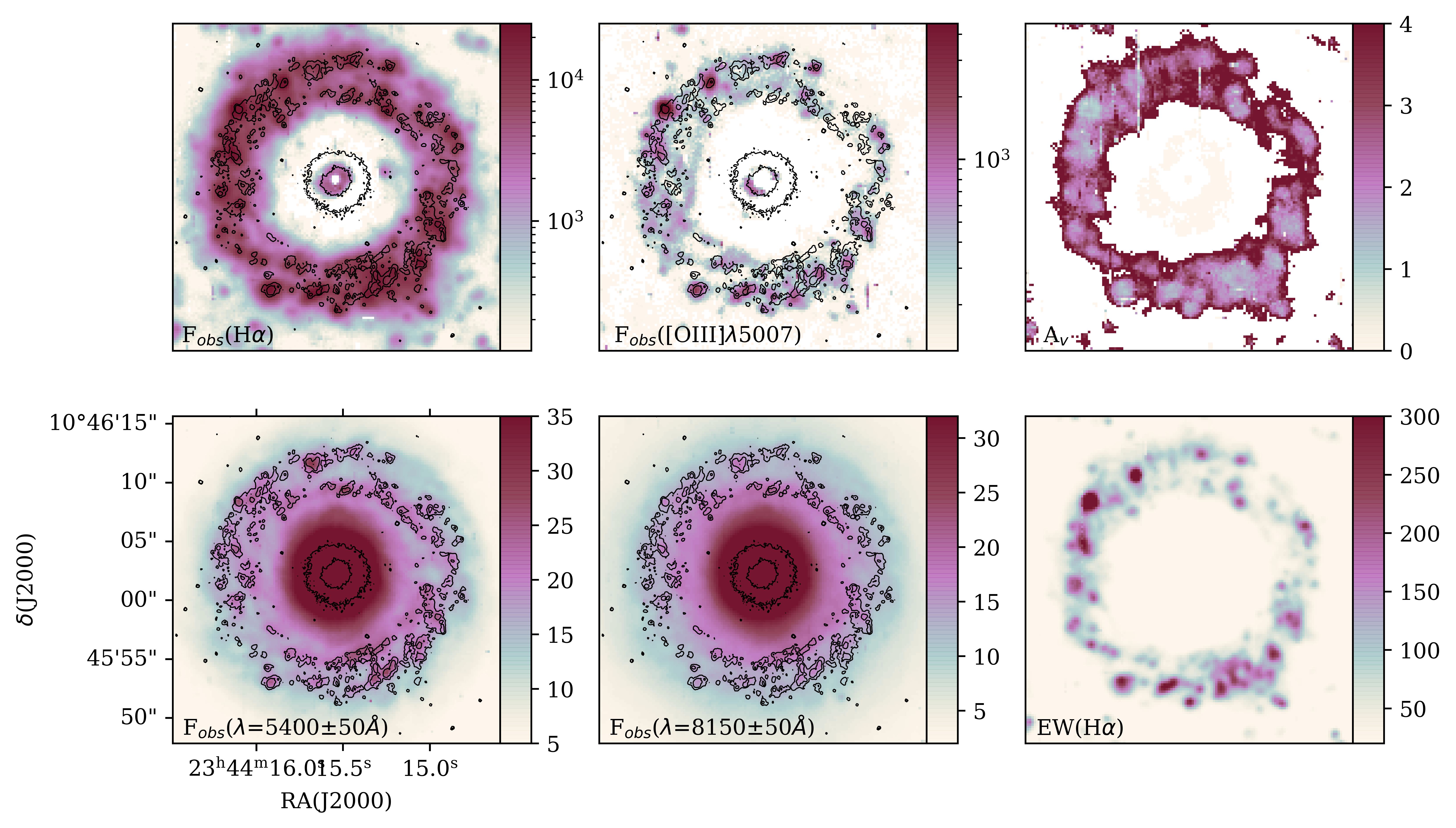}
\caption  {From left to right and top to bottom: Maps of the observed H$\alpha$ and [OIII]$\lambda$5007 \r{A} fluxes (in units of 10$^{-20}$ erg/s/cm$^2$ and logarithmic scale); A$_v$ extinction (in magnitudes); maps for the observed continuum in the blue and red spectral bands (5400 \AA\ and 8150\AA\ respectively, in units of 10$^{-17}$ erg/s/cm$^2$ and lineal scale); and EW($H\alpha$) in \AA . Upper and bottom left and centre images show superimposed contours of the HST-UV image which has been described in the text. Orientation is north up, east to the left.}
\label{fig:Ha_OIII_map}
\end{figure*}
%%%%%%%%%%%%%%%%%%%%%%%%
%%%%%%%%%%%%%%%%%%%%%%%%
\begin{table}
\centering
\caption{Extraction parameters for emission line maps.}
\label{tab:line ranges}
\begin{tabular}{lcccc}
\hline
Line &  $\lambda_c$ (\r{A})  & $\Delta\lambda$ (\r{A}) & $\Delta\lambda_{left}$ (\r{A}) & $\Delta\lambda_{right}$ (\r{A}) \\ 
\hline
 H$\alpha$ & 6563 & 8& 6531.5 - 6539.5 & 6597.0 - 6605.0 \\
 H$\beta$  & 4861 & 8& 4811.0 - 4819.0 & 4901.0 - 4909.0 \\
 $[OIII]$ & 5007 & 15& 4902.5 - 4917.5 & 5092.5 - 5107.5 \\
 $[NII]$ & 6583& 15 &6593.5-6608.5 & 6542.5-6527.5\\
\hline
\end{tabular}
\begin{tablenotes}
\item All wavelengths are in rest frame.
\end{tablenotes}
\end{table}
%%%%%%%%%%%%%%%%%%%%%%%%%

From the observed data cubes we have constructed 2D maps which are presented in Figure \ref{fig:Ha_OIII_map}. For the different emission lines we have assumed a linear behavior of the continuum emission in the region of interest choosing side-bands around each line of a given   width. Table \ref{tab:line ranges} give the identification of each line in column 1, its centre wavelength, $\lambda_c$ in \AA\ in column 2, its width, $\Delta\lambda$, in \AA\ in column 3, and the limits of the two continuum side-bands, in \AA\ in columns 4 and 5. The H$\alpha$ and H$\beta$ maps have been combined to produce an extinction map. We have also produced maps in two continuum bands of 100 \r{A} width centered at 5400 \r{A} (blue) and 8150 \r{A} (red).
All wavelengths are in rest frame.

The two top left panels of Fig. \ref{fig:Ha_OIII_map} show the spatial distribution of the observed H$\alpha$ and [OIII] fluxes in logarithmic scale. Superimposed on these maps we have represented the contours of HST data from WFC2 in the F336W filter where young star clusters would be more conspicuous. This provides a good comparison between the spatial resolution provided by the two instrumental configurations. The coincidence between the young clusters identified by the HST contours with the MUSE maps, specially in the H$\alpha$ one ensures that we are actually detecting ionising star clusters. On this H$\alpha$ map the diffuse gas of the circumnuclear ring is also clearly seen. On the other hand, the [OIII]$\lambda $5007 \r{A} emission, traces the different excitation conditions of the ionised gas across the ring.  The extinction $A_V$, in magnitudes, is shown in the top right panel of the figure and  has been calculated by adopting the Galactic extinction law of \citet{reddening}, with a specific attenuation of R$_v$ = 2.97 and the theoretical ratio H$\alpha $/H$\beta $ = 2.87 from \citet{Osterbrock2006} (n$_e$ = 100 cm$^{-3}$, T$_e$ = 10$^4$ K, case B recombination) (see also Section \ref{sec:line measurements}). The distribution of the gas extinction seems to be smooth inside clumps and typically low, with a median value of 1.053 mag in the pixel by pixel analysis peaking around 1.83 mag.  The apparently higher extinction values at the edges of the HII regions could be due to the low S/N ratio in the H$\beta$ emission line that produces an artificially high H$\alpha$/H$\beta$ ratio with a large uncertainty. Alternatively, it could be the result of dust created by intense star formation and accumulated by the gas expansion.  The two bottom left panels of Fig. \ref{fig:Ha_OIII_map} show maps of observed continuum fluxes at blue and red wavelengths, 5400\AA\ and 8150\AA\ respectively, on top of which the contours of HST-WFC3 data in the F336W filter. Exponential fits to the stellar surface brightness show two different components with different scale-lengths \citet{2006AJ....131.1336S}. In both maps, HII regions contrast over the galaxy profile and seem to follow the clusters identified in UV emission. Finally, in the bottom right panel we can see the map of the equivalent width (EW) of $H\alpha$ (in \AA ). All circumnuclear regions present in the ring have EW(H$\alpha$) > 20 \AA . This value is consistent with the presence of star formation occurred less than 10 Ma ago.

\subsection{HII region selection}\label{sec:segmentation}

%%%%%%%%%%%%%%%%%%%%%%%%%%%%
\begin{figure}
\centering
 \includegraphics[width=\columnwidth]{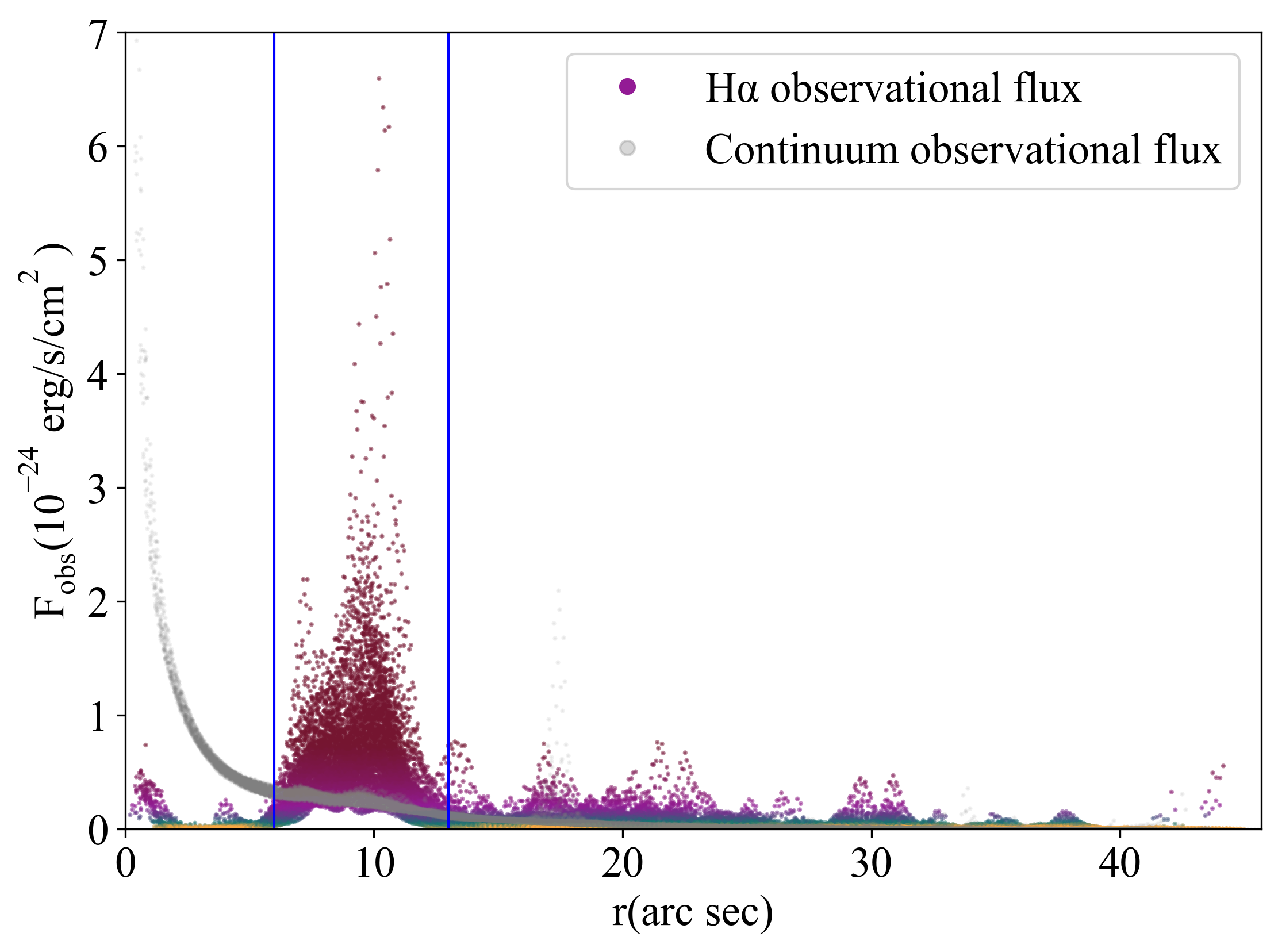}
 \includegraphics[width=0.87\columnwidth]{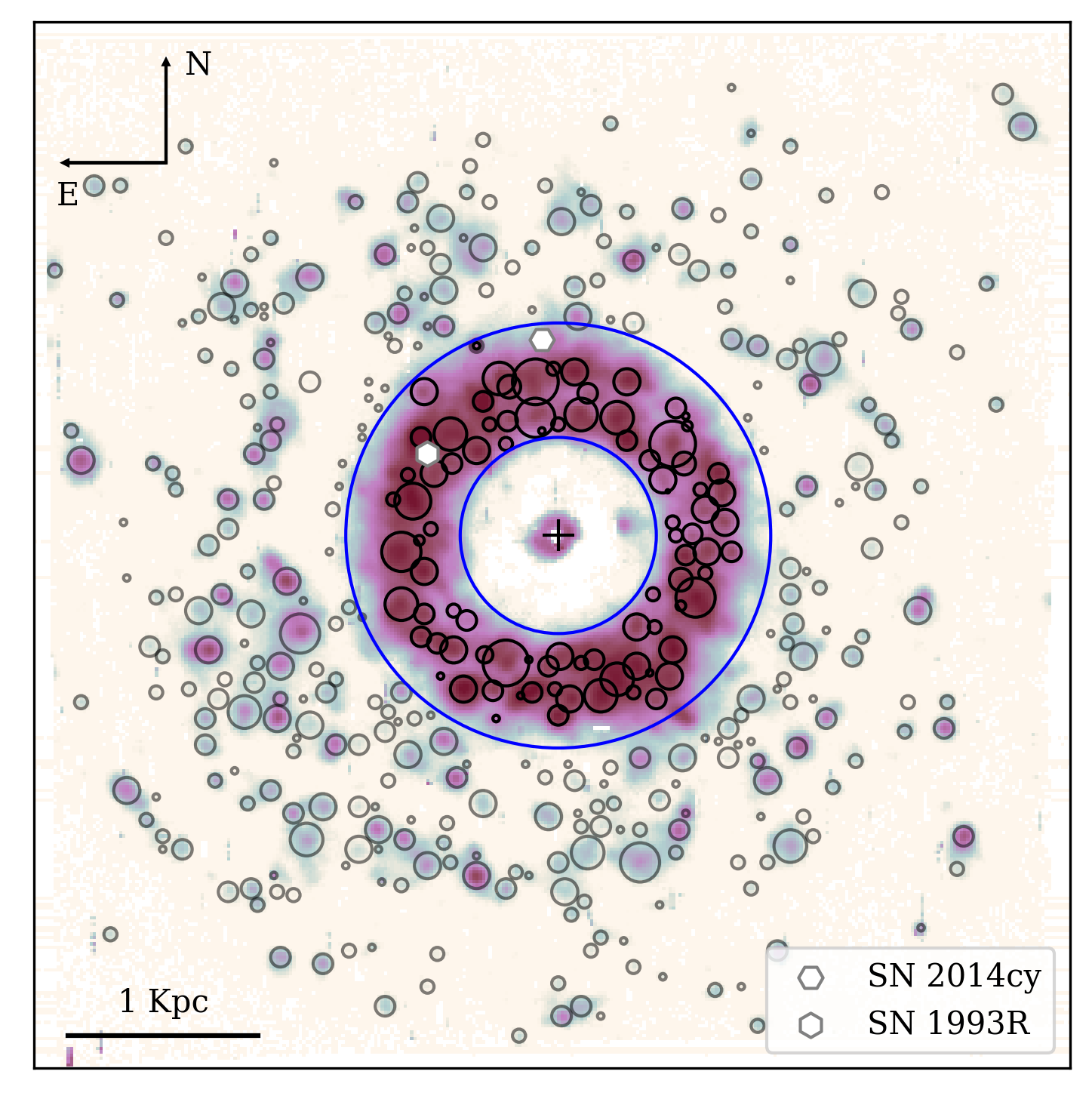}
 \caption{Upper panel: Flux in individual spaxels as a function of radius: continuum near H$\alpha$ with grey dots, integrated flux of this line with colored dots and limits of the ring marked with blue vertical lines. Lower panel: HII regions selected with our segregation program. SN 2014cy and SN 1993R are also plotted (see text). The logarithmic color scale is equivalent with colours presented in the top panel. Orientation is north up, east to the left. The physical scale is represented at the bottom left corner of the map. The limits of the ring are marked with blue circles.}
 \label{fig:ring_ha_profile}
\end{figure}
%%%%%%%%%%%%%%%%%%%%%%%%%%%%

Reddening spaxel-by-spaxel analysis has a high uncertainty due to the low S/N of hydrogen H$\beta$ emission (see the top right panel of Fig. \ref{fig:Ha_OIII_map}). However, we can observe that the reddening is very similar for all ring regions. Thus, we have decided to use the observed H$\alpha$ flux map, i. e. without correcting for reddening, to select ionised regions. In this case, binning is not necessary since a higher S/N is not required and we can preserve the spatial resolution. Furthermore, we do not introduce additional errors in our subsequent analysis.
We have selected the spatial extension of the ring on the basis of the spaxel-by-spaxel radial distribution of the observed H$\alpha$ flux shown in the top panel of Figure \ref{fig:ring_ha_profile}. The area that belongs to the circumnuclear ring can be seen as a bump, in dark-red colour, with H$\alpha$ emission excess over the adjacent continuum, in grey. The limits of the ring are marked on the figure by vertical lines.  It has an inner radius of 6 arcsec, 0.75 kpc at the chosen distance of 22.2 Mpc, and an outer radius of 13 arcsec, 1.63 kpc (see Tab. \ref{tab:galaxy characteristics}). Previous studies report similar radial ring limits \citep[$\sim$ 1 kpc;][]{Comeron2010,Contrarrotante}.

Our selection method for the HII regions has been made on the basis of the existing \href{http://www.caha.es/sanchez/HII_explorer/}{HII$_{EXPLORER}$} package, proposed by \citet{Sanchez2012}. This program works on a line emission map, usually H$\alpha$, and detects high intensity clumps, starting with the brightest pixel and adding adjacent ones following specific criteria. It requires several input parameters: the maximum size of the regions, the absolute flux intensity background for all of them (diffuse gas emission level) and the relative flux intensity at maximum to the background for each of the regions. This procedure has already been used in a series of IFS studies, related to CALIFA and MANGA data, but it tends to select regions with similar sizes \citep{2016AA...591A..48G}. However, HII regions exhibit different sizes, something that shows up in the higher spatial resolution MUSE data, and hence the HII$_{EXPLORER}$ package is not appropriate. This fact has already been remarked by \citet{Galbany2016} and here below we describe the implementations we have made to the original package in order to tackle this problem. 

We have developed a specific software with the same iterative procedure as the one followed in HII$_{EXPLORER}$ but with some additional requirements. We have selected the maximum extent of regions according to their typical projected size at z $\sim$ 0.016 \citep[][500pc]{Gonzalez-Delgado1997,Lopez2011}, setting a minimum for it according to the point spread function (PSF) value of the input map. We have assumed spherical symmetry and we have modulated the radius of the different regions according to their brightness. We have set an individual flux intensity threshold to each region setting a limit of 10\% with respect to the emission of its centre since an asymptotic behavior was found by \citet{Angeles2000} in H$\alpha$ intensity in their study of CNSFRs. Finally, we have tried different values for the absolute flux intensity background of the complete map adopting that resulting of the best fit of the program, i.e. the one that minimises the dispersion of the spatial residuals in the map.

Apart from the study of the ring HII regions as described above, we have also analysed HII regions external to it for comparison purposes. In order to do that we have ran our segmentation two times, a first one for circumnuclear regions and a second one for those in the outer limit to the ring. The two procedures are slightly different since the absolute flux intensity background is much larger in the further (the diffuse gas emission in the ring is higher). The second panel of Fig.\ref{fig:ring_ha_profile} shows the HII regions selected with the use of the methodology described above. Two supernovae, SN 1993R and SN 2014cy, have also been plotted at their respective positions \citep[][respectively]{1993IAUC.5812....1T,2014CBET.3964....1N}. SN 1993R is a peculiar supernova, similar to SN Ia class 1991bg, but with stronger CaII triplet lines, a weak emission of [OI]$\lambda$630nm  and detected in X-Ray emission \citep{1993IAUC.5842....2F,2003ApJ...596..323B}. Also, it is superimposed on a very bright HII region which has the highest values of H$\alpha$ emission, [OIII]$\lambda$5007 emission, A$_v$ and EW(H$\alpha$) (see Fig. \ref{fig:Ha_OIII_map}). SN 2014cy was classified as SN II and, in the figure, is on top of a region with characteristics similar to the rest of the sample. 

Finally, in order to further select data with high quality and also discard failures from the method, we have imposed the following requirements to the integrated spectrum extracted from each selected region:
(i) to be certain that the emission has a star formation origin, EW(H$\alpha$) must to be higher than 6 \AA\ \citep{CidFernandez2010,Sanchez2015};
(ii) to claim the extracted spectrum has physical meaning, the ratio H$\alpha$/H$\beta$ must be between 2.7 and 6.0, which corresponds to the theoretical values from \citet{Osterbrock2006} (assuming an electron density of n$_e$ = 100 cm$^{-3}$ and an electron temperature of T$_e$ = 10$^4$ K) and an extinction up to 2.3 mag respectively. 

%%%%%%%%%%%%%%%%%%%%%%%
\begin{table*}
\centering
\caption{Selection characteristics for observed CNSFRs. The complete table is available online; here only a part is shown as an example.}
\label{tab:seleccion}
\begin{tabular}{cccc}
\hline
Region ID & \begin{tabular}[c]{@{}c@{}} Area \\ (arcsec$^2$)\end{tabular} & \begin{tabular}[c]{@{}c@{}}Offsets from galaxy center $^a$ \\ (arcsec)\end{tabular} & \begin{tabular}[c]{@{}c@{}}F(H$\alpha$)\\ (10$^{-15}$ erg$\cdot s^{-1}\cdot cm^{-2}$)\end{tabular} \\ \hline
R1*	&	1.48	&	-8.4, 6.0	&	17.621 $\pm$ 0.023\\
R2	&	1.40	&	-4.6, 8.2	&	10.138 $\pm$ 0.019\\
R3	&	4.28	&	-8.9, 2.1	&	19.753 $\pm$ 0.053\\
R4	&	2.28	&	-5.8, -9.4	&	10.306 $\pm$ 0.024\\
R5	&	1.44	&	-1.6, -9.6	&	8.012 $\pm$ 0.017\\
R6	&	0.20	&	-2.3, -9.8	&	1.426 $\pm$ 0.003\\
R7	&	2.72	&	7.0, -7.0	&	11.355 $\pm$ 0.026\\
R8	&	4.12	&	8.4, -3.8	&	15.943 $\pm$ 0.044\\
R9	&	4.32	&	2.6, -9.8	&	16.294 $\pm$ 0.040\\
R10	&	2.88	&	3.6, -8.8	&	14.017 $\pm$ 0.030\\
\hline
\end{tabular}
\begin{tablenotes}
\centering
\item $^a$ Offsets from centre of the galaxy to the centre of each individual region.
\item * Region near SN explosion.
\end{tablenotes}
\end{table*}

%%%%%%%%%%%%%%%%%%%%%%%

At the end of the entire procedure, we have obtained a total of 88 HII regions in the ring and 158 regions outside it. Table \ref{tab:seleccion} shows the position of each HII region in the ring, with respect to that of the galaxy centre, its size and its integrated observed H$\alpha$ emission flux. The identification of each region is given in column 1 of the table. SN 1993R lies close to R1.

\subsection{Emission line measurements and uncertainties}\label{sec:line measurements}

%%%%%%%%%%%%%%%%%%%%
\begin{figure*}
 \includegraphics[width=2\columnwidth]{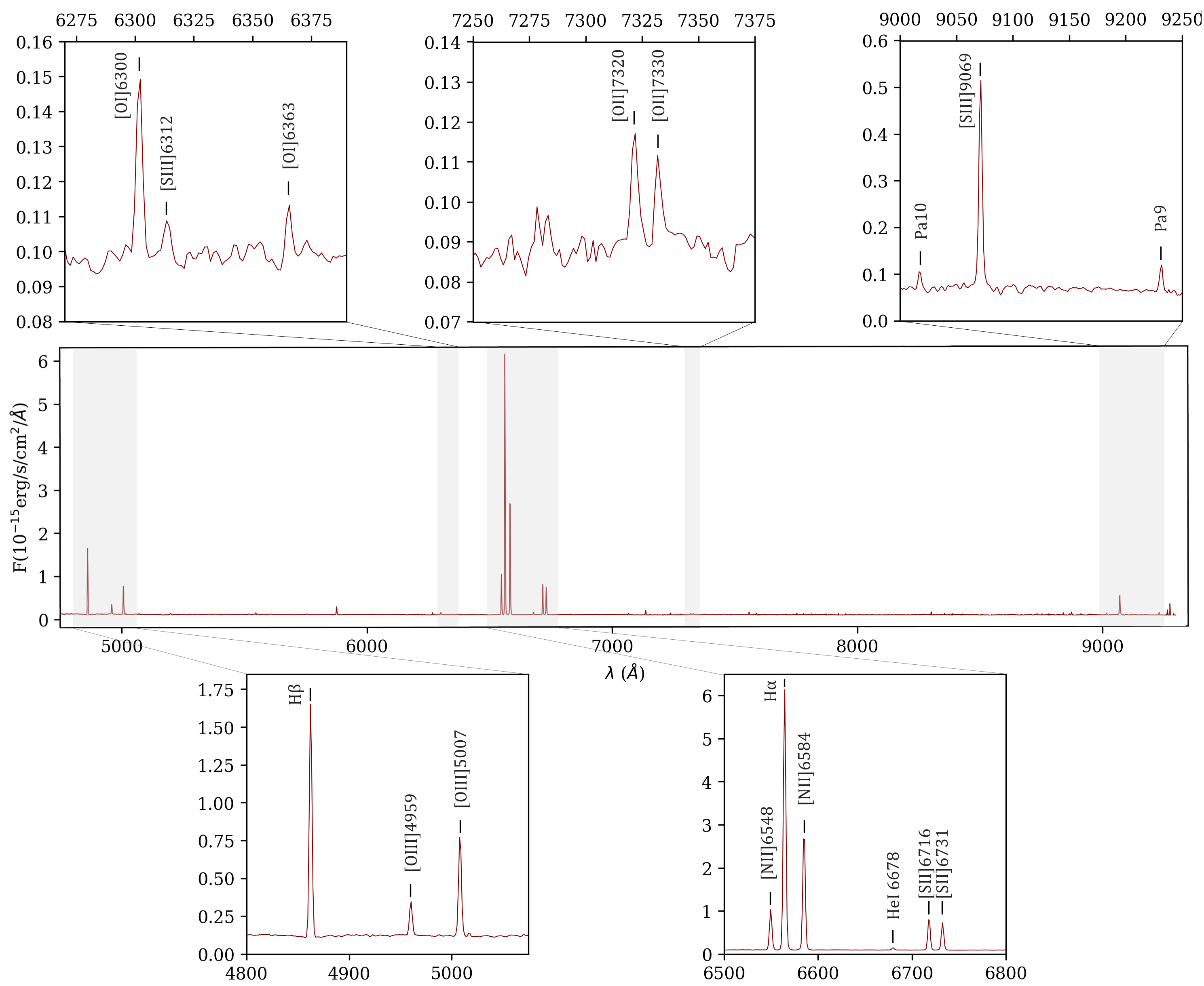}
 \caption{Extracted and reddening corrected spectrum of region R1. Flux is expressed in units of $10^{-15} erg/s/cm^2/$\AA . }
 \label{fig:espectro_muse}
\end{figure*}
%%%%%%%%%%%%%%%%%%%%

We have extracted each region spectrum, integrating its corresponding flux in every single aperture produced by the segregation, except in the case of the weak  [SIII]$\lambda$ 6312 \AA\ line for which we have integrated only pixels with S/N > 1.0, as described below. Fig. \ref{fig:espectro_muse} shows one of these spectra. An underlying stellar population is slightly appreciable in some of our spectra. In order to correct for this effect, we have fitted a Gaussian to the underlying absorption both in the H$\beta$ and H$\alpha$ lines and subtracted it from the extracted spectra. For the brightest region, this correction has been found to be less than 3\% of the observed flux which reflects in a contribution to the measured  H$\beta$ flux within the observational errors. 

For each region spectrum, a global continuum has been estimated by fitting a second order polynomial, $F_c(\lambda)$ = $a\lambda^2+b\lambda +c$ after masking nebular and stellar features. The masks have been built assuming a width of 8\r{A} at each side the central wavelength of the lines involved. To obtain an accurate measure of line fluxes, we have estimated the standard deviation of the residuals of the global continuum fit ($\sigma _c$). After subtracting the global continuum, the measurement of fluxes is performed using a single Gaussian fit plus a linear term:
\begin{equation}
f(\lambda)=  A_g\cdot e^{-\frac{(\lambda -\lambda _g)^2}{2\sigma_g^2}}+A_c
\end{equation}
\noindent where A$_g$, $\lambda_g$ and $\sigma_g$ are the amplitude, central wavelength and width of the fitted Gaussian. The linear term, $A_c$, appears as a correction to the global continuum value close to each line. It can take values between $[-\sigma_c$, $+\sigma_c]$ taking a given value for each measured line. To determine the error of this measurement, and impose quality conditions to it, we have calculated the local standard deviation of the residuals of the Gaussian fit ($\sigma _l$) in 30\r{A} around each line.  The wavelength window selected for this modelling is adjusted for each line at the first point compatible with $\sigma _c$ at each side of its central wavelength.

Using this procedure, we have measured the most prominent emission lines in our spectra: H$\beta$ and H$\alpha$ Balmer lines; [OIII]$\lambda\lambda$ 4959,5007 \AA , [NII]$\lambda\lambda$ 6548,84 \AA , [SII]$\lambda\lambda$ 6716,31 \AA , [ArIII]$\lambda$ 7136 \AA\ an [SIII]$\lambda$ 9069 \AA\ forbidden lines. We have taken into account only fluxes of the lines that meet the requirement: $A_g >3\sigma_l$, thus discarding the most uncertain values. Additionally, we have measured the weak HeI$\lambda$ 6678 \AA\ line with $A_g >1\sigma_l$. In the case of [SIII]$\lambda$ 6312 \AA\ and [OII]$\lambda\lambda$ 7320,30 \AA , their fluxes have been measured on the spectrum extracted by integration of the pixels where these lines are detected with a tolerance larger than 1$\sigma_l$. The [SIII] line has been finally measured with sufficient accuracy in 40 ring HII regions while only 13 regions allowed accurate measurements for the [OII] lines. They have been finally measured with sufficient accuracy in 40 and 13 ring HII regions respectively.

The errors in the observed fluxes have been calculated from the expression given in \citet{Gonzalez-Delgado1994}:
\begin{equation}
\Delta [F_\lambda] =\sigma _l \cdot N^{1/2}[1+EW/(N \Delta)]^{1/2}
\end{equation}
where $\Delta [F]$ is the error in the line flux, $\sigma _l$ represents the standard deviation in the local continuum, N is the number of pixels used in the Gaussian fit, $\Delta$ is the wavelength dispersion (1.25 \r{A}/pix) and EW is the line equivalent width.  The mean value of continuum fluxes, $F_c(\lambda)+A_c$, in the wavelength range  [$\lambda _{line}-\sigma _g$, $\lambda _{line}+\sigma _g$] has been used to compute this latter value. 

Regarding the effects of reddening, we have used a simple screen distribution of the dust and have assumed the same extinction for emission lines and the stellar continuum. The measured line intensities have been corrected using a reddening constant c(H$\beta$), derived from the observed H$\alpha $/H$\beta $ ratio, adopting the Galactic extinction law of \citet{reddening}, with a specific attenuation of R$_v$ = 2.97. A theoretical value for the H$\alpha $/H$\beta $ ratio of 2.87 has been assumed \citep[][for n$_e$ = 100 cm$^{-3}$ and T$_e$ = 10$^4$ K]{Osterbrock2006}. Given the wavelengths of the lines involved in our study, we have measured their intensities with respect to  H$\alpha$ which has a high S/N and is less affected by underlying absorption and reddening effects, thus providing a more precise measurement. Therefore the extinction correction has been applied using the following equation:

\begin{equation}
 log \left(\frac{I(\lambda)}{I(H\alpha)}\right) =   log \left( \frac{F(\lambda)}{F(H\alpha)}\right)+c(H\beta)\cdot (f(\lambda) - f({H\alpha}))
\end{equation}
where c(H$\beta$) is the reddening constant, f($\lambda$) gives the value of the logarithmic extinction normalised to H$\beta$, and F$_\lambda$ and I$_\lambda$ are the observed and corrected emission line fluxes at wavelength $\lambda$ respectively. Then the I($\lambda$)/I(H$\alpha$) ratio has been translated to I($\lambda$)/I(H$\beta$) assuming the aforementioned value of 2.87 for the theoretical H$\alpha$/H$\beta$ Balmer decrement. The corresponding errors have been propagated in quadrature. 

\begin{table*}
\centering
\caption{Reddening corrected emission line intensities. The complete table is available online; here only a part is shown as an example.}
\label{tab:lines}
\begin{tabular}{cccccccccccc}
\cline{2-12}
& Line & H$\beta$ & [OIII] & [OIII] &  [NII] & H$\alpha$ & [NII] & HeI & [SII] & [SII] & [SIII]\\
& $\lambda$ & 4861 & 4959 & 5007  & 6548 & 6563 & 6584 & 6678 & 6717 & 6731 & 9069 \\
& f($\lambda$) & 0.000 & -0.024 & -0.035 & -0.311 & -0.313 & -0.316 & -0.329 & -0.334 & -0.336 & -0.561  \\ \hline
\multicolumn{1}{c}{Region ID} & \multicolumn{1}{c}{c(H$\beta$)} & \multicolumn{1}{c}{I(H$\beta$)$^a$}& \multicolumn{9}{c}{I($\lambda $)$^b$} \\ \hline
R1* & 0.44 $\pm$ 0.01 & 12.33 $\pm$ 0.24 & 158 $\pm$ 3 & 444 $\pm$ 4 & 438 $\pm$ 2 & 2870 $\pm$ 24 & 1342 $\pm$ 3 & 22 $\pm$ 0 & 334 $\pm$ 2 & 284 $\pm$ 2 & 246 $\pm$ 3\\ 
R2 & 0.58 $\pm$ 0.01 & 8.85 $\pm$ 0.24 & 147 $\pm$ 4 & 422 $\pm$ 5 & 461 $\pm$ 3 & 2870 $\pm$ 33 & 1423 $\pm$ 5 & 23 $\pm$ 1 & 367 $\pm$ 2 & 289 $\pm$ 2 & 210 $\pm$ 4\\ 
R3 & 0.47 $\pm$ 0.02 & 14.53 $\pm$ 0.57 & 85 $\pm$ 6 & 224 $\pm$ 7 & 421 $\pm$ 4 & 2870 $\pm$ 48 & 1295 $\pm$ 6 & 16 $\pm$ 1 & 460 $\pm$ 4 & 339 $\pm$ 3 & 113 $\pm$ 4\\ 
R4 & 0.39 $\pm$ 0.01 & 6.62 $\pm$ 0.22 & 101 $\pm$ 5 & 283 $\pm$ 6 & 442 $\pm$ 3 & 2870 $\pm$ 41 & 1372 $\pm$ 6 & 18 $\pm$ 1 & 525 $\pm$ 4 & 389 $\pm$ 3 & 119 $\pm$ 5\\ 
R5 & 0.46 $\pm$ 0.01 & 5.79 $\pm$ 0.19 & 103 $\pm$ 5 & 270 $\pm$ 6 & 439 $\pm$ 3 & 2870 $\pm$ 40 & 1359 $\pm$ 5 & 19 $\pm$ 1 & 367 $\pm$ 3 & 274 $\pm$ 3 & 151 $\pm$ 4\\ 
R6 & 0.66 $\pm$ 0.02 & 1.42 $\pm$ 0.06 & 91 $\pm$ 7 & 248 $\pm$ 8 & 442 $\pm$ 3 & 2870 $\pm$ 52 & 1373 $\pm$ 6 & 21 $\pm$ 1 & 364 $\pm$ 3 & 259 $\pm$ 2 & 150 $\pm$ 3\\ 
R7 & 0.44 $\pm$ 0.02 & 7.97 $\pm$ 0.29 & 81 $\pm$ 6 & 231 $\pm$ 7 & 387 $\pm$ 3 & 2870 $\pm$ 45 & 1182 $\pm$ 5 & 18 $\pm$ 1 & 415 $\pm$ 3 & 299 $\pm$ 3 & 141 $\pm$ 5\\ 
R8 & 0.42 $\pm$ 0.02 & 10.72 $\pm$ 0.44 & 82 $\pm$ 7 & 227 $\pm$ 7 & 411 $\pm$ 4 & 2870 $\pm$ 49 & 1263 $\pm$ 7 & 18 $\pm$ 1 & 435 $\pm$ 4 & 323 $\pm$ 4 & 121 $\pm$ 5\\ 
R9 & 0.56 $\pm$ 0.02 & 13.73 $\pm$ 0.6 & 102 $\pm$ 8 & 266 $\pm$ 8 & 419 $\pm$ 4 & 2870 $\pm$ 53 & 1292 $\pm$ 6 & 15 $\pm$ 1 & 445 $\pm$ 4 & 323 $\pm$ 3 & 114 $\pm$ 5\\ 
R10 & 0.41 $\pm$ 0.01 & 9.37 $\pm$ 0.34 & 58 $\pm$ 5 & 170 $\pm$ 6 & 367 $\pm$ 3 & 2870 $\pm$ 44 & 1124 $\pm$ 5 & 14 $\pm$ 1 & 399 $\pm$ 3 & 291 $\pm$ 3 & 129 $\pm$ 4\\ 
\hline                                
\end{tabular}
\begin{tablenotes}
\centering
\item $^a$ In units of 10$^{-15}$ erg/s/cm$^2$.\\
\item $^b$ Values normalized to I(H$\beta$) 10$^{-3}$. 
\item * Region near SN explosion.
\end{tablenotes}
\end{table*}

Table \ref{tab:lines} shows, for each selected ring HII region, the reddening corrected emission line intensities of strong lines relative to H$\beta$, and its corresponding reddening constant. 

\subsection{Integrated magnitudes}\label{sec:int_flux}

We have calculated integrated fluxes inside the \href{http://svo2.cab.inta-csic.es/theory/fps/index.php?mode=browse&gname=SLOAN&asttype=}{Sloan Digital Sky Survey (SDSS) filters} for each region using the following expression \citep{Fukugita1995}:

\begin{equation}
f_\lambda = \int _{\Delta \lambda} F(\lambda) \cdot \lambda \cdot T_\lambda \cdot d\lambda
\end{equation}

\noindent where T$_\lambda$ denotes the response curves of the r and i bands and $F(\lambda)$ denotes the extracted spectrum of each region. The origin of the additional $\lambda$ term in this expression lies on the assumption of a photon-counting detector whose response is proportional to the photon-count rate, $\lambda/(hc)$. Before computation, all nebular emission lines present in the spectra have been masked assuming a width of $\pm$ 8\r{A} around the line central wavelength. The spectra have also been corrected for reddening using the c(H$\beta$) values calculated from the H$\alpha$/H$\beta$ Balmer decrement (see Sec. \ref{sec:line measurements}).

The apparent magnitudes of the selected HII regions have been calculated from their integrated fluxes using:
\begin{equation}
m_\lambda = -2.5\cdot log\left(\frac{f_\lambda}{F_0^\lambda \cdot \int_{\Delta \lambda} \lambda \cdot T_\lambda \cdot d\lambda }\right)
\end{equation}
where $F_0^\lambda$ is the constant flux density per unit wavelength in the AB System, 2.88637 $\times$ $10^{-9}$ erg/cm$^2$/s/\AA\ for r band (Zero Point = 21.35 mag) and 1.95711 $\times$ $10^{-9}$ erg/cm$^2$/s/\AA\ for i band (Zero Point = 21.77 mag). We have assumed a distance of 22.2 Mpc \citep[][see Tab. \ref{tab:galaxy characteristics}]{1988cng..book.....T} to translate these values into absolute magnitudes. Finally, we have calculated the r-i colours of the regions. 

Integrated flux errors have been calculated from the global continuum dispersion and the width of each filter as:
\begin{equation}
\Delta [f_\lambda] = \sigma_c \cdot W_{eff}= \sigma_c \cdot \frac{\int T_\lambda \cdot d\lambda }{ T_\lambda ^{max}}
\end{equation}
\noindent where $\Delta [f]$ is the error in the integrated flux in one band, $\sigma _c$ represents the standard deviation in the global continuum flux and T$_\lambda$ is the response curve of each filter. W$_{eff}$ represents the width of a rectangle with the same area covered by the filter and a height equal to its maximum transmission. The calculated errors have been propagated in quadrature for the rest of the derived quantities (apparent magnitudes, absolute magnitudes and r-i colours).

\begin{table*}
\centering
\caption{Colour and magnitude results. The complete table is available online; here only a part is shown as an example.}
\label{tab:magnitudes}
\begin{tabular}{cccccc}
\hline
Region ID & m$_i$ (mag) & m$_r$ (mag) & M$_i$ (mag) & M$_r$ (mag) & r-i (mag)\\ \hline

R1* & 17.96 $\pm$ (0.23 $ \times$ 10$^{-4}$) & 18.01 $\pm$ (0.15 $ \times$ 10$^{-4}$) & -13.77 $\pm$ (0.23 $ \times$ 10$^{-4}$) & -13.72 $\pm$ (0.28 $ \times$ 10$^{-4}$) & 0.049 $\pm$ (0.275 $ \times$ 10$^{-4}$)\\ 
R2 & 17.92 $\pm$ (0.21 $ \times$ 10$^{-4}$) & 17.97 $\pm$ (0.14 $ \times$ 10$^{-4}$) & -13.81 $\pm$ (0.21 $ \times$ 10$^{-4}$) & -13.76 $\pm$ (0.25 $ \times$ 10$^{-4}$) & 0.050 $\pm$ (0.247 $ \times$ 10$^{-4}$)\\ 
R3 & 16.76 $\pm$ (0.21 $ \times$ 10$^{-4}$) & 16.85 $\pm$ (0.14 $ \times$ 10$^{-4}$) & -14.97 $\pm$ (0.21 $ \times$ 10$^{-4}$) & -14.89 $\pm$ (0.25 $ \times$ 10$^{-4}$) & 0.085 $\pm$ (0.254 $ \times$ 10$^{-4}$)\\ 
R4 & 17.69 $\pm$ (0.26 $ \times$ 10$^{-4}$) & 17.77 $\pm$ (0.18 $ \times$ 10$^{-4}$) & -14.04 $\pm$ (0.26 $ \times$ 10$^{-4}$) & -13.96 $\pm$ (0.32 $ \times$ 10$^{-4}$) & 0.081 $\pm$ (0.319 $ \times$ 10$^{-4}$)\\ 
R5 & 18.00 $\pm$ (0.22 $ \times$ 10$^{-4}$) & 18.08 $\pm$ (0.15 $ \times$ 10$^{-4}$) & -13.73 $\pm$ (0.22 $ \times$ 10$^{-4}$) & -13.65 $\pm$ (0.26 $ \times$ 10$^{-4}$) & 0.085 $\pm$ (0.263 $ \times$ 10$^{-4}$)\\ 
R6 & 19.90 $\pm$ (0.18 $ \times$ 10$^{-4}$) & 19.90 $\pm$ (0.12 $ \times$ 10$^{-4}$) & -11.83 $\pm$ (0.18 $ \times$ 10$^{-4}$) & -11.83 $\pm$ (0.22 $ \times$ 10$^{-4}$) & 0.003 $\pm$ (0.215 $ \times$ 10$^{-4}$)\\ 
R7 & 17.37 $\pm$ (0.24 $ \times$ 10$^{-4}$) & 17.42 $\pm$ (0.16 $ \times$ 10$^{-4}$) & -14.36 $\pm$ (0.24 $ \times$ 10$^{-4}$) & -14.31 $\pm$ (0.29 $ \times$ 10$^{-4}$) & 0.049 $\pm$ (0.285 $ \times$ 10$^{-4}$)\\ 
R8 & 16.88 $\pm$ (0.23 $ \times$ 10$^{-4}$) & 16.97 $\pm$ (0.16 $ \times$ 10$^{-4}$) & -14.85 $\pm$ (0.23 $ \times$ 10$^{-4}$) & -14.76 $\pm$ (0.27 $ \times$ 10$^{-4}$) & 0.093 $\pm$ (0.275 $ \times$ 10$^{-4}$)\\ 
R9 & 16.76 $\pm$ (0.21 $ \times$ 10$^{-4}$) & 16.79 $\pm$ (0.14 $ \times$ 10$^{-4}$) & -14.97 $\pm$ (0.21 $ \times$ 10$^{-4}$) & -14.95 $\pm$ (0.25 $ \times$ 10$^{-4}$) & 0.024 $\pm$ (0.249 $ \times$ 10$^{-4}$)\\ 
R10 & 17.27 $\pm$ (0.22 $ \times$ 10$^{-4}$) & 17.32 $\pm$ (0.15 $ \times$ 10$^{-4}$) & -14.46 $\pm$ (0.22 $ \times$ 10$^{-4}$) & -14.41 $\pm$ (0.27 $ \times$ 10$^{-4}$) & 0.051 $\pm$ (0.266 $ \times$ 10$^{-4}$)\\

\hline 
\end{tabular}
\begin{tablenotes}
\centering
\item * Region near SN explosion.
\end{tablenotes}
\end{table*}

Table \ref{tab:magnitudes} shows the integrated magnitudes and derived quantities for each HII region within the ring and lists in columns 1 to 6: (1) the region ID, (2) the apparent magnitude for the i band, (3) the apparent magnitude for the r band, (4) the absolute magnitude for the i band, (5) the absolute magnitude for the r band and (6) the r-i colour.

\begin{figure}
 \includegraphics[width=\columnwidth]{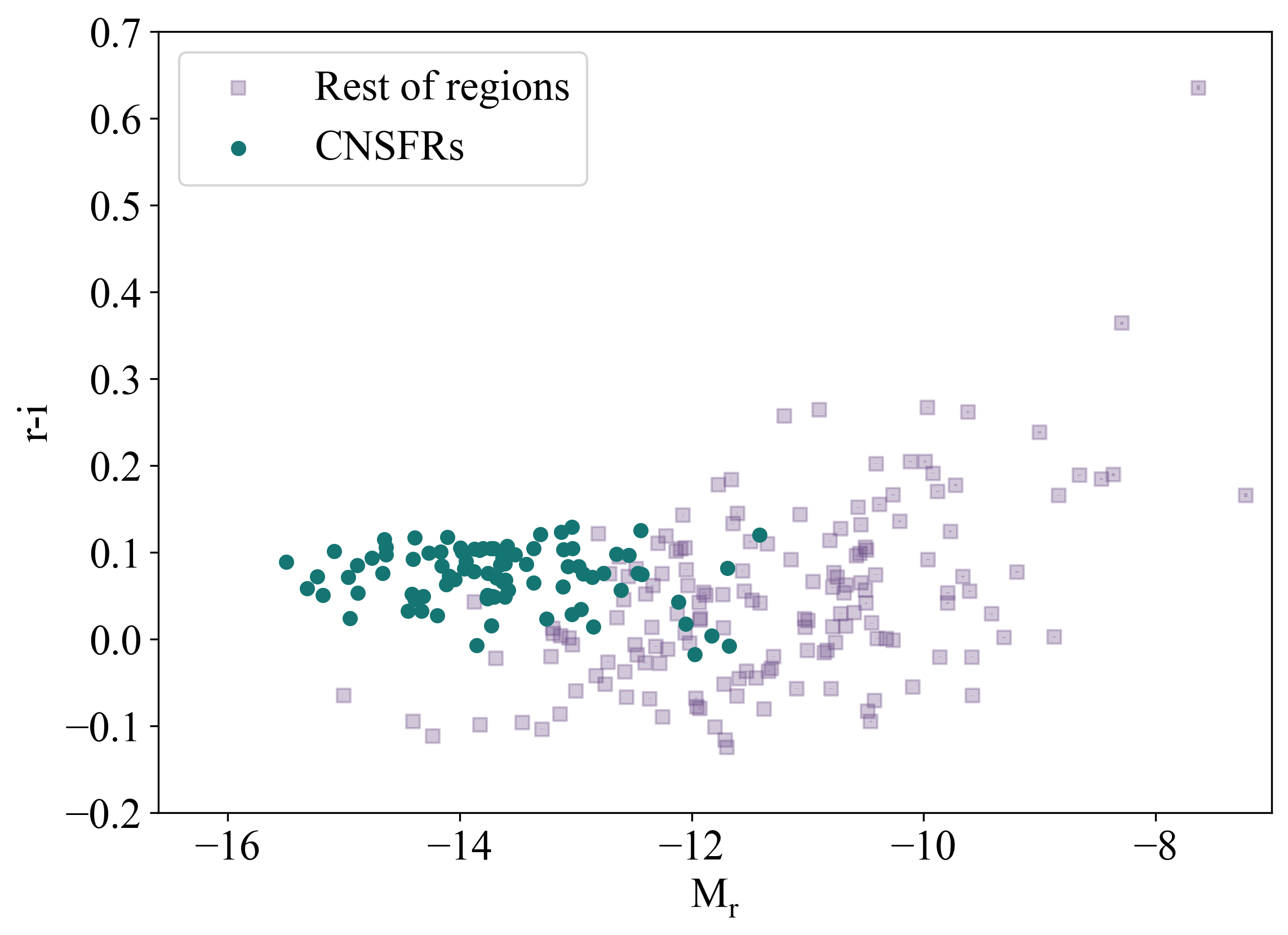}
 \caption{Colour-magnitude diagram for HII regions inside (blue dots) and outside (purple squares) the galaxy ring.}
 \label{fig:color_mag}
\end{figure}

Figure \ref{fig:color_mag} shows the colour-magnitude diagram of the studied ionised regions that represents a first approach to the properties of their stellar populations underneath.  Ring HII regions show r-band luminosities larger than the rest. In regions outside the ring  there is a trend of redder r-i colours for lower r-band luminosities that seems to be real given the small size of the observational errors involved (inside the symbols in the graph).    

\subsection{Chemical abundance determinations}
\label{sec:abundances}

HII region metallicities have been traced by their sulphur abundances following the methodology described in \citep{2022MNRAS.511.4377D}, based on red-to-near infrared spectroscopy. The wavelength range used includes the [SIII]$\lambda$ 6312 \AA\ and [SII]$\lambda\lambda$ 6717,31 \AA\ lines in the red part of the spectrum and the [SIII]$\lambda\lambda$ 9069,9532 \AA\ in the far-red wavelengths. These lines are analogous to the [OII] and [OIII] lines, commonly used to derive oxygen abundances in nebulae. Since sulphur and oxygen are both produced in massive stars both elements are supposed to be proportional to each other. Due to the longer wavelengths of the sulphur lines, reddening effects are less important, which is rather interesting for the study of our observed regions since they are located in the central part of the galaxy. Also, [SII] and [SIII] lines can be measured relative to nearby hydrogen recombination lines (H$\alpha$, P$_9$) in order to minimise the uncertainties. In addition, sulphur, contrary to oxygen, does not seem to be depleted in diffused clouds \citep[][]{2009ApJ...700.1299J} and, due to the lower energies of the involved transitions, the electron temperature sensitive line of [SIII] at  $\lambda$ 6312 \AA\ can be detected and measured up to, at least, solar abundances \citep{2007MNRAS.382..251D}. This methodology is ideal to deal with data from MUSE which do not include the [OII] lines at $\lambda\lambda$ 3727,29 \AA . Using this approach, the [SIII] electron temperature, $T_e([SIII])$, can be derived using the ratio of the nebular to auroral lines of [SIII] which originate from different upper levels with different excitation energies and hence depend strongly on temperature:

\begin{equation}
R_{S3} = \frac{I (\lambda 9069,9532 \AA )}{I(\lambda 6312 \AA)}
\end{equation}

\noindent where I($\lambda $9069,9532 \AA ) denotes the sum of the two near infrared [SIII] lines. MUSE data cover only from 4800 to 9300\r{A}) and therefore do not include the [SIII]$\lambda$ 9532 \AA\ line; we have used the theoretical relation  [SIII]$\lambda$ 9532 \AA / [SIII]$\lambda$ 9069 \AA = 2.44 in order to account for this fact. The following expression has been used  \citep[see ][]{2022MNRAS.511.4377D}:

\begin{equation}
t_{e}([SIII]) = 0.5597-1.808\cdot 10^{-4} R_{S3}+\frac{22.66}{R_{S3}} 
\label{eq_1}
\end{equation}

\noindent where $t_{e}([SIII]) = 10^{-4} \cdot T_e([SIII]) [K]$. This expression has been calculated using PyNeb \citep{pyneb} for values of electron temperatures, T$_e$([SIII]), between 5000 K and 25000 K, an electron density of n$_e$ = 100 cm$^{-3}$ and the atomic data references listed in Table \ref{tab:atomic data}. This equation has a very weak dependence on electron density, increasing by about 3\% for values of n$_e$ between 100 and 1000 cm$^{-3}$ \citep{2017PASP..129d3001P}. We have used PyNeb \citep{pyneb} to calculate the HII region electron densities, $n_e$, from the [SII]$\lambda$ 6717 \AA\ / [SII]$\lambda$ 6731 \AA\ using the atomic coefficients listed in Tab. \ref{tab:atomic data}. This ratio has a slight temperature dependence; a value of T$_e$ = 10000 K has been assumed. This value is close to the mean value of  9197 K derived from the relation between T$_e$([SIII]) and the sulphur abundance given below \citep[see][]{2022MNRAS.511.4377D}: 

\begin{equation} \label{eq:S-te}
\begin{split}
t_e([SIII]&) = (19.226 \pm 0.028) + (-4.7274\pm 0.0081) \cdot \\
&(12+log(S/H)) + (0.29879\pm 0.00058) \cdot \\ &(12+log(S/H))^2
\end{split}
\end{equation}

The derived electron densities of the HII regions within the ring are found to be low and within a narrow range of values centred around 55 cm$^{-3}$, with a median value of 61 cm$^{-3}$ and a standard deviation of 37 cm$^{-3}$, in the lower limit for derived densities using these lines. Only eight regions (R1, R2, R16, R19, R51, R54, R66 and R70) have values higher than 100 cm$^{-3}$, and only three of them (R1, R75 and R66) are significantly different from the median value (>3$\sigma$). About 50\% of the regions show electron density values lower than 50 cm$^{-3}$ and hence are undetermined.  \citet{2006ApJ...649L..79M} also found similar results, concluding that the ring is predominantly populated by clouds of very low electronic density, which are typical of extragalactic HII regions, but lower than those derived for CNSFR \citep{2007MNRAS.382..251D}. 

The weak [SIII]$\lambda$ 6312 \AA\ line has been measured with a S/N higher than 1 in $\sim$45 \% (40 out of 88) of the HII regions within the ring.
For these regions total sulphur abundances have been derived directly using the described method.  

\begin{table}
\centering
\caption{Atomic data sources.}
\label{tab:atomic data}
\begin{tabular}{ccc}
\hline
Ionisation state & Collisional strengths &  Transition probabilities\\ \hline
$S^{+}$ &\citet{Tayal2010} & \citet{Podobedova2009}\\
$S^{2+}$ &\citet{Hudson2012} & \citet{Podobedova2009}\\ $O^{+}$ &\begin{tabular}[c]{@{}c@{}}\citet{2006MNRAS.366L...6P} \\ \citet{Tayal_2007}\end{tabular}& \begin{tabular}[c]{@{}c@{}}\citet{1982MNRAS.198..111Z} \\ \citet{1996atpc.book.....W}\end{tabular}\\
$O^{2+}$ &\citet{1999ApJS..123..311A} & \begin{tabular}[c]{@{}c@{}}\citet{2000MNRAS.312..813S} \\ \citet{1996atpc.book.....W}\end{tabular}\\
\hline
Ion & \multicolumn{2}{c}{Atomic data }\\
\hline
H & \multicolumn{2}{c}{\citet{Storey1995}}\\ \hline
\end{tabular}
\end{table}

In the ring HII regions of our sample, located in the central part of the galaxy and relatively close to its nucleus, most of the sulphur is expected to be in the form of S$^+$ and S$^{++}$. Given the low ionisation potential for sulphur, the contribution by S$^0$ can be neglected and no contribution from S$^{3+}$ is expected in HII regions of moderate to high metallicity \citep[][]{2022MNRAS.511.4377D}.  Therefore, we have assumed an only zone in which $T_e(S^+) \approx T_e(S^{++}) = T_e([SIII])$, the characteristic electron temperature of the region where both the S$^+$ and S$^{++}$ S ions overlap and that encompasses almost the whole nebula. Abundances have been calculated using the expressions derived using the PyNeb package using the atomic coefficients listed in Tab. \ref{tab:atomic data} and are given below:

\begin{equation}
\begin{split}
12+log\left(\frac{S^{+}}{H^{+}}\right)=&log\left(\frac{I(\lambda 6717,31\AA)}{I(H_{\beta})}\right)+5.516+ \\&
+\frac{0.884}{t_{e}([SIII]) }-0.480\cdot log(t_{e}([SIII]))
\end{split}
\end{equation}

\begin{equation}
\begin{split}
12+log\left(\frac{S^{2+}}{H^{+}}\right)=&log\left(\frac{I(\lambda 9069,9532 \AA)}{I(H_{\beta})}\right)+6.059+\\&
+\frac{0.608}{t_{e}([SIII]) }-0.706\cdot log(t_{e}([SIII])) 
\end{split}
\label{eq_4}
\end{equation}

\noindent where I($\lambda$ 6717,31 \AA\ ) denotes the sum of the intensities of the two red [SII] lines, I($\lambda$9069,9532 \AA) denotes the sum of the intensities of the two near infrared [SIII] lines (i.e. 3.44 times the intensity of the [SIII]$\lambda$ 9069 \AA\ line), I(H$\beta$) denotes the H$\beta$ intensity and t$_e$([SIII]) denotes the electron temperature in units of 10$^{-4}$ K. The errors involved in the fitting procedures used for the derivation of the above expressions are lower than observational errors, thus we have propagated the latter to calculate the final emission line intensity uncertainties.

The total abundance of sulphur has then been calculated as: 

\begin{equation}
12+log\left(\frac{S}{H}\right) = 12+log\left(\frac{S^{+}}{H^{+}}+\frac{S^{++}}{H^{+}} \right)  
\label{eq_5}
\end{equation}

\noindent and are given in Table \ref{tab:sulfur_measurements} for the 39 HII regions with measurements of the [SIII]$\lambda$ 6312 \AA\ line. The table lists in columns 1 to 7: (1)  the region ID; (2) the measured  [SIII]$\lambda$ 6312 \AA\ emission line intensity; (3) the R$_{S3}$ line ratio; (4) the [SIII] electron temperature; (5) and (6)  the ionic abundances of S$^+$ and S$^{++}$ relative to H$^+$; and (7) the total S/H abundance. These regions show sulphur abundance values, that range from 6.52 to 7.49 in units of 12+log(S/H) \citep[12+log(S/H)$_{\odot}$= 7.12,][]{2009ARA&A..47..481A}, with errors between 0.007 to 0.049 dex.

\begin{table*}
\centering
\caption{Ionic and total sulphur abundances derived by the direct method for the CNSFRs with measured [SIII]$\lambda$ 6312 \AA\ line intensities. The complete table is available online; here only a part is shown as an example.}
\label{tab:sulfur_measurements}
\begin{tabular}{ccccccc}
\hline
Region ID & I([SIII]$\lambda $6312)$^a$& R$_{S3}$& t$_e$([SIII])$^b$ & 12+log(S$^{+}$/H$^{+}$)&	12+log(S$^{++}$/H$^{+}$) & 12+log(S/H)\\ \hline
R1*	 & 	59.2 $\pm$ 1.3	 & 	176.2 $\pm$ 1.6	 & 	0.6564 $\pm$ 0.0014	 & 	6.7413 $\pm$ 0.0034	 & 	7.0415 $\pm$ 0.0098	 & 	7.218 $\pm$ 0.007\\ 
R2	 & 	39.6 $\pm$ 1.4	 & 	161.1 $\pm$ 2.8	 & 	0.6712 $\pm$ 0.0030	 & 	6.7331 $\pm$ 0.0062	 & 	6.9454 $\pm$ 0.0180	 & 	7.153 $\pm$ 0.011\\ 
R3	 & 	29.1 $\pm$ 1.5	 & 	193.6 $\pm$ 2.9	 & 	0.6417 $\pm$ 0.0023	 & 	6.8886 $\pm$ 0.0057	 & 	6.7310 $\pm$ 0.0212	 & 	7.118 $\pm$ 0.009\\ 
R4	 & 	13.6 $\pm$ 0.6	 & 	199.3 $\pm$ 2.4	 & 	0.6374 $\pm$ 0.0018	 & 	6.9577 $\pm$ 0.0045	 & 	6.7637 $\pm$ 0.0196	 & 	7.172 $\pm$ 0.008\\ 
R5	 & 	21.4 $\pm$ 0.8	 & 	140.1 $\pm$ 2.2	 & 	0.6961 $\pm$ 0.0029	 & 	6.6678 $\pm$ 0.0060	 & 	6.7580 $\pm$ 0.0189	 & 	7.016 $\pm$ 0.011\\ 
R6	 & 	6.5 $\pm$ 0.3	 & 	113.5 $\pm$ 1.5	 & 	0.7388 $\pm$ 0.0029	 & 	6.5699 $\pm$ 0.0055	 & 	6.6885 $\pm$ 0.0166	 & 	6.934 $\pm$ 0.010\\ 
R9	 & 	30.1 $\pm$ 1.6	 & 	178.6 $\pm$ 2.4	 & 	0.6543 $\pm$ 0.0021	 & 	6.8409 $\pm$ 0.0053	 & 	6.7115 $\pm$ 0.0212	 & 	7.082 $\pm$ 0.010\\ 
R11	 & 	10.4 $\pm$ 0.6	 & 	236.4 $\pm$ 2.7	 & 	0.6128 $\pm$ 0.0016	 & 	6.8517 $\pm$ 0.0053	 & 	6.8636 $\pm$ 0.0203	 & 	7.159 $\pm$ 0.011\\ 
R13	 & 	14.8 $\pm$ 1.0	 & 	320.8 $\pm$ 3.0	 & 	0.5723 $\pm$ 0.0012	 & 	7.0735 $\pm$ 0.0050	 & 	7.1241 $\pm$ 0.0181	 & 	7.401 $\pm$ 0.010\\ 
R14	 & 	11.9 $\pm$ 0.8	 & 	172.0 $\pm$ 2.7	 & 	0.6603 $\pm$ 0.0025	 & 	6.7984 $\pm$ 0.0070	 & 	6.7218 $\pm$ 0.0247	 & 	7.063 $\pm$ 0.012\\ 
\hline
\end{tabular}
\begin{tablenotes}
\centering
\item $^a$ In units of 10$^{-18}$ erg/s/cm$^2$.\\
\item $^b$ In units of 10$^{4}$ K.\\
\item * Region near SN explosion.
\end{tablenotes}
\end{table*}

For the rest of the regions no reliable detection of the [SIII]$\lambda$ 6312 \AA\ line could be made and hence we had to rely on empirical calibrations to derive their sulphur abundances. This has been done through the use of the S$_{23}$ parameter, analogous to the commonly used R$_{23}$ for the case of oxygen and can be defined as: 

\begin{equation}
S23 = \frac{\left({[SII]\lambda 6717,6731+[SIII]\lambda 9069,9532}\right)}{H\alpha}\cdot\frac{H\alpha}{H\beta}
\label{eq_6}
\end{equation}

This parameter has little dependence on reddening effects or calibration uncertainties since the lines involved can be measured relative to nearby hydrogen recombination lines. Also, the lines are observable even at over-solar abundances given their lower dependence with electron temperature.

A recent calibration of the S$_{23}$ parameter has been presented in \citet{2022MNRAS.511.4377D} and has the following expression:

 \begin{equation}
\begin{split}
12+log \left(\frac{S}{H}\right)=(6.636\pm 0.011)+\\+ (2.202\pm 0.050)\cdot log S_{23} +(1.060\pm 0.098) \cdot (log S_{23})^2
\end{split}
\label{eq_7}    
\end{equation}

The sulphur abundances derived from this calibration for all the objects in our sample are given in Table \ref{tab3}.

\begin{table}
\centering
\caption{Sulphur abundances of the observed CNSFRs derived by empirical methods. The complete table is available online; here only a part is shown as an example.}
\label{tab3}
\begin{tabular}{ccc}
\hline
Region ID & S23 &  12+log(S/H)\\ \hline
R1 & 1.463 $\pm$ 0.011 & 7.029 $\pm$ 0.016\\ 
R2 & 1.378 $\pm$ 0.014 & 6.963 $\pm$ 0.017\\ 
R3 & 1.187 $\pm$ 0.015 & 6.806 $\pm$ 0.018\\ 
R4 & 1.324 $\pm$ 0.016 & 6.920 $\pm$ 0.018\\ 
R5 & 1.158 $\pm$ 0.015 & 6.781 $\pm$ 0.017\\ 
R6 & 1.140 $\pm$ 0.013 & 6.765 $\pm$ 0.016\\ 
R7 & 1.199 $\pm$ 0.017 & 6.817 $\pm$ 0.019\\ 
R8 & 1.173 $\pm$ 0.018 & 6.794 $\pm$ 0.020\\ 
R9 & 1.160 $\pm$ 0.016 & 6.782 $\pm$ 0.018\\ 
R10 & 1.135 $\pm$ 0.015 & 6.761 $\pm$ 0.018\\ 
\hline
\end{tabular}
\end{table}

\section{Discussion}
\subsection{Ionisation nature}
\label{sec:nature}

\begin{figure*}
\centering
\includegraphics[width=\textwidth, valign=c]{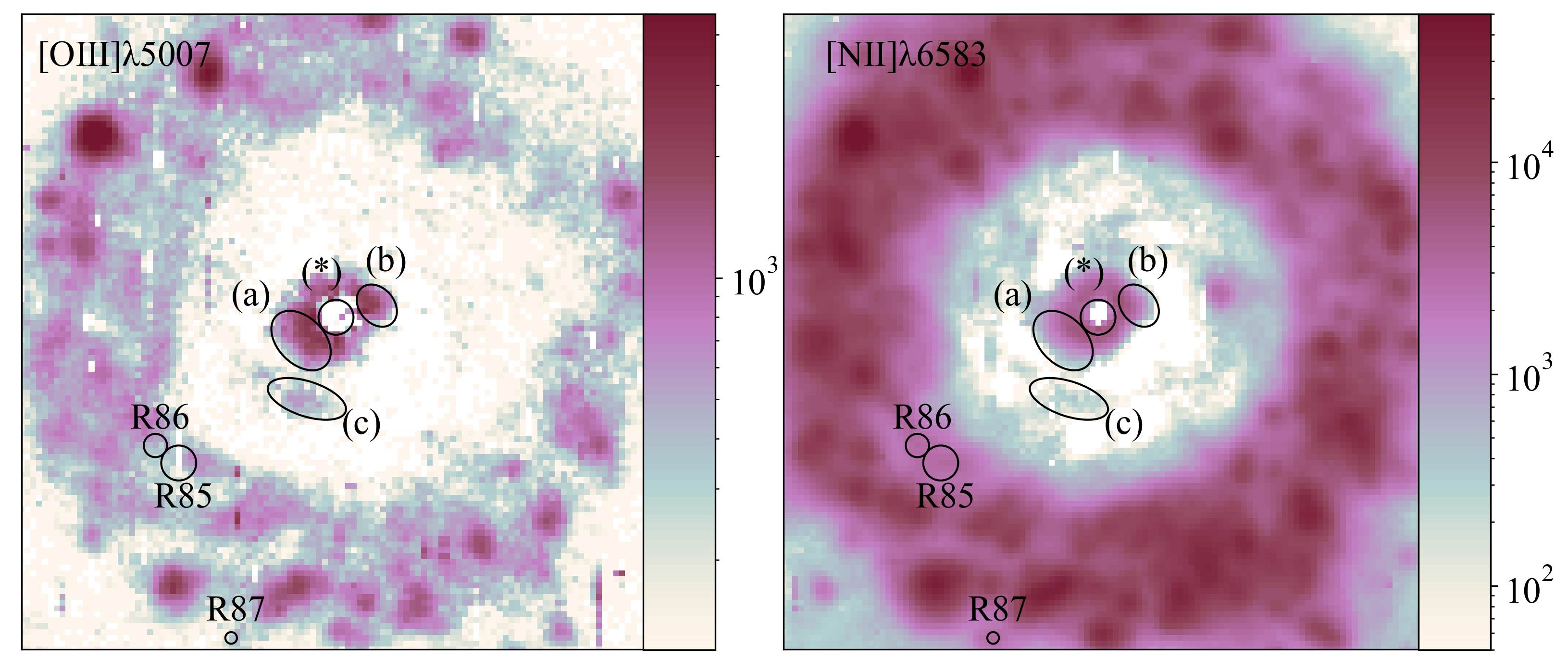}
\includegraphics[width=\textwidth, valign=c]{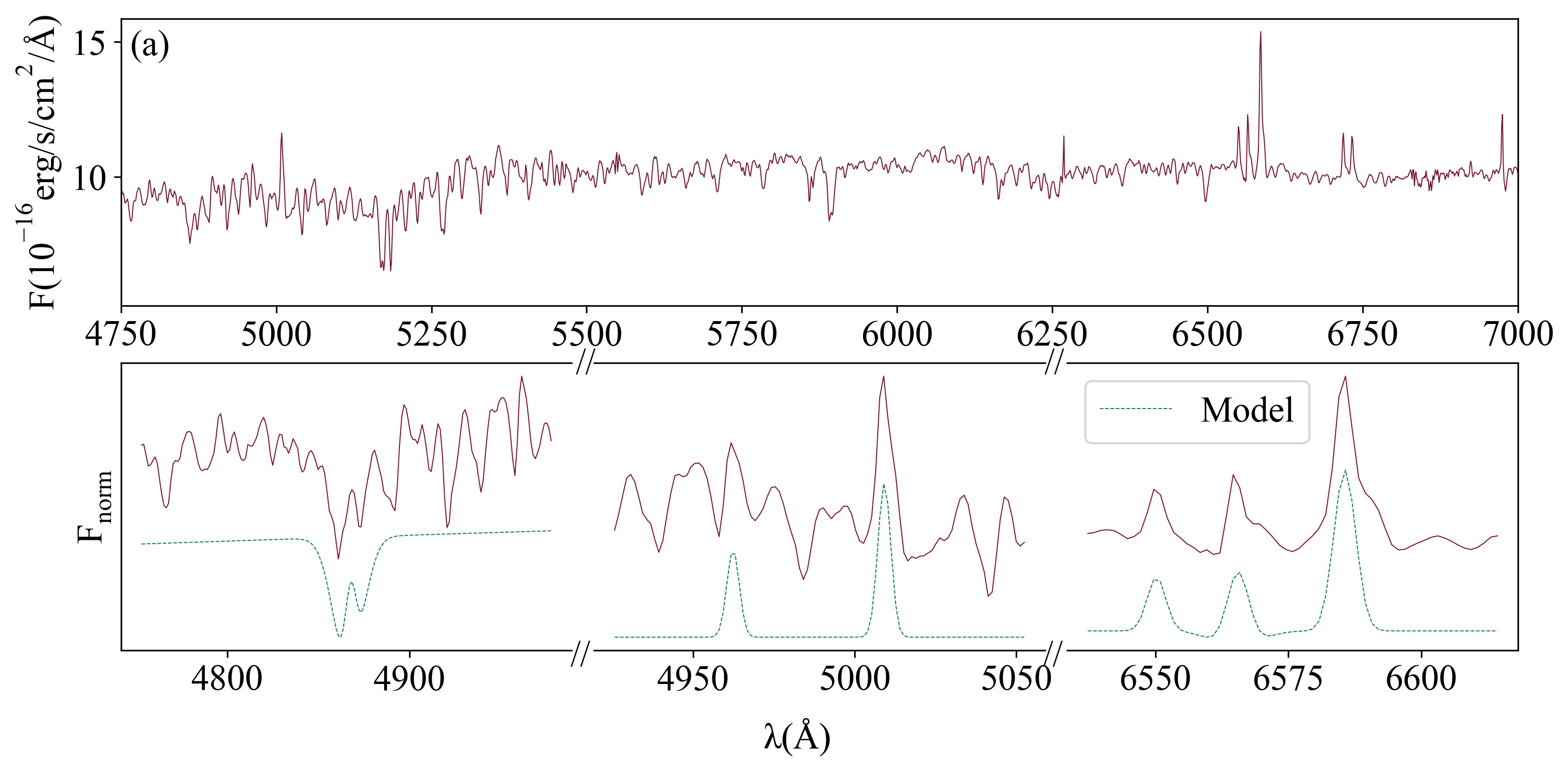}

\caption{Upper panels: Maps of the observed [OIII]$\lambda$ 5007 \AA\ (left) and [NII]$\lambda$ 6583 \AA\ (right) fluxes (in units of 10$^{-20}$ erg/s/cm$^2$)showing the nuclear environment of NGC~7742. Selected  regions are labelled. Bottom panel: Spectrum of region (a) and Gaussian fits to the H$\beta$, [OIII], H$\alpha$ and [NII] emission lines.}
\label{fig:nuclear_region}
\end{figure*}

\begin{figure*}
 \includegraphics[width=\columnwidth]{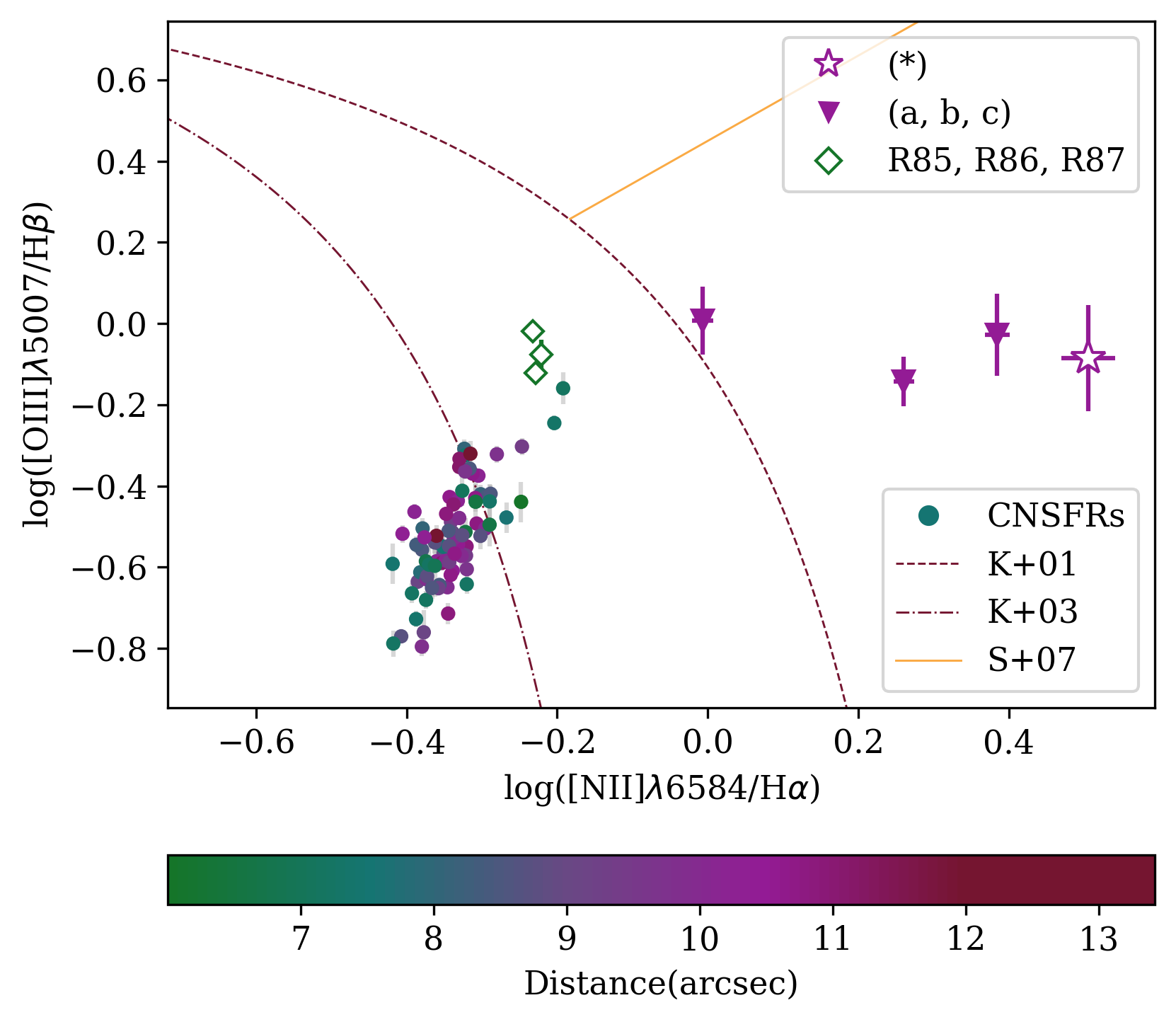}
 \includegraphics[width=0.9\textwidth]{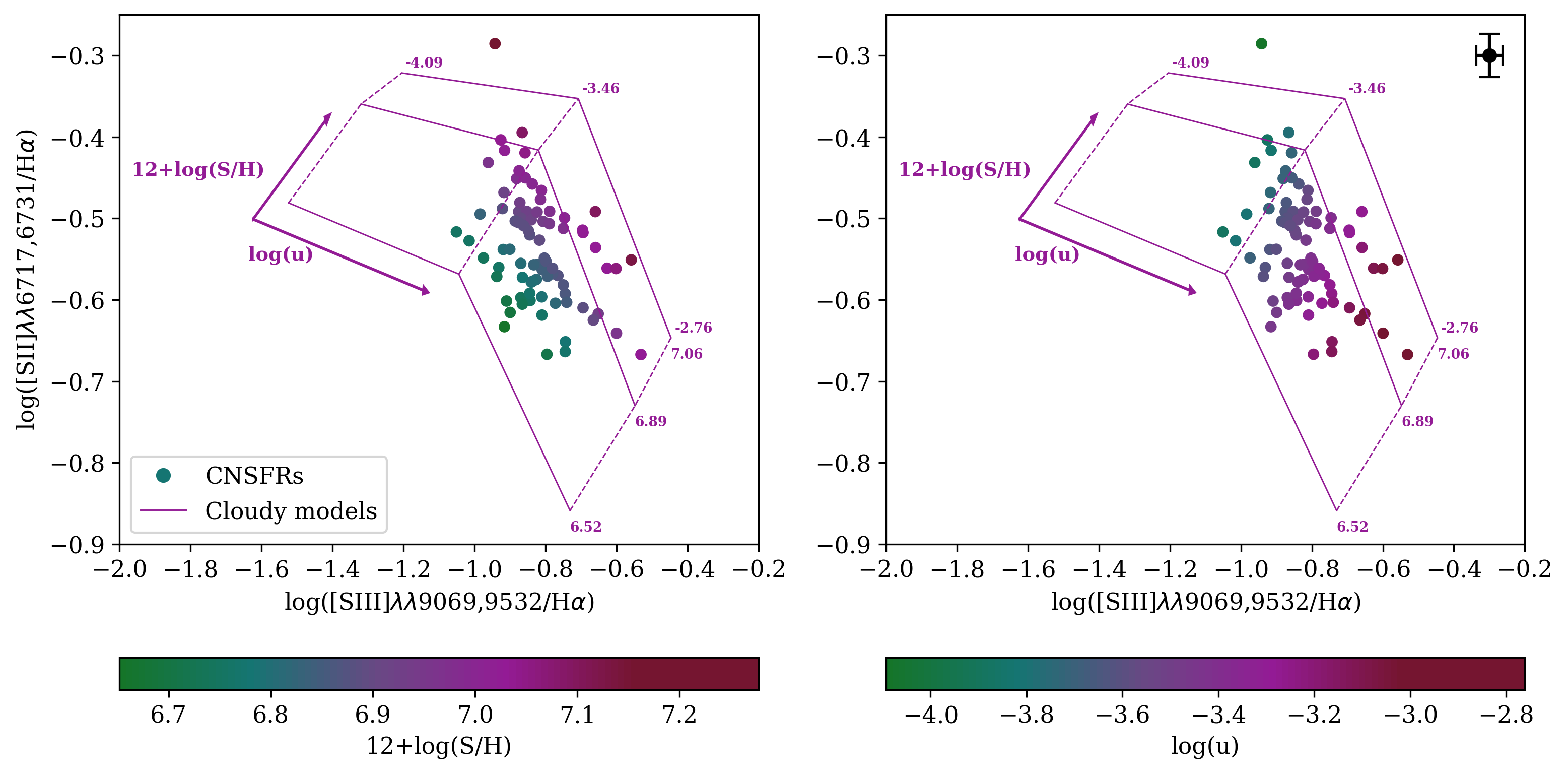}
 \caption{Upper panel: the [OIII]/H$\beta$ vs [NII]/H$\alpha$ diagnostic diagram for the selected ring HII regions, colour coded according to their distance to the galaxy nucleus. Lower panel, left: the [SII]/H$\alpha$ - [SIII]/H$\alpha$ diagnostic diagram, colour coded for metallicity. Lower panel, right: The [SII]/H$\alpha$ - [SIII]/H$\alpha$ diagnostic diagram, colour coded for ionisation parameter. Mean error bars 
 are shown in the upper right corner of the panel. Over-plotted, derived separations between LINER/Seyfert \citep[S+07,][]{2007MNRAS.382.1415S} and HII regions \citep[K+01 and K+03,][]{2001ApJ...556..121K, 2003MNRAS.346.1055K}}
 \label{fig:OIII_HB_NII_Ha}
\end{figure*}

According to \citet{2006ApJ...649L..79M}, the emission line ratios within the ring are consistent with the predictions of star forming models, although regions near the inner edge of the ring are compatible with the ionisation being produced by shocks or an AGN component, something that could be related to the LINER nature of the galaxy nucleus.
Figure \ref{fig:OIII_HB_NII_Ha} shows in the upper panel the classical BPT \citep{1981PASP...93....5B} diagnostic diagram involving the [NII] $\lambda$ 6584 / H$\alpha$ and [OIII] $\lambda$ 5007 / H$\beta$ emission line intensity ratios for the observed regions. The star-forming regions within the ring are shown to lie on the moderate to high metallicity end of the empirical star forming sequence defined by \citet{2003MNRAS.346.1055K} from observations of a large sample of SDSS emission line galaxies. On the other hand, the data analysed here allow to get some inside on the nature of the nuclear ionised regions. The upper panels of Figure \ref{fig:nuclear_region} show maps of the central part of NGC~7742 in the [OIII]$\lambda$ 5007 \AA\ (left) and [NII]$\lambda$ 6583 \AA\ (right) emission line ratios where a small circumnuclear ring at about 200 pc from the galaxy nucleus is seen very distinctively.
We have extracted the spectra corresponding to the galaxy nucleus and three of the circumnuclear regions: (a), (b) and (c) and measured their [OIII]/H$\beta$ and [NII]/H$\alpha$ ratios. The lower panel of the figure shows the spectrum of region (a) where the large [NII]/H$\alpha$ characteristic of LINER-type spectra can be seen. Below the optical spectrum we show the Gaussian fits to the H$\beta$, [OIII], H$\alpha$ and the [NII] lines.

The location of the galaxy nucleus and the three circumnuclear selected regions in the BPT diagram marked with a star and purple inverted triangles respectively, lie below and to the right of the yellow line that marks the division between Seyfert and LINER spectra \citep{2007MNRAS.382.1415S}, thus showing that the emission is probably dominated by shocks or an AGN non-thermal component of low activity. Three of our segmented ring regions to the South-East, R85, R86 and R87 (see Fig. \ref{fig:nuclear_region}), lie on the BPT diagram in the zone between the \citet{2003MNRAS.346.1055K} empirical sequence and the one derived by \citet{2001ApJ...556..121K} from theoretical models of starburst galaxies and may be somewhat affected by the radiation from the galaxy nucleus; consequently, they have not be considered in our analysis.

The BPT diagram shown is the diagnostic more commonly used, mostly due to the fact that is almost insensitive to reddening effects; however, it is sensitive to the N/O ratio, which is difficult to estimate for nuclear and circumnuclear ionised regions. On the other hand, the near-infrared sulphur emission lines constitute a powerful diagnostic to distinguish between shock and photo-ionisation mechanism \citep[][see]{1985MNRAS.212..737D}; it is independent of relative abundances and little sensitive to reddening. The two lower panels of  Figure \ref{fig:OIII_HB_NII_Ha} show the location on the diagram of the observed ring HII regions colour coded for sulphur abundance (left) and ionisation parameter (right) where a segregation in these two parameters can clearly be seen. This can be further interpreted with the help of photo-ionisation models. Using the Cloudy \citep{cloudy} code we have computed models for ionisation-bounded nebulae assuming a plane parallel geometry. The computed models have ionisation parameter values log(u) = -4.09, -3.46 and -2.76 and metallicity values 12+log(S/H) = 6.52, 6.89, and 7.06, with the rest of the elements in solar proportions. These parameters cover the range of derived values for the regions. A constant value of the electron density of 100 cm$^{-3}$ has been assumed. The nebula is ionised by a young star cluster synthesised using the PopStar code  \citep{Popstar} with Salpeter's IMF \citep{Salpeter1955} IMF with lower and upper mass limits of 0.85 and 120 M$_\odot$ respectively, including the nebular continuum in a self-consistent way. We have selected an age of 4.7 Myr to represent the simulated clusters. In the right panel we can see that, within the range of our derived values, regions with high ionisation parameters tend to occupy the lower right zone of the diagram while in the left panel regions with low metallicities lie in the upper left zone. No correlation between these two parameters: ionisation parameter and metallicity, is apparent. 

\subsection{Characteristics of the observed CNSFRs}
\label{sec:CNSFR}
The measured H$\alpha$ fluxes for the selected ring HII regions, prior to extinction correction, are between (226.4 $\pm$ 2.7) $\times$ 10$^{-18}$ erg/s/cm$^2$ and (1762.1 $\pm$ 2.3) $\times $ 10$^{-17}$ erg/s/cm$^2$.

We can compare these results with those obtained by \citet{Mazzuca2008} for this galaxy using narrow band photometry data obtained with the Auxiliary Port camera of the William Herschel Telescope (WHT) at a spatial resolution comparable to that of the MUSE data. For this comparison we have identified 12 matching regions from Fig. 3 of \citet{Mazzuca2008} and converted our measured fluxes to their tabulated values by adding to them the continuum fluxes in a wavelength band 50 \AA\ wide centred at H$\alpha$ and the fluxes of the [NII]$\lambda \lambda $6548,84 \AA\ emission lines also included within this band. Although a linear correlation seems to exist between both sets of measurements, Mazzuca et al.'s H$\alpha$ fluxes are found to be between $\sim$ 6.8 and $\sim$ 37.6 times our flux values. Only 38 HII regions within the ring were selected by these authors, hence their sizes could be larger than those selected in this work.

\begin{figure}
 \includegraphics[width=\columnwidth]{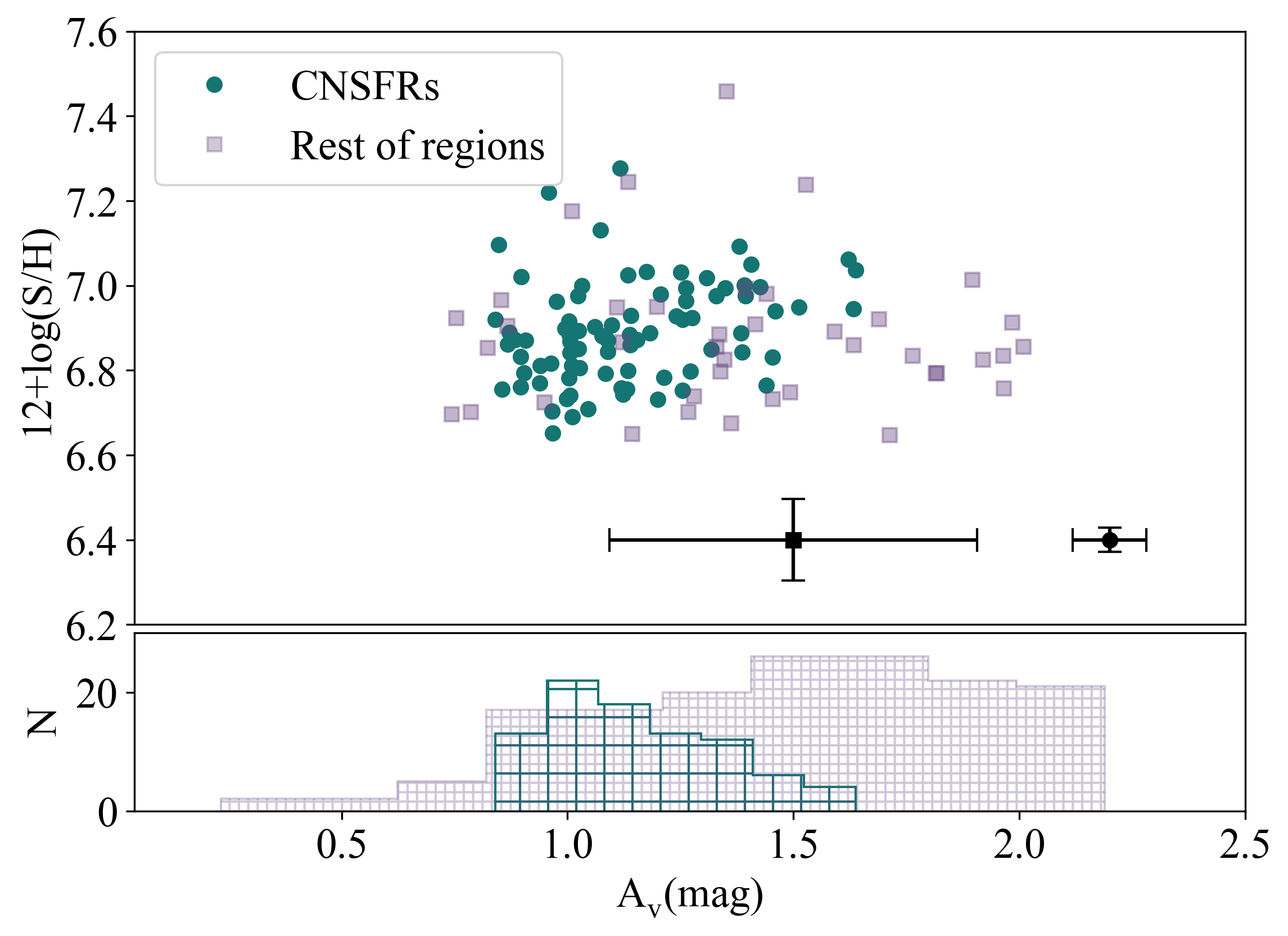}
 \caption{Visual extinction in magnitudes, A$_V$ as a function of sulphur abundance. Average error bars for CNSFRs (right) and the rest of studied regions (left) are shown at the bottom right corner of the panel.}
 \label{fig:aproximaciones} 
 \end{figure}
 
In our analysis, we have used a simple screen distribution to the dust and we have assumed that extinction affects lines and stellar continuum in a similar way and we have derived the extinction from the observed Balmer  H$\alpha $/H$\beta$ as compared with the theoretical one. Figure \ref{fig:aproximaciones} shows that no relation is apparent between the derived extinction values, A$_V$, in magnitudes with the derived sulphur abundances tracing the global metal content, hence we can conclude that dust is distributed along the line of sight and is not an intrinsic characteristic of each particular region. The fact that the extinction values are very similar both in the regions within the ring and outside it supports our assumption. 

The number of hydrogen ionising photons, Q(H$_0$), in each of the HII regions can be calculated from their extinction corrected H$\alpha$ fluxes, F(H$\alpha$), once translated into luminosities. Using a local universe typical galaxy distance of 10 Mpc we can write L(H$\alpha$) in erg s$^{-1}$ as: 
\begin{equation}\label{eq:Halfalum}
L\left(H\alpha\right)=1.2\cdot 10^{38}\left(\frac{F(H\alpha)}{10^{-14}}\right)\left(\frac{D}{10}\right)^2
\end{equation}
\noindent where F(H$\alpha$) is expressed in erg s$^{-1}$ cm$^{-2}$ and D is the distance to NGC~7742 which has been taken as 22.2 Mpc (see Tab. \ref{tab:galaxy characteristics}). The corresponding number of hydrogen ionising photons per second is: 

\begin{equation}\label{eq:Q(H0)}
Q\left(H_0\right)= 7.31 \cdot 10^{11}L(H\alpha) s^{-1}
\end{equation}

\noindent where L(H$\alpha$) is expressed in erg s$^{-1}$ \citep[see for example,][]{1995ApJ...439..604G}. This equation has been derived using the recombination coefficient of the H$\alpha$ line assuming a constant value of electron density of 100 cm$^{-3}$, a temperature of 10$^4$ K and case B recombination \citep{Osterbrock2006}.

\begin{figure*}
\centering
 \includegraphics[width=0.66\textwidth]{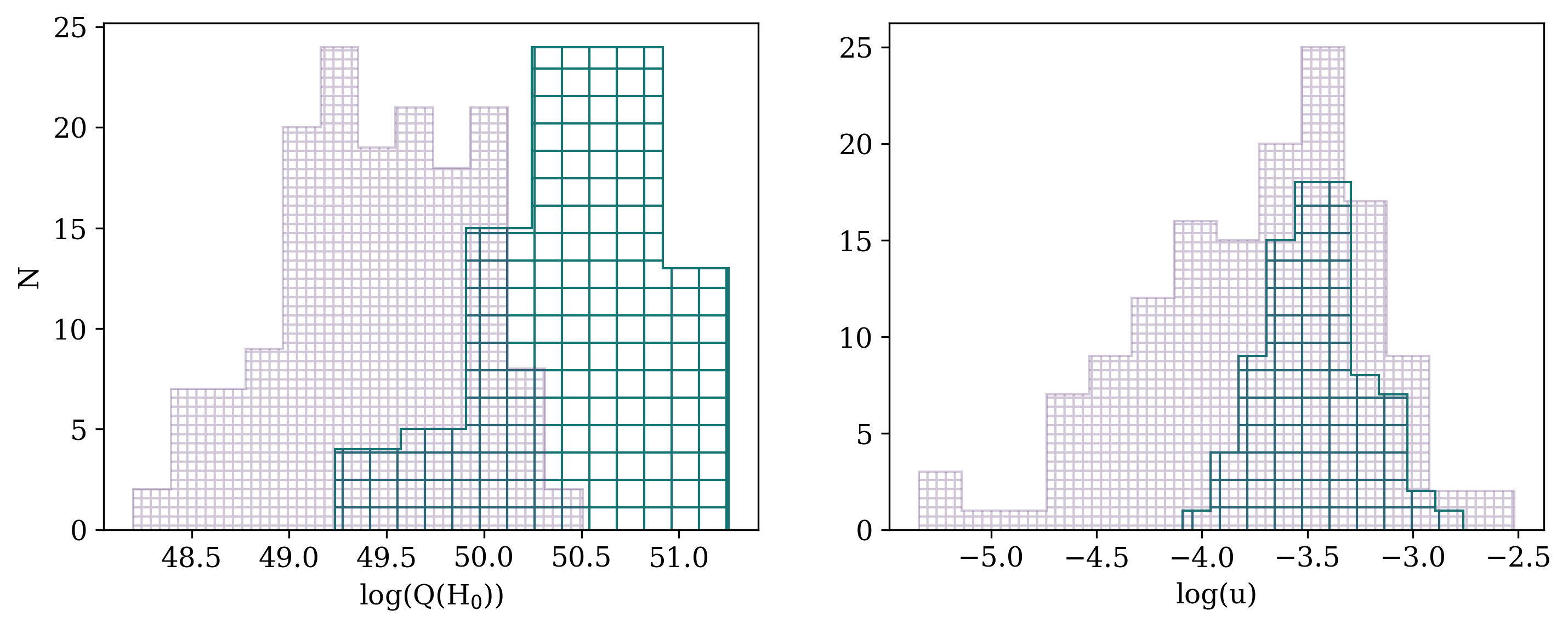}
 \includegraphics[width=\textwidth]{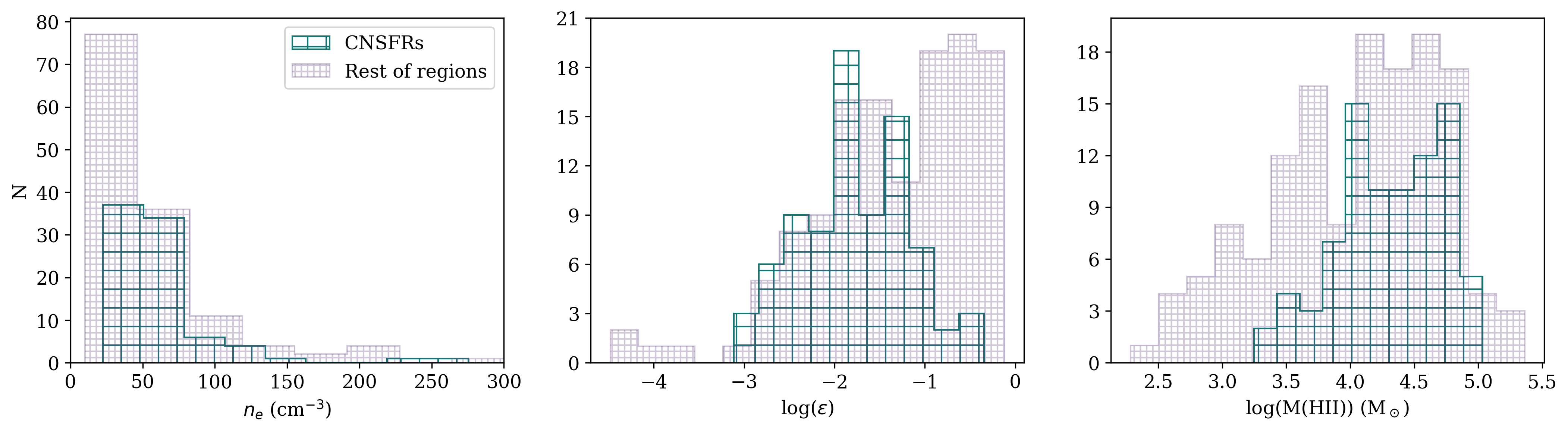}
 \caption{The different histograms in the figure show for the ring HII regions, in green, and regions outside, in purple, the distributions of: the number of hydrogen ionising photons (upper left), the ionisation parameter (upper right), the electron density (bottom left), the filling factor (bottom centre) and the mass of ionised hydrogen (bottom right).}
 \label{fig:hist_r_ne_Q}
\end{figure*}

For the ring regions this number is between 1.7 $\times$ 10$^{49}$ and 1.8 $\times$ 10$^{51}$ photons$\cdot$s$^{-1}$, corresponding to logarithmic H$\alpha$ luminosities between 37.36 and 39.40, on the lower side of the distribution found by \citet{2015MNRAS.451.3173A} for a large sample of CNSFRs, but similar to those of the disc HII regions analysed in \citet{2002MNRAS.337..540C}. According to \citet{Mazzuca2008} disc HII regions in the outer side of the ring have H$\alpha$ luminosities, and therefore a number of ionising photons, lower than the ring regions by about 2 orders of magnitude, implying a higher star formation rate SFR in the galaxy ring as compared to the disc. 

On the other hand, the dimensionless ionisation parameter, u, as estimated from the [SII]/[SIII] ratio \citep{1991MNRAS.253..245D}, ranges from 5.5 $\times$ 10$^{-5}$ to 1.1 $\times$ 10$^{-3}$ for the regions within the ring, with a median value of 2.2 $\times$ 10$^{-4}$. This procedure could be used only for 39 out of 158 of the regions outside the ring. For the rest, the [SIII]$\lambda$ 9069 \AA\ line could be measured with too large errors due to poor signal to noise, hence we have calculated the u values from its definition below (see Eq. \ref{eq:17}).

The upper left panel of Figure \ref{fig:hist_r_ne_Q} shows these results: HII regions within the ring, with values of Q(H$_0$) larger than those in the outer side of it, show ionisation parameters centred at about log(u) = -3.5 with a relatively narrow distribution. For the regions outside the ring, the distribution is also centred at the same u value but looks wider extending to lower values. In a previous work analysing a large sample of disc HII regions in more than 200 nearby galaxies from the CALIFA sample, \citet{2019AA...631A..23R} found that inner disc regions showing a larger number of Lyman continuum photons showed  ionisation parameters lower than their outer disc counterparts. They tentatively attributed this to a selection effect due to the lack of spatial resolution close to the galaxy bulges. However, this was possibly due to the fact that they were missing the population of low H$\alpha$ luminosity with very noisy spectra.

Some light can be shed on this issue by estimating the filling factors of the observed regions which can be done by comparing their sizes, as estimated using the definition of the ionisation parameter, with the actually measured ones. According to its definition:
\begin{equation}\label{eq:17}
u=\frac{Q(H_0)}{4\pi c n_e R^2}
\end{equation}
\noindent where R stands for the radius of the ionised nebulae, provided they have reached their maximum expansion \citep[see][]{2010MNRAS.403.2012M}. 

\begin{figure}
 \includegraphics[width=\columnwidth]{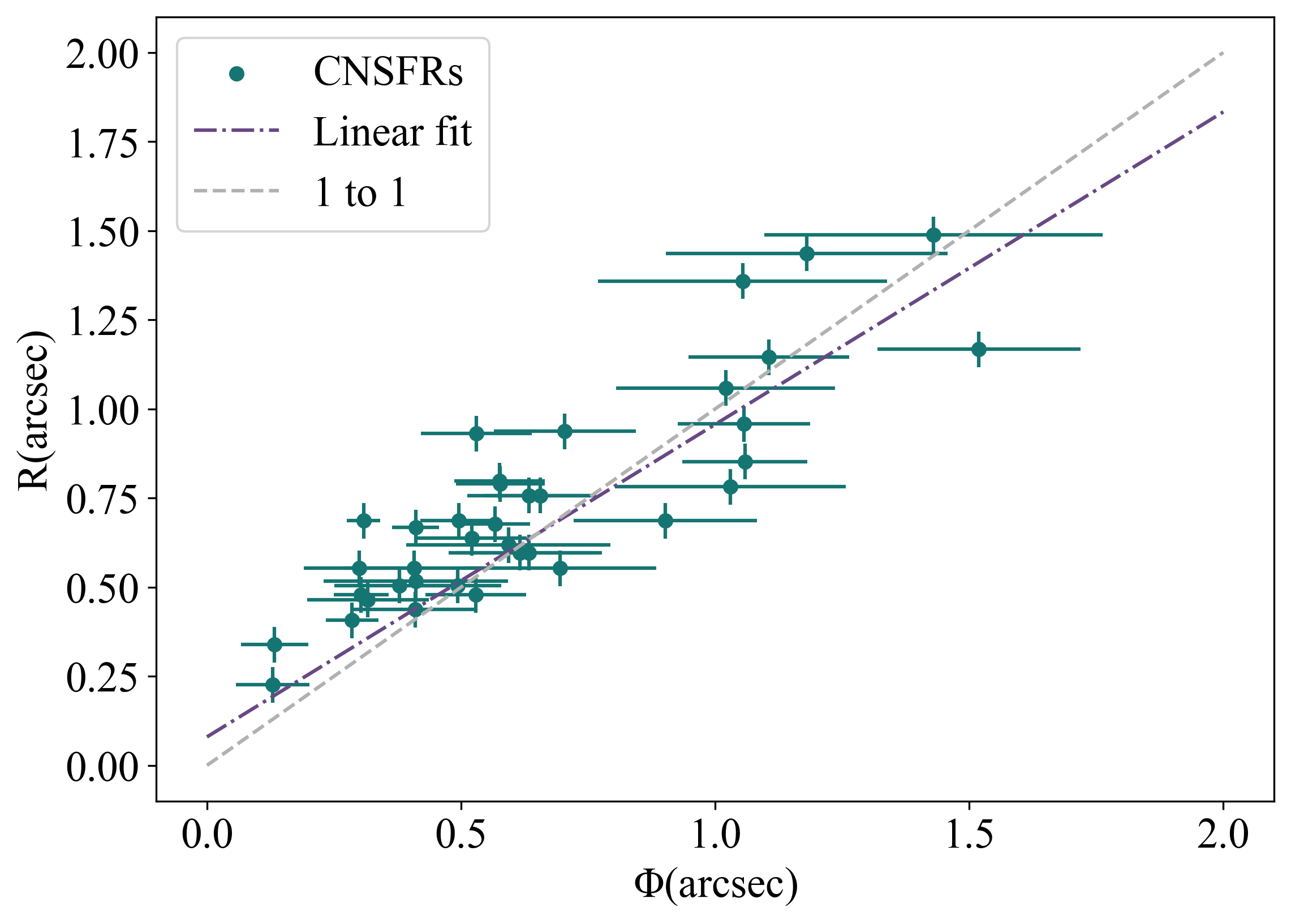}
 \caption{The ionisation derived angular radius  against the angular radius measured from the HII region segmentation (see Sec. \ref{sec:segmentation}).}
 \label{fig:radios}
\end{figure}

Using the expressions from \citet{2002MNRAS.337..540C} we have estimated the angular radii of the observed ring HII regions that have derived electron densities larger than 50 cm$^{-3}$ as:

\begin{equation}
\phi = 0.51\left(\frac{F(H\alpha)}{10^{-14}}\right)^{1/2}\left(\frac{u}{10^{-3}}\right)^{-1/2}\left(\frac{n_e}{40}\right)^{-1/2}
\end{equation}

\noindent where $\phi$ is the angular radius in arcsec. The ionisation parameter predicted radii together with the measured ones, are given in Table \ref{tab:characteristics CNSFR} and compared in Figure \ref{fig:radios}. As can be seen from the figure predicted and measured radii are in very good agreement within the errors and are found to be between 0.39 (corresponding to the element resolution) and 1.5 arcsec which correspond to linear values between 34 to 130 pc. This range of values is similar to those found by \citet{2000MNRAS.318..462D} and \citet{2002MNRAS.337..540C} for disc HII regions and is also consistent with the ones calculated by \citet{2013MNRAS.432.2746G} from PopStar models.

The electron density can be calculated from the [SII]$\lambda$ 6717 \AA\ / [SII]$\lambda$ 6731 \AA\ ratio only for n$_e$ > 50 cm$^{-3}$. For the regions within the ring for which only upper limits could be estimated, the electron density has been derived from the observed region sizes as:

\begin{equation}
\frac{n_e}{10} = \left(\frac{F(H\alpha)}{10^{-14}}\right)\left(\frac{10^{-3}}{u}\right)\left(\frac{1}{\phi}\right)^2
\end{equation}

Once the ionisation parameter and the angular radius of each observed HII  region have been estimated, the filling factor can be derived using the expression:

\begin{equation}
\epsilon = 0.165 \left(\frac{10^{-14}}{F(H\alpha)}\right)\left(\frac{u}{10^{-3}}\right)^2\left(\frac{1}{\phi}\right)\left(\frac{10}{D}\right)
\end{equation}

\noindent \citep[see][]{1991MNRAS.253..245D}. The Filling factors for the ring HII regions are low, ranging from (7.7 $\pm$ 1.5) $\times$ 10$^{-4}$ to 0.45 $\pm$ 0.15, with a mean value of 0.043. These values are similar to those estimated for high metallicity disc HII regions \citep[between 0.008 and 0.52][]{1991MNRAS.253..245D,2000MNRAS.318..462D,2002MNRAS.337..540C} and CNSFRs \citep[1 $\times$ 10$^{-3}$ to 6 $\times$ 10$^{-4}$][]{2007MNRAS.382..251D}.

Finally, the mass of ionised hydrogen, in solar masses, can be derived as \citep[see][]{1991MNRAS.253..245D}:

\begin{equation}
M(HII) = 2.69 \times 10^4  \left(\frac{u}{10^{-3}}\right) \phi^2 \left(\frac{D}{10}\right)^2
\end{equation}

\noindent These values range from (1.77 $\pm$ 0.55)$\times$10$^3$ M$_\odot$ to (1.08 $\pm$ 0.16)$\times$10$^5$ M$_\odot$, with a mean value of 3.07$\times$10$^4$ M$_\odot$ for the HII regions within the ring and are similar to what is found in disc HII regions. 

The bottom panels of Figure \ref{fig:hist_r_ne_Q} show, in panels from left to right, the distribution of electron density, filling factor and ionised hydrogen mass for the ring HII regions as compared with the ones outside the ring. In general, although the electron density shows similar distributions in the two HII region populations, regions within the ring seem to be more diffuse and showing lower filling factors than the regions outside the ring. Given that the size distribution of these latter regions is concentrated around a smaller mean value, this result is expected \citep[][]{2013ApJ...765L..24C}. Exception is made of a population of outside ring HII regions with low H$\alpha$ luminosity and high filling factor that may correspond to regions ionised by a single star \citep[][]{1996ApJ...460..914V}.

\begin{table*}
\centering
\caption{Characteristics of the observed CNSFRs. The complete table is available online; here only a part is shown as an example.}
\label{tab:characteristics CNSFR}
\begin{tabular}{cccccccccc}
\hline 
Region ID &\begin{tabular}[c]{@{}c@{}}L(H$\alpha$) \\ (erg s$^{-1}$)\end{tabular}& \begin{tabular}[c]{@{}c@{}}Q(H$_0$) \\ (photons s$^{-1}$)\end{tabular}&\begin{tabular}[c]{@{}c@{}}log(u) \\ \end{tabular}&\begin{tabular}[c]{@{}c@{}}$\phi$ \\ (arcsec)\end{tabular} & \begin{tabular}[c]{@{}c@{}}R \\ (arcsec)\end{tabular} & \begin{tabular}[c]{@{}c@{}}n$_e$ \\ (cm$^{-3}$)\end{tabular}& \begin{tabular}[c]{@{}c@{}}log($\epsilon$) \\ \end{tabular} & \begin{tabular}[c]{@{}c@{}}M(HII) \\ (M$_\odot$)\end{tabular}\\ \hline

R1 & (208.8 $\pm$ 4.1) $\times$ 10$^{37}$ & (152.8 $\pm$ 3.0) $\times$ 10$^{49}$ & -2.761 $\pm$ 0.009 & 0.31 $\pm$ 0.03 & 0.69 $\pm$ 0.05 & 224 $\pm$ 48 & -1.04 $\pm$ 0.04 & (10.8 $\pm$ 1.6) $\times$ 10$^{4}$\\ 
R2 & (15.0 $\pm$ 1.2) $\times$ 10$^{38}$ & (109.7 $\pm$ 3.0) $\times$ 10$^{49}$ & -2.920 $\pm$ 0.014 & 0.41 $\pm$ 0.05 & 0.67 $\pm$ 0.05 & 130 $\pm$ 29 & -1.20 $\pm$ 0.04 & (7.1 $\pm$ 1.1) $\times$ 10$^{4}$\\ 
R3 & (24.6 $\pm$ 2.0) $\times$ 10$^{38}$ & (180.0 $\pm$ 7.1) $\times$ 10$^{49}$ & -3.518 $\pm$ 0.028 & 1.52 $\pm$ 0.20 & 1.17 $\pm$ 0.05 & 62 $\pm$ 16 & -2.85 $\pm$ 0.06 & (54.8 $\pm$ 5.8) $\times$ 10$^{3}$\\ 
R4 & (112.0 $\pm$ 9.0) $\times$ 10$^{37}$ & (82.0 $\pm$ 2.8) $\times$ 10$^{49}$ & -3.575 $\pm$ 0.028 & 1.06 $\pm$ 0.12 & 0.85 $\pm$ 0.05 & 66 $\pm$ 15 & -2.49 $\pm$ 0.06 & (25.6 $\pm$ 3.4) $\times$ 10$^{3}$\\ 
R5 & (98.1 $\pm$ 7.8) $\times$ 10$^{37}$ & (71.8 $\pm$ 2.4) $\times$ 10$^{49}$ & -3.144 $\pm$ 0.020 & 0.57 $\pm$ 0.07 & 0.68 $\pm$ 0.05 & 75 $\pm$ 18 & -1.47 $\pm$ 0.05 & (43.6 $\pm$ 6.8) $\times$ 10$^{3}$\\ 
R6 & (24.0 $\pm$ 2.0) $\times$ 10$^{37}$ & (175.5 $\pm$ 7.5) $\times$ 10$^{48}$ & -3.125 $\pm$ 0.017 & - & 0.25 $\pm$ 0.05 & 89 $\pm$ 36 & -0.39 $\pm$ 0.09 & (6.3 $\pm$ 2.5) $\times$ 10$^{3}$\\ 
R7 & (13.5 $\pm$ 1.1) $\times$ 10$^{38}$ & (98.7 $\pm$ 3.6) $\times$ 10$^{49}$ & -3.272 $\pm$ 0.025 & - & 0.93 $\pm$ 0.05 & 39 $\pm$ 14 & -2.00 $\pm$ 0.06 & (61.4 $\pm$ 7.5) $\times$ 10$^{3}$\\ 
R8 & (18.1 $\pm$ 1.5) $\times$ 10$^{38}$ & (132.8 $\pm$ 5.4) $\times$ 10$^{49}$ & -3.429 $\pm$ 0.031 & 1.11 $\pm$ 0.16 & 1.15 $\pm$ 0.05 & 70 $\pm$ 19 & -2.53 $\pm$ 0.07 & (64.7 $\pm$ 7.3) $\times$ 10$^{3}$\\ 
R9 & (23.2 $\pm$ 2.0) $\times$ 10$^{38}$ & (170.1 $\pm$ 7.4) $\times$ 10$^{49}$ & -3.481 $\pm$ 0.030 & - & 1.17 $\pm$ 0.05 & 48 $\pm$ 13 & -2.76 $\pm$ 0.06 & (60.2 $\pm$ 6.6) $\times$ 10$^{3}$\\ 
R10 & (15.9 $\pm$ 1.3) $\times$ 10$^{38}$ & (116.1 $\pm$ 4.2) $\times$ 10$^{49}$ & -3.311 $\pm$ 0.024 & 1.06 $\pm$ 0.13 & 0.96 $\pm$ 0.05 & 51 $\pm$ 12 & -2.16 $\pm$ 0.06 & (59.4 $\pm$ 7.0) $\times$ 10$^{3}$\\ 
\hline

\end{tabular}
\end{table*}

Tab. \ref{tab:characteristics CNSFR} shows the characteristics of each HII region within the ring and lists in column 1 to 9: (1) the region ID; (2) the extinction corrected H$\alpha$ luminosity; (3) the number of hydrogen ionising photons; (4) the ionisation parameter; (5) the estimated angular radius; (6)  the measured linear radius; (7) the electron density; (8) the filling factor; and (9) the mass of ionised hydrogen.

\subsection{Chemical abundances} \label{sec_dis_abundances}
\subsubsection{Sulphur abundance determinations}

Reliable measurements of the weak electron temperature sensitive [SIII] line at $\lambda$ 6312 \AA\ have been obtained for 38 ring HII regions out of the 88 originally selected; for these regions sulphur abundances have been derived by the direct method described in section \ref{sec:abundances} and their distribution can be seen in Figure\ref{fig:hist_S} as the histogram filled with oblique lines. Total sulphur abundances are between 6.525 $\pm$ 0.007 and 7.50 $\pm$ 0.01 in units of 12+log(S/H), that is, between 0.25 and 2.40 times the solar value with a median value of 7.01, slightly below solar. The two ionic species present, S$^+$ and S$^{++}$, contribute approximately 50\% each to the total abundance.

\begin{figure}
\centering
\includegraphics[width=\columnwidth]{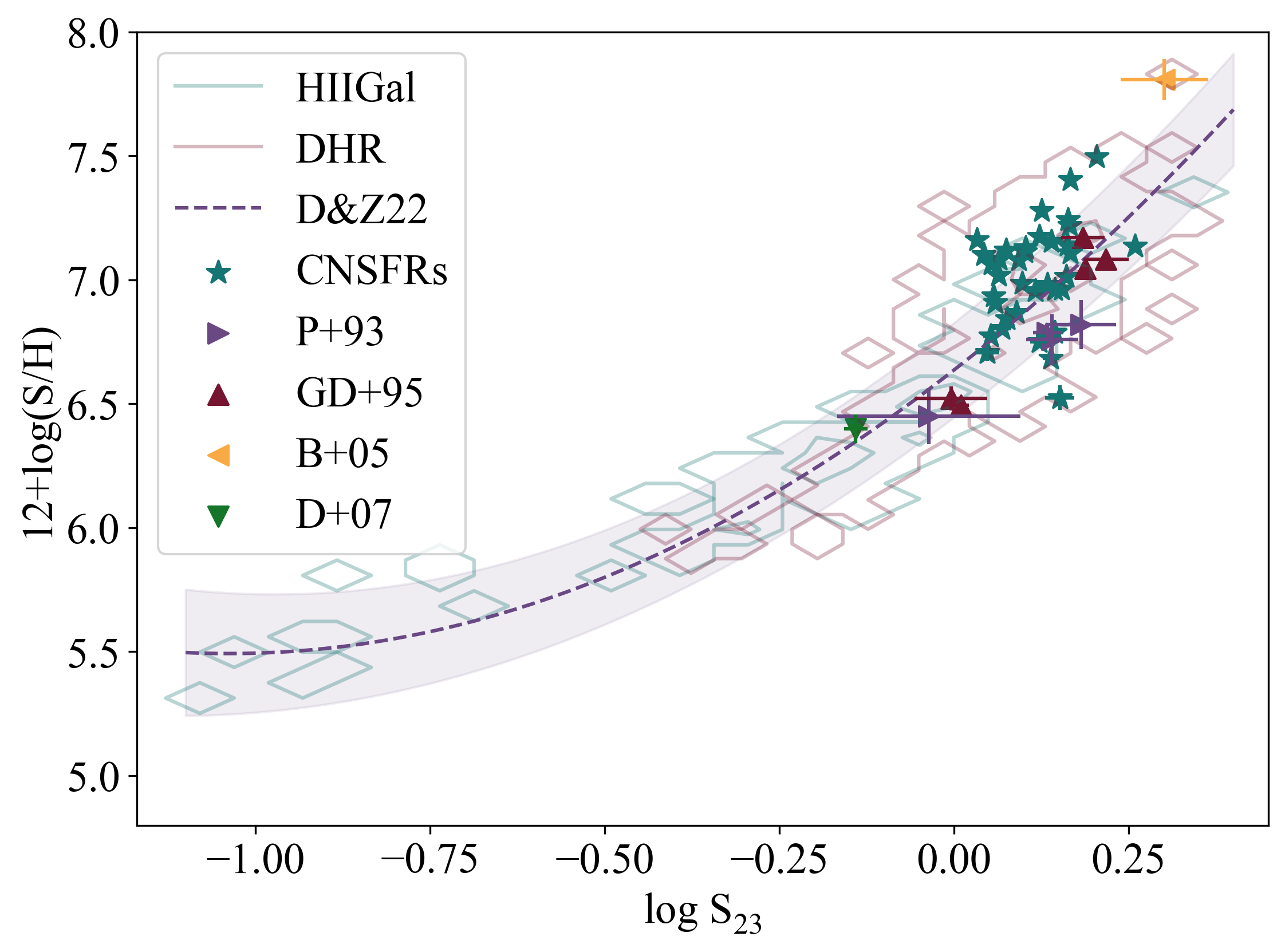}
\caption{The S$_{23}$abundance calibration from \citet{2022MNRAS.511.4377D}. Red contours correspond to disc HII regions while blue contours correspond to HII galaxies. Green stars show the values found for the 38 HII regions within the galaxy ring with the [SIII]$\lambda$ 6312 \AA\ line measured. Observational errors for these data are inside the symbols in the graph. Circumnuclear regions from  the works by \citet[][P+93]{1993MNRAS.260..177P}, \citet[][GD+95]{1995ApJ...439..604G} and \citet[][B+05]{2005AandA...441..981B} are also shown as labelled in the figure.} 
 \label{fig:calibracion}
\end{figure}

%\textbf{\textcolor{blue}{Indirect abundances}}
For the rest of the regions, we have resorted to the empirical calibration by the S$_{23}$ parameter as described in section \ref{sec:abundances} which is single valued up to, at least, solar metallicity. This calibration is shown in Figure \ref{fig:calibracion} where red and blue contours correspond to data for disc HII regions and HII galaxies respectively. Superimposed are the directly derived abundances for the 38 analysed regions with measurements of the [SIII]$\lambda$ 6312 \AA\ line, shown by green stars, and data on regions classified as circumnuclear in the literature as labelled in the figure  \citep[][from NGC~3310, NGC~7714 respectively]{1993MNRAS.260..177P,1995ApJ...439..604G} and are all found to lie at the tip of the curve. The yellow triangle shows the position of HII region 11 from M83 observed by \citet{2005AandA...441..981B}. This region is also part of the sample analysed by \citet{2022MNRAS.511.4377D} and its derived sulphur abundance is in full agreement with the results of the former authors although the ionisation structure is slightly different.

\begin{figure}
 \includegraphics[width=\columnwidth]{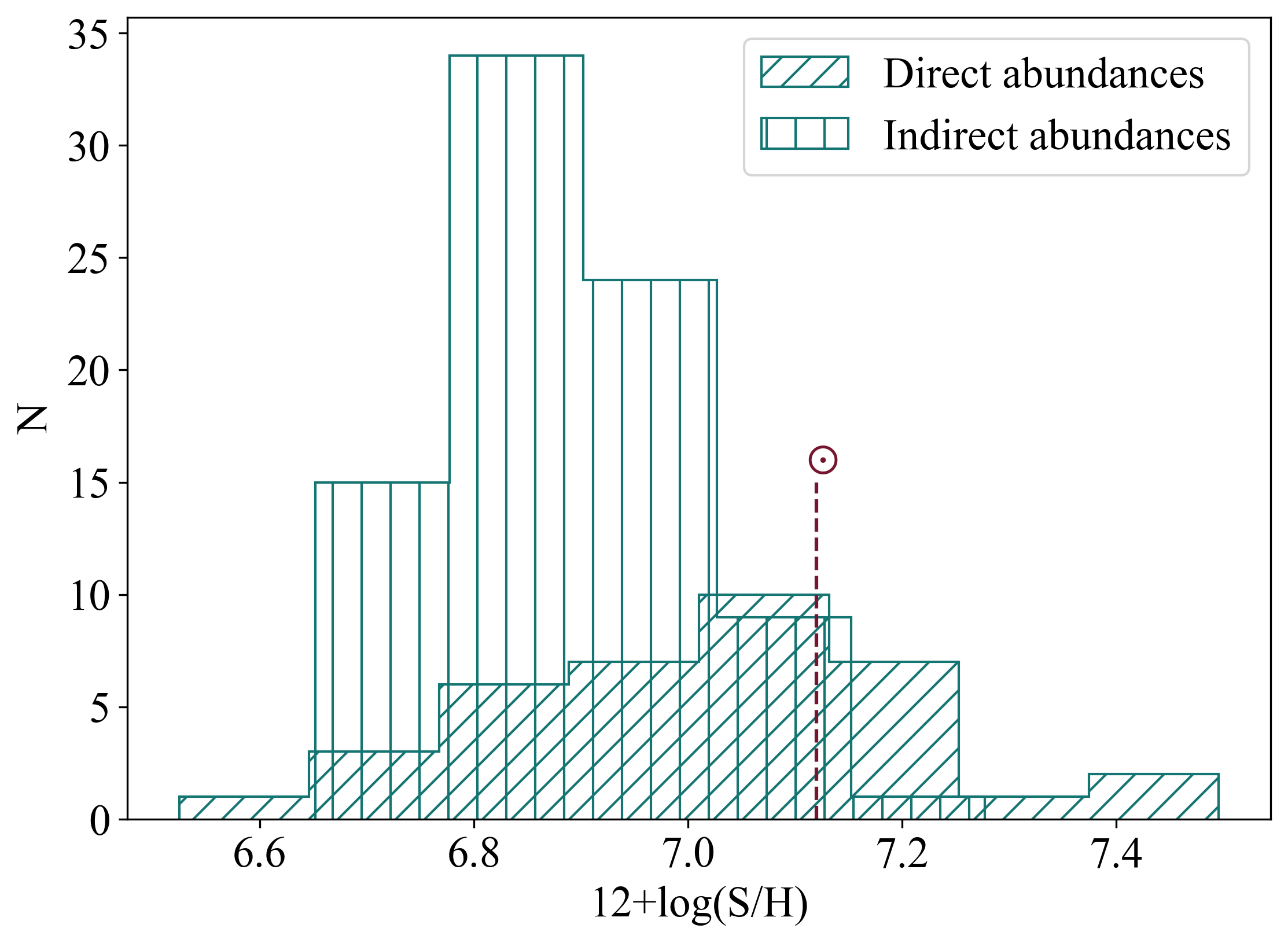}
 \caption{Distribution of the total sulphur abundances for the ring HII regions. The dashed line corresponds to the solar value \citep[12+log(S/H)$_{\odot}$= 7.12,][]{2009ARA&A..47..481A}.} %(En las indirectas están todas las regiones, con y sin 6312).
 \label{fig:hist_S}
\end{figure}

We have therefore considered jointly the sulphur abundances derived by the two methods. Their distribution can be seen in Figure \ref{fig:hist_S} as the histogram filled with vertical lines. We can see that sulphur abundances derived by the direct method look higher than those calculated from S$_{23}$ parameter calibration. This might reflect the fact that, as in the case of the R$_{23}$ parameter, along the low branch of the calibration the intensity of the [SIII] nebular lines increases with metallicity reaching its maximum at the point where the cooling starts to be dominated by sulphur (about two times the solar value), where the calibration bends to lower values of S$_{23}$ starting to show its bi-valued nature \citep[see Fig. of][]{2005MNRAS.361.1063P}. In those cases our empirically derived sulphur abundances could be somewhat underestimated.

\subsubsection{The S/O abundance}

For 13 regions out of the 88 selected within the ring the [OII]$\lambda\lambda$ 7320,30 \AA\ emission lines have been detected and measured, and 10 of them also show the [SIII]$\lambda$ 6312 \AA\ line. For those 10 regions it was possible to derive the O$^+$/H$^+$ ionic abundance. We have assumed an only zone in which $T_e(O^+) \approx T_e(S^{++}) = T_e([SIII])$, n$_e$ = 100 cm$^{-3}$ and we have used the following equation derived using the PyNeb package \citep{pyneb} and the atomic coefficients listed in Tab. \ref{tab:atomic data}:

\begin{equation}
\begin{split}
12+log\left(\frac{O^{+}}{H^{+}}\right)=&log\left(\frac{I(\lambda 7320,30)}{I(H_{\beta})}\right)+6.952+ \\
&+\frac{2.433}{t_{e}([SIII])}-0.571\cdot log(t_{e}([SIII]))
\end{split}
\label{eq_11}
\end{equation}

The [OII] auroral line intensities might present some contribution by recombination emission that can be estimated as shown in \citet{2001MNRAS.327..141L}:

\begin{equation}
\left[\frac{I(\lambda 7320 + \lambda 7330)}{H\beta} \right]_R = 9.36\cdot t_e^{0.44}\frac{O^{++}}{H^+}
\end{equation}
\noindent where $t_e$ is the temperature of the O$^+$ ion in units of 10$^4$ K and takes values between 0.5 and 1.0 (T$_e$ = 5000 - 10000 K).

For our ring regions this contribution takes values from 0.00016 to 0.0011 and represents between 1.5\% and 4.5 \% of the emission line intensity, within measurement uncertainties. 

The $O^{++}/H^+$ abundance ratios have been calculated using the expression:

\begin{equation}
\begin{split}
12+log\left(\frac{O^{++}}{H^{+}}\right)=log\left(\frac{I(\lambda 4959+\lambda 5007)}{I(H_{\beta})}\right)+6.249+\\
+\frac{1.184}{t_{e}([OIII])}-0.708\cdot log(t_{e}([OIII])) 
 \end{split}
\label{eq_12}
\end{equation}
\noindent also derived using PyNeb \citep{pyneb} and the atomic coefficients listed in Tab. \ref{tab:atomic data}.

\begin{figure}
\centering
\includegraphics[width=\columnwidth]{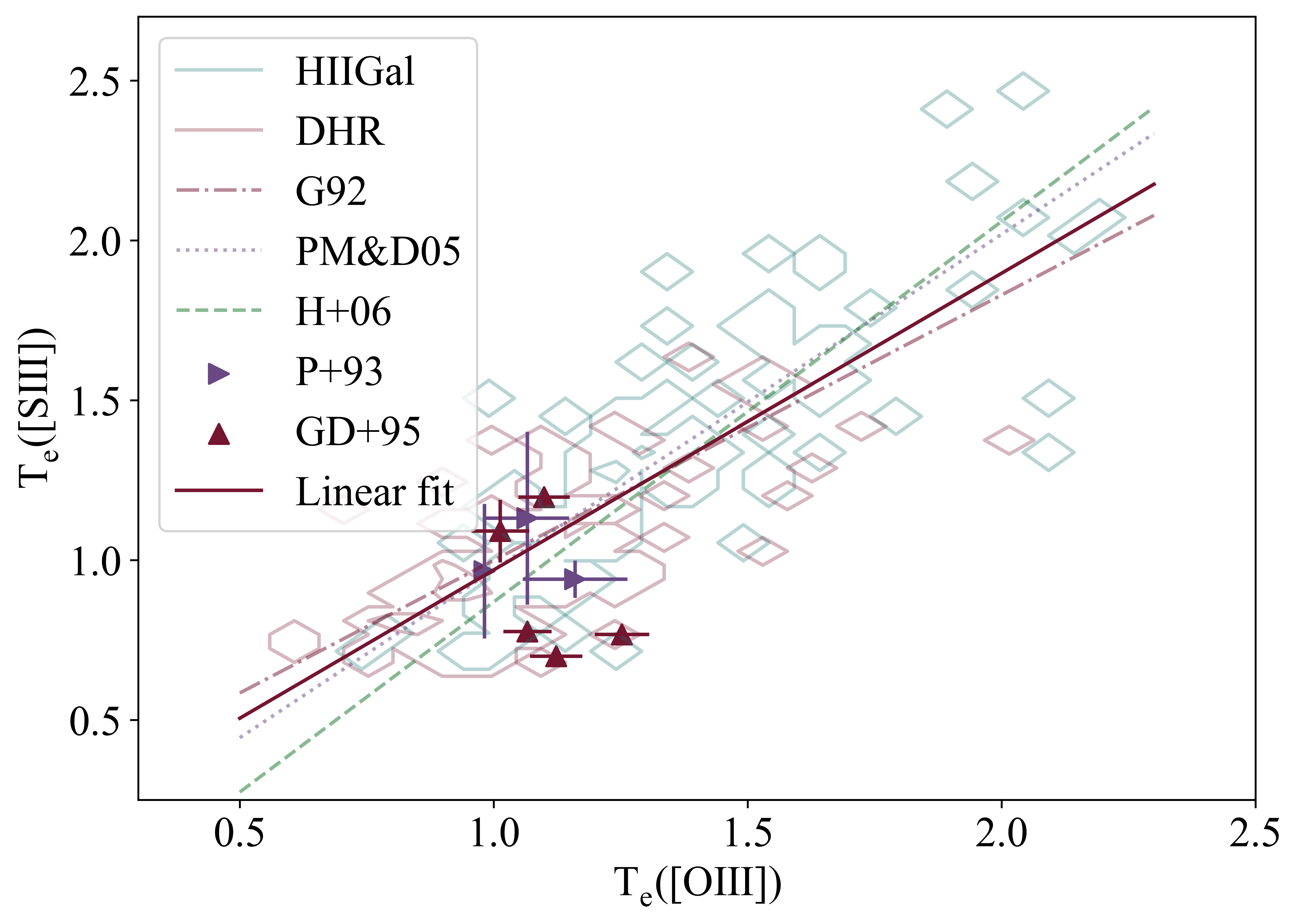}
\caption{The T$_e$([SIII]) - T$_e$([OIII]) relation derived by different authors \citep[][(G92), (PM\&D05) and (H+06) respectively]{1992AJ....103.1330G,2005MNRAS.361.1063P,2006MNRAS.372..293H}. Red and blue contours are from disc HII regions and HII galaxies data from \citet{2022MNRAS.511.4377D} respectively. Triangles are electron temperatures measured for CNSFRs from \citet[][P+93]{1993MNRAS.260..177P} and \citet[][GD+95]{1995ApJ...439..604G}.} 
\label{fig:teOIII}
\end{figure}

We have assumed that the temperature of the region where the $O^{++}$ ion is originating , T$_e([OIII])$, can be derived from $T_e([SIII])$. Figure \ref{fig:teOIII} shows data from \citet{2022MNRAS.511.4377D} for disc HII regions and HII galaxies (contours in red and blue respectively). Superimposed are different $T_e([OIII])$ - $T_e([SIII])$ relations proposed in the literature \citep[see][]{1992AJ....103.1330G,2005MNRAS.361.1063P,2006MNRAS.372..293H} as well as other circumnuclear regions with direct determination of these two temperatures \citep[][]{1993MNRAS.260..177P,1995ApJ...439..604G}. For high abundances the difference between the plotted relations increases and for T$_e$([OIII]) = 5000 K, T$_e$([SIII]) varies from 2740 K to 5850 K for those given by \citet{2006MNRAS.372..293H} and \citet{1992AJ....103.1330G} respectively. Due to these differences and for consistency with the present work, we have decided to fit the data from \citet{2022MNRAS.511.4377D} and we have obtained and used the following equation:

\begin{equation}
t_e([SIII]) = (0.928\pm 0.053)\cdot t_e([OIII]) +(0.042\pm 0.057)
\end{equation}

The total abundance of oxygen has then been calculated as:
\begin{equation}
12+log\left(\frac{O}{H}\right) = 12+log\left(\frac{O^{+}}{H^{+}}+\frac{O^{++}}{H^{+}} \right)  
\label{eq_25}
\end{equation}

Table \ref{tab:oxygen abundances} shows the oxygen ionic and total abundances and the relative sulphur to oxygen abundance for the 10 regions refered above and lists in column 1 to 8: (1) the region ID; (2,3) the [OII]$\lambda \lambda$ 7320,30 \AA\ auroral lines fluxes and its recombination emission correction; (4) the temperature of the O$^{++}$ ion; (5,6) the ionic oxygen abundances; (7) the total oxygen abundance in units of 12+log(O/H); and (8) the relative sulphur to oxygen abundance in units of log(S/O). The derived total oxygen abundances, 12+log(O/H), are found to be between 8.16 and 9.5 (mean value of 8.84), corresponding to 0.30 and 6.46 times solar \citep[12+log(O/H)$_{\odot}$= 8.69,][]{2009ARAA..47..481A}, this latter value having a rather high error ($\pm$ 0.15). For region R42, that shows the highest directly derived sulphur abundance, the [OIII] auroral lines are not detected. The second highest sulphur abundance, 12+log(S/H) = 7.40 ($\sim$ 2 times the solar value) is found for region R13  which also shows the highest value of the oxygen abundance. For all these regions very high values of the O$^+$/O ionic fraction have been found, between 87\% and 95\% with a mean value of 92\%. Similar ratios have been reported for other high metallicity objects: circumnuclear regions NGC~7714-A and NGC~7714-N110 with ratios $\sim$ 92\% \citep{1995ApJ...439..604G} and region NGC~5236-R11 with a value of 92\% \citep{2005AandA...441..981B}. This could be partly due to the highest metallicity regions being ionised by metal rich stars, relatively cool, thus implying a certain lack of O$^+$ ionising photons decreasing the [OIII] emission line intensity and increasing the [OII] one \citep[e.g.][]{1978ApJ...222..821S}. 

\begin{figure}
\centering
\includegraphics[width=\columnwidth]{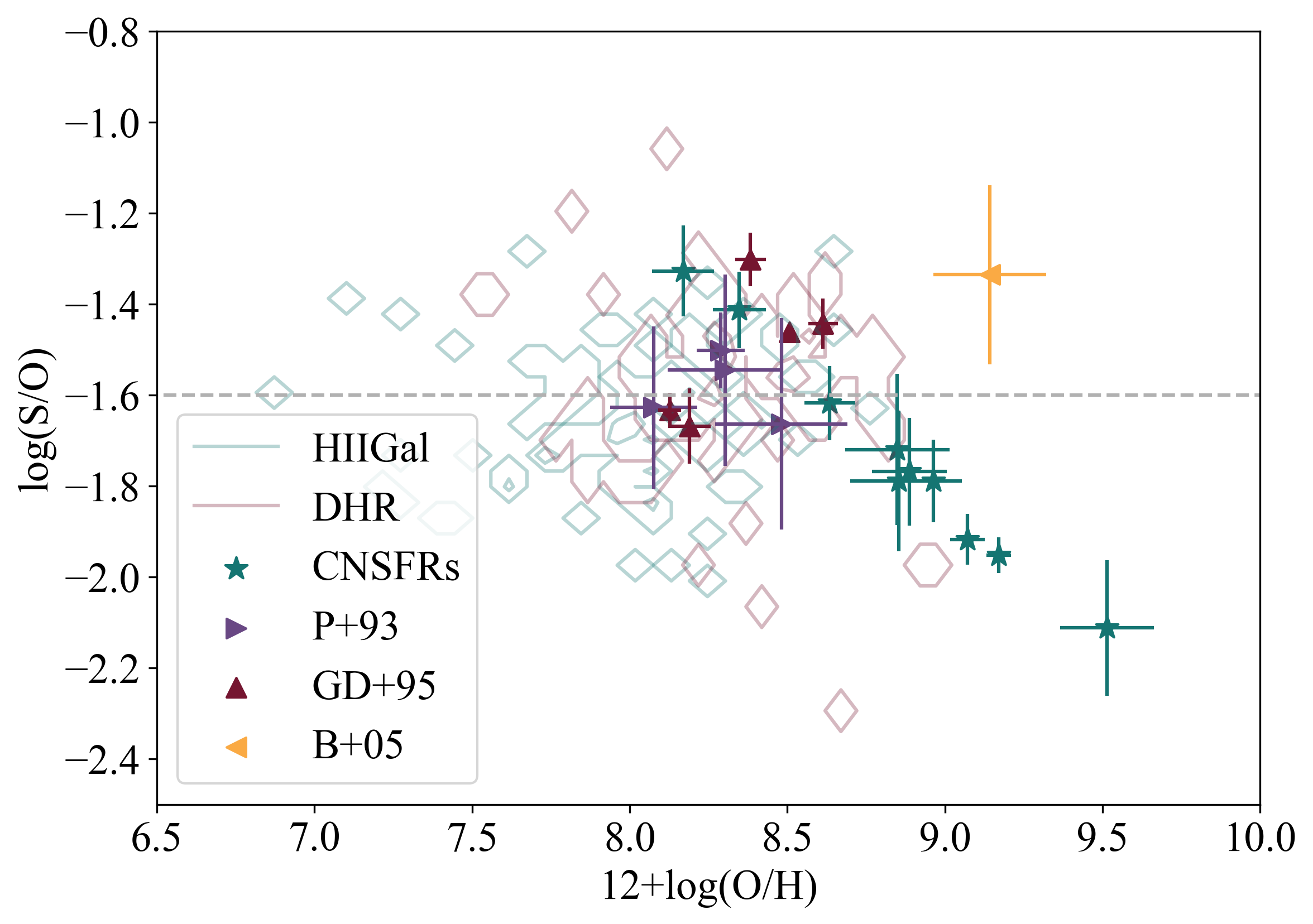}
\includegraphics[width=0.98\columnwidth]{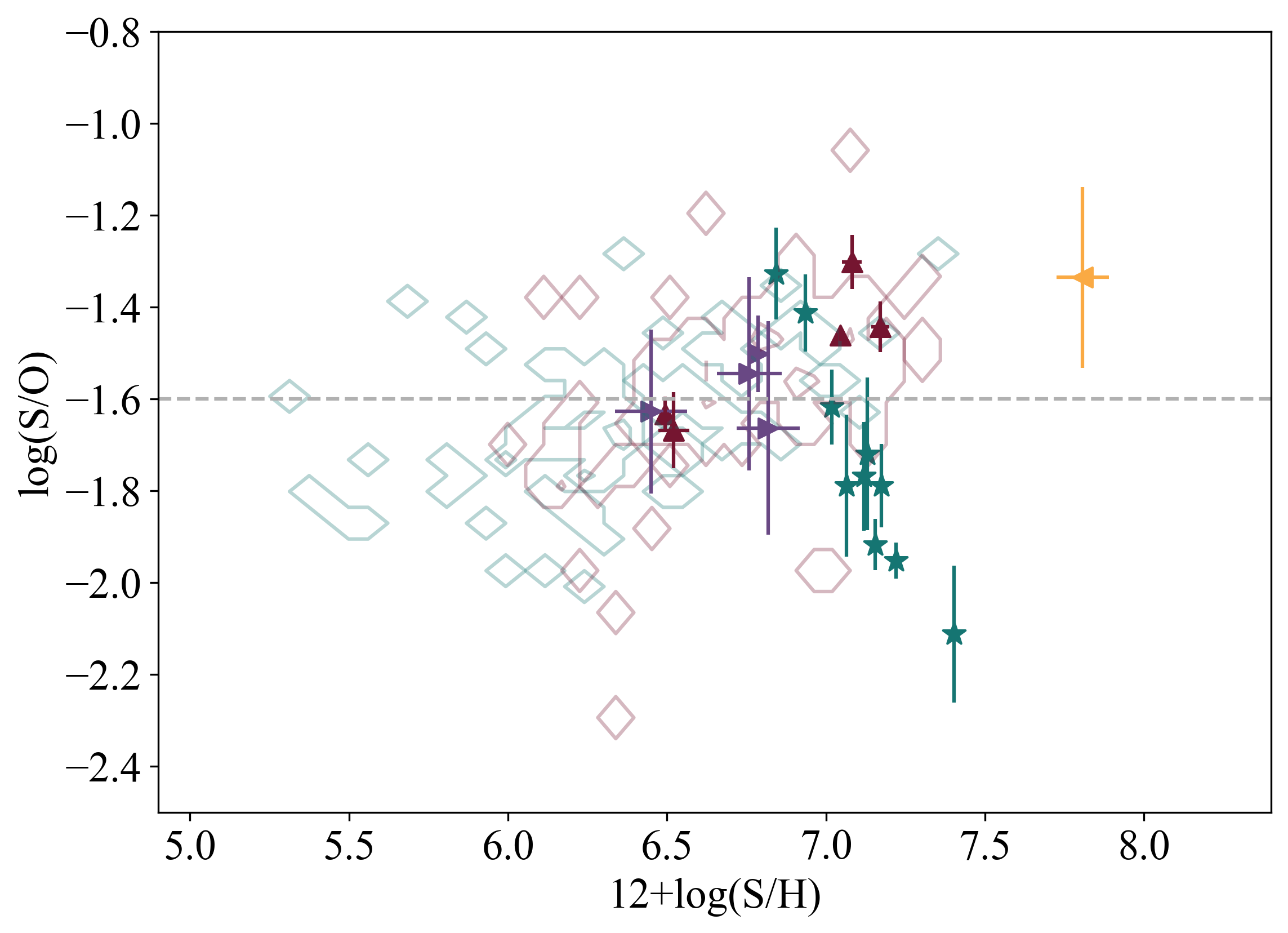}
\caption{S/O relation against the total abundances of oxygen (upper panel) and sulphur (lower panel) for regions included in Table \ref{tab:oxygen abundances} (green stars) and data from the literature as labelled \citep[][(P+93), (GD+95) and (B+05) respectively]{1993MNRAS.260..177P,1995ApJ...439..604G,2005AandA...441..981B}. The black dashed line in each panel marks the solar S/O ratio.}
 \label{fig:S_O}
\end{figure}

In order to compare our results with other circumnuclear regions from the literature \citep{1993MNRAS.260..177P,1995ApJ...439..604G,2007MNRAS.382..251D,2005AandA...441..981B} we have used published emission line measurements and calculated their abundances following the analysis proposed in this work. The $O^{+}/H^{+}$ has been determined from the [OII]$\lambda \lambda$ 3727,29 \AA\ lines and the equation given in \citet{2022MNRAS.511.4377D}. For a few objects we could verify that O$^+$/H$^+$ ratios calculated using the  blue and red lines of [OII] are compatible within the errors.

Finally, we have calculated the S/O ratios that are plotted in Figure \ref{fig:S_O} as a function of both sulphur and oxygen abundance in the upper and lower panel respectively. Both graphs show a clear tendency: lower S/O ratios for higher metallicities, for abundances larger than solar. This effect has already been noticed in other works \citep{1991MNRAS.253..245D,1997A&A...322...41C,2002astro.ph.11148G,2002A&A...391.1081V,2006MNRAS.367.1139P,2022MNRAS.511.4377D}. This could be due to an overestimation of the derived oxygen total abundances. In fact, as shown by \citet{2005A&A...434..507S} for metallicities larger than solar, directly derived oxygen abundances using photoionisation models deviate greatly from input abundances. However, if this is not the case, the observed S/O lower values for high metallicity regions should be explained almost only by stellar nucleosynthesis \citep{1988A&A...197...33T}.

\begin{table*}
\centering
\caption{Oxygen abundances and sulphur to oxygen ratios for the observed CNSFRs. }
\label{tab:oxygen abundances}
\begin{tabular}{cccccccc}
\hline 
Region ID & [OII]$\lambda \lambda$ 7320,30/H$\beta ^a$ & [I($\lambda $ 7320,30)/H$\beta$]$_R ^a$&t$_e$([OIII])$^b$ & 12+log(O$^+$/H$^+$) &  12+log(O$^{++}$/H$^+$) & 12+log(O/H) & log(S/O)\\ \hline

R1* & 23.9 $\pm$ 2.2 & 0.69 & 0.661 $\pm$ 0.072 & 9.145 $\pm$ 0.041 & 7.946 $\pm$ 0.009 & 9.171 $\pm$ 0.039 & -1.952 $\pm$ 0.039 \\ 
R2 & 23.1 $\pm$ 3.0 & 0.58 & 0.677 $\pm$ 0.073 & 9.043 $\pm$ 0.059 & 7.872 $\pm$ 0.017 & 9.072 $\pm$ 0.055 & -1.917 $\pm$ 0.056 \\ 
R3 & 10.0 $\pm$ 2.9 & 0.39 & 0.646 $\pm$ 0.072 & 8.858 $\pm$ 0.126 & 7.708 $\pm$ 0.019 & 8.888 $\pm$ 0.118 & -1.768 $\pm$ 0.118 \\ 
R4 & 11.1 $\pm$ 2.5 & 0.50 & 0.641 $\pm$ 0.072 & 8.931 $\pm$ 0.097 & 7.818 $\pm$ 0.014 & 8.963 $\pm$ 0.090 & -1.789 $\pm$ 0.090 \\ 
R5 & 11.3 $\pm$ 2.3 & 0.32 & 0.704 $\pm$ 0.073 & 8.592 $\pm$ 0.090 & 7.610 $\pm$ 0.018 & 8.635 $\pm$ 0.081 & -1.617 $\pm$ 0.082 \\ 
R6 & 9.3 $\pm$ 2.0 & 0.23 & 0.750 $\pm$ 0.075 & 8.290 $\pm$ 0.095 & 7.447 $\pm$ 0.019 & 8.348 $\pm$ 0.084 & -1.412 $\pm$ 0.084 \\ 
R13 & 14.0 $\pm$ 5.0 & 1.14 & 0.571 $\pm$ 0.070 & 9.493 $\pm$ 0.156 & 8.194 $\pm$ 0.015 & 9.514 $\pm$ 0.148 & -2.112 $\pm$ 0.149 \\ 
R14 & 11.9 $\pm$ 4.6 & 0.42 & 0.666 $\pm$ 0.072 & 8.819 $\pm$ 0.167 & 7.728 $\pm$ 0.022 & 8.853 $\pm$ 0.154 & -1.789 $\pm$ 0.155 \\ 
R17 & 11.0 $\pm$ 2.8 & 0.17 & 0.813 $\pm$ 0.077 & 8.107 $\pm$ 0.114 & 7.294 $\pm$ 0.024 & 8.169 $\pm$ 0.099 & -1.327 $\pm$ 0.100 \\ 
R26 & 11.0 $\pm$ 4.5 & 0.37 & 0.660 $\pm$ 0.072 & 8.818 $\pm$ 0.178 & 7.678 $\pm$ 0.024 & 8.849 $\pm$ 0.166 & -1.720 $\pm$ 0.166 \\ 

\hline
\end{tabular}
\begin{tablenotes}
\centering
\item $^a$ Values normalized to I(H$\beta$) 10$^{-3}$. 
\item $^b$ In units of 10$^{4}$ K.\\
\item * Region near SN explosion.
\end{tablenotes}
\end{table*}

\subsection{Ionising cluster properties}
\label{clusters}
The temperature of the ionising stars can be mapped using the $\eta$ parameter defined by \citet{1988MNRAS.231..257V} as: 
\begin{equation}
    \eta = \frac{O^{+}/O^{++}}{S^{+}/S^{++}}
\end{equation}

We can calculate this parameter for only 10 ring HII regions (see Sec. \ref{sec_dis_abundances}) and their logarithmic values are between 0.78 $\pm$ 0.12 and 1.498 $\pm$ 0.043, on the higher side of the distribution found by \citet{1995ApJ...439..604G}. A direct relation seems to exist between this parameter and the metallicity for a given region: $\eta$ is greater (and therefore the temperature of the ionising stars is lower) for regions with higher abundances. This behaviour has already been reported for a sample of HII galaxies by \citet[][and references therein]{2006A&A...457..477K} and for a large sample of HII galaxies and HII regions of different metallicity by \citet{2022MNRAS.511.4377D}. On the theoretical side it was already introduced by \citet{1991ApJ...380..140M} as a moderator in the empirical calibration of the R$_{23}$ parameter. 

\begin{figure}
\centering
\includegraphics[width=0.98\columnwidth]{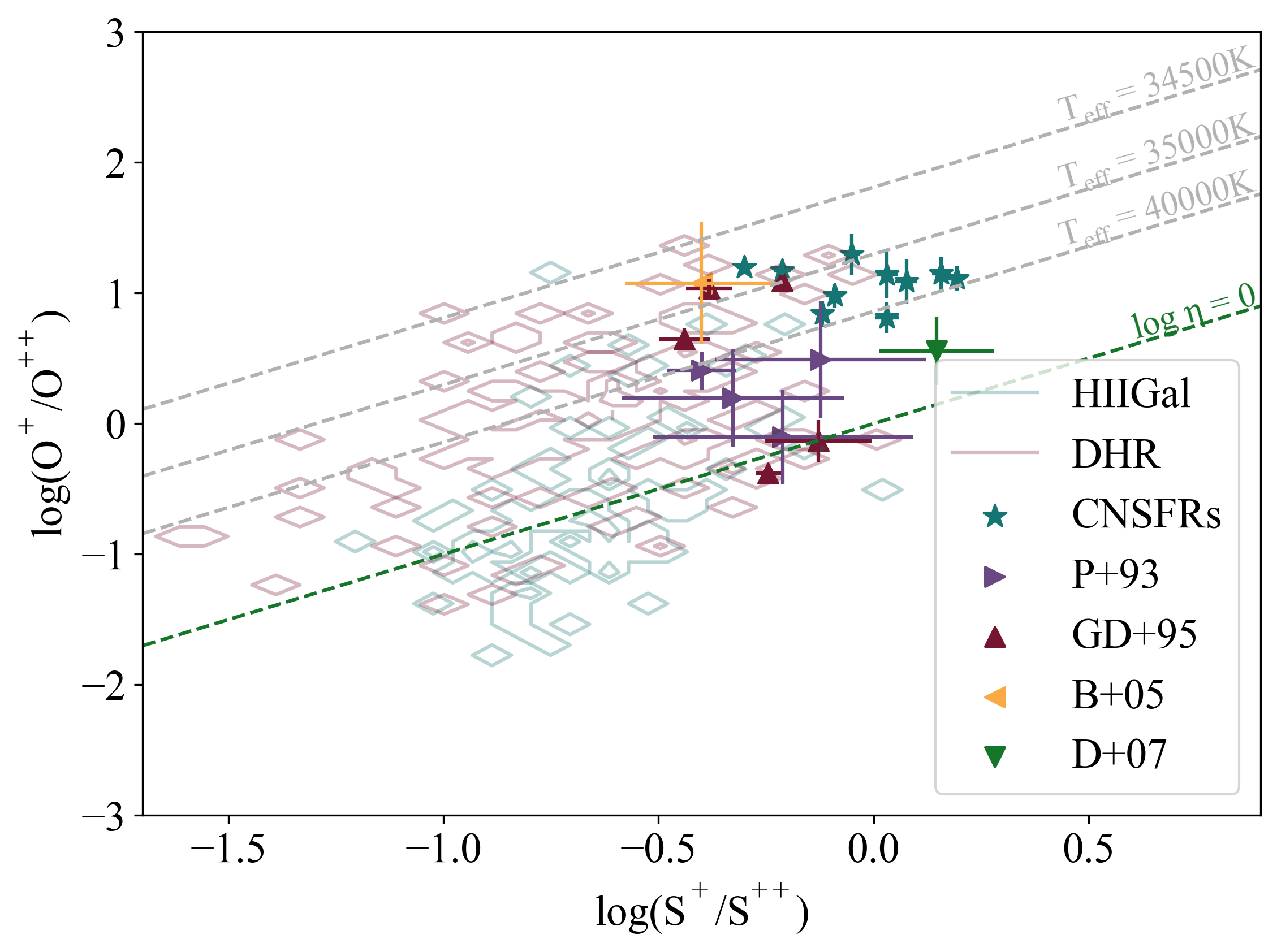}
\caption{Relation between the ionic ratios of oxygen and sulphur for the objects included in Table \ref{tab:oxygen abundances} (green stars) and data from the literature. Red and blue contours correspond to disc HII regions and HII galaxies respectively from \citet{2022MNRAS.511.4377D}. Circumnuclear regions are from \citet{1993MNRAS.260..177P} (P+93),  \citet{1995ApJ...439..604G} (GD+95), \citet{2005AandA...441..981B} (B+05) and \citet{2007MNRAS.382..251D} (D+07).} 
 \label{fig:eta}
\end{figure}

Fig. \ref{fig:eta} shows the relationship between the ionic ratios S$^+$/S$^{++}$ and O$^+$/O$^{++}$ for our ring regions and for other circumnuclear regions from the literature \citep{1993MNRAS.260..177P,1995ApJ...439..604G,2007MNRAS.382..251D,2005AandA...441..981B}. Superimposed are dotted diagonal lines which show the location of ionised regions with constant values of $\eta$. These values have been correlated with stellar effective temperatures of stars using the Cloudy code \citep[][log(u) = -4.0 - -2.5, Z$_\odot$, n$_e$ = 100 cm$^{-3}$]{cloudy} with stellar atmospheres from \citet[][non-LTE models for B and O stars, log(g) = 4 and T$_{eff}$ from 30000 K to 55000 K]{1978stat.book.....M}. Our circumnuclear regions are located at the high end of S$^+$/S$^{++}$ ratio distribution and show high values of the $\eta$ parameter, corresponding to relatively low stellar temperatures similar to high metallicity disc HII regions. This location corresponds to effective temperatures between 34700 K and 40000 K.

The spectral energy distribution of the ionising radiation implied by the $\eta$ parameter would correspond to an ionising star cluster equivalent temperature that can be estimated from the quotient between the number of helium and hydrogen ionising photons, Q(He$_0$)/Q(H$_0$). We have calculated the number of ionising He$_0$ photons from the observed luminosity in the HeI$\lambda$ 6678 \AA\ emission line using the expression:

\begin{equation}
Q(He_0)=1.21\cdot 10^{49}\left(\frac{F(HeI 6678 \AA)}{10^{-14}}\right)\left(\frac{D}{10}\right)^2
\end{equation}

\noindent where F(HeI6678), the flux of the HeI$\lambda$ 6678 \AA\ line, is expressed in erg s$^{-1}$ cm$^{-2}$ and D is the distance to NGC~7742 which has been taken as 22.2 Mpc (see Tab. \ref{tab:galaxy characteristics}). This equation has been derived using the recombination coefficient of HeI$\lambda$ 6678 \AA\ line assuming a constant value of electron density of 100 cm$^{-3}$, a temperature of 10$^4$ K and  case B recombination \citep{Osterbrock2006}.

\begin{figure}
\centering
\includegraphics[width=0.98\columnwidth]{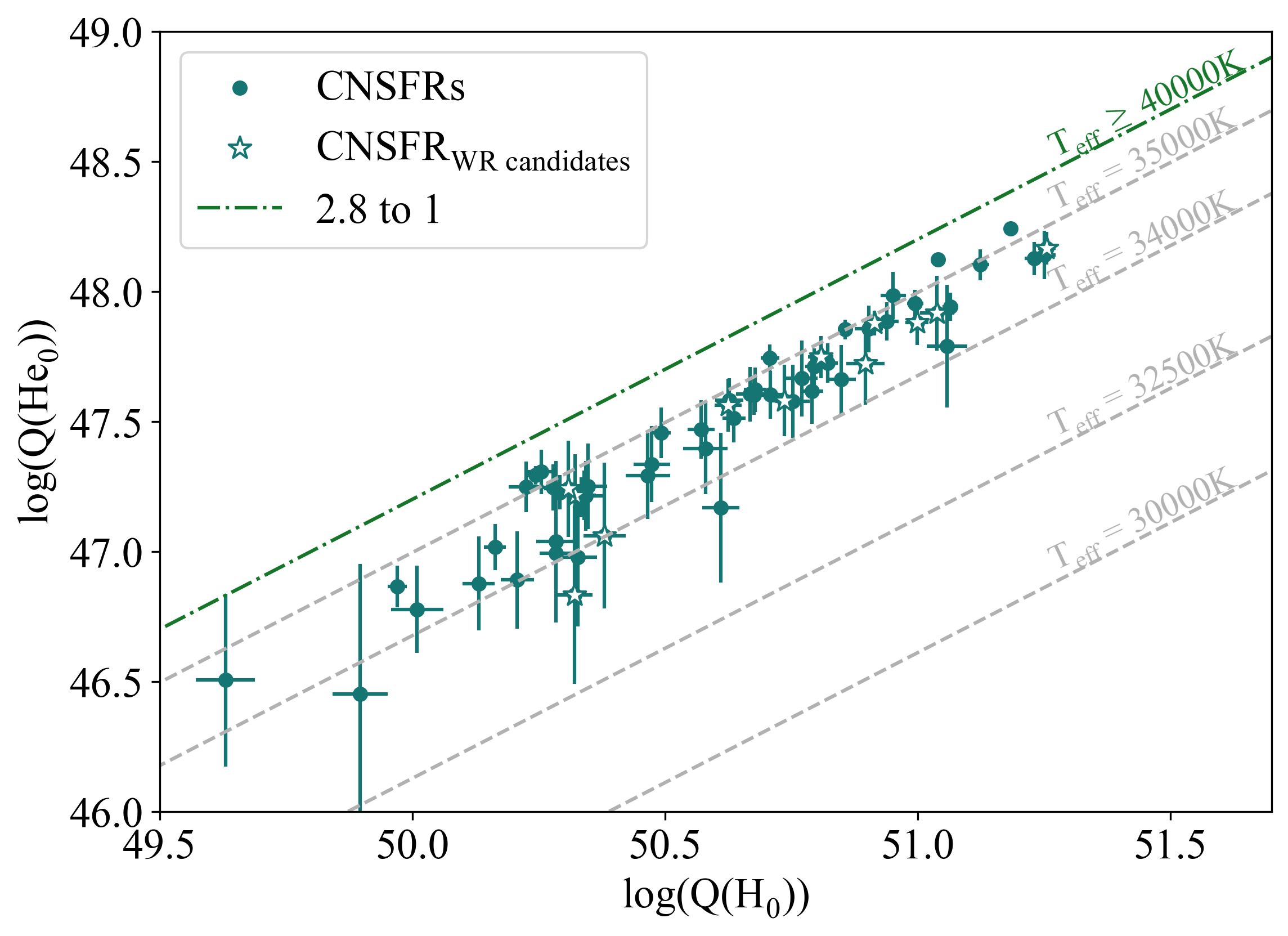}
\caption{Relation between the logarithmic numbers of HeI and HI ionising photons (see text for details).} 
 \label{fig:teff}
\end{figure}

We have detected and measured the HeI$\lambda$ 6678 \AA\ line in 63 ring regions. The corresponding values range from 2.8 $\times$ 10$^{46}$ to 1.7 $\times$ 10$^{48}$ with a mean value of 4.5 $\times$ 10$^{47}$ photons s$^{-1}$.

Using the Cloudy \citep{cloudy} code we have computed models with ionisation parameter values from -4.0 to -2.5, solar metallicity and a constant value of the electron density of 100 cm$^{-3}$. The nebula is ionised by stellar atmospheres from \citet[][non-LTE models for B and O stars, log(g) = 4 and T$_{eff}$ from 30000 K to 55000 K]{1978stat.book.....M}. For temperatures lower than 40000 K, the nebular zone of He$^+$ is smaller than that of H$^+$ and hence we can use the number of ionising hydrogen and helium photons to estimate the effective temperature of our ionising star clusters for which we have derived the following equation:
\begin{equation}
\begin{split}
log\left(\frac{Q(H_0)}{Q(He_0)}\right) = (1.944\pm0.284)\cdot 10^{-8}\cdot T_{eff}^2\\
(-1.527\pm 0.199)\cdot 10^{-3}\cdot T_{eff}+(32.77\pm3.46)
\end{split}
\end{equation}
This relation can be used only for logarithmic values of Q(H$_0$)/Q(He$_0$) greater than 2.8 since, for higher temperatures, the ionisation zones of helium and hydrogen coincide and the relationship between them remains constant. Fig. \ref{fig:teff} shows the number of  He$_0$ ionising photons as a function of the number of H$_0$ ionising photons. Superimposed are 
 the lines corresponding to different temperatures as obtained with the last equation. According to these models, we can deduce that the He$^+$ nebular zone is much smaller, approximately in a ratio r$_{He}$/r$_H$ $\sim $ 0.73, than that of H$^+$ in all these star clusters and they seem to have similar effective temperatures, around 34600 K, a result similar with that obtained with the $\eta$ parameter.

The equivalent width (EW) of Balmer lines can be understood as an estimator of the age of a young cluster single stellar population \citep{1981Ap&SS..80..267D} reflecting the ratio between the present and past star formation rates. Following the notation used in Sec. \ref{sec:line measurements}, the equivalent width has been calculated as: $EW(\lambda) = F_\lambda/(F_c(\lambda)+A_c(\lambda))$. The equivalent widths of the H$\beta$ emission line for the selected ring HII regions are between 2.5 to 44.0 \AA\ with a mean value of 10.9 \AA\ corresponding to regions of active star formation. Although the Balmer emission line luminosities are higher for ring regions, on average, the H$\beta$ equivalent widths for regions outside the ring are comparable within the errors with a mean value of 12.4 \AA\ covering values from 2.03  to 140.5 \AA . This might seem to point out to HII regions outside the ring being on the same evolutionary stage with similar percentages of young populations, it should be noted that the H$\beta$ equivalent width depends also on the underlying continuum and the area covered by the ring shows an additional blue population that could be affecting the results (see lower left panel of Fig. \ref{fig:Ha_OIII_map}).

In principle, we would expect that the regions inside the ring, being closer to the galactic nucleus, were redder, with a higher metal content and with older ages \citep{2018A&A...609A.102R} than the regions outside the ring; however their r-i colours are comparable (see Sec. \ref{sec:int_flux}, Fig. \ref{fig:color_mag}). Thus, all the regions analysed in our sample seem to have similar ages and metallicities, in spite of its distance to the centre of the galaxy. In fact, this can be seen by looking at the radial r and i magnitude profiles shown by \citet{1996ASPC...91...83W} which follow each other.

A linear regression can be performed between the EW(H$\beta$)  and the number of ionising photons in order to estimate the ionising masses of our circumnuclear HII regions. We have used single stellar population (SSP) PopStar models \citep{Popstar} to fit the following equation:
\begin{equation}
\label{ion-mass}
log\left[Q(H)/M_\odot\right] = a+b\cdot log\left[EW(H\beta)\right]
\end{equation}
where $Q(H)/M_\odot$ is the total number of ionising photons per unit mass.
This relation is well established for ages under 10 Ma and metallicities between 0.004 and 0.02. Its linearity can be lost: (i) for higher metallicities, because the effective temperature decreases and hence stars of the same spectral type with more metals have fewer ionising photons; (ii) for lower metallicities, because there are more massive stars and clusters are hotter, showing a significant nebular continuum; and (iii) due to the presence of Wolf Rayet (WR) stars. The slope of the initial mass function (IMF) and the lower mass limit also affect the numbers of stars of different types (and therefore their number of ionising photons). We have used the different IMFs listed in Tabé \ref{tab:masa_ionizante_salp2}. Finally, we have selected ages lower than 7 Ma since, as explained above, star clusters older than that do not produce significant ionising radiation.

\begin{figure}
\includegraphics[width=\columnwidth]{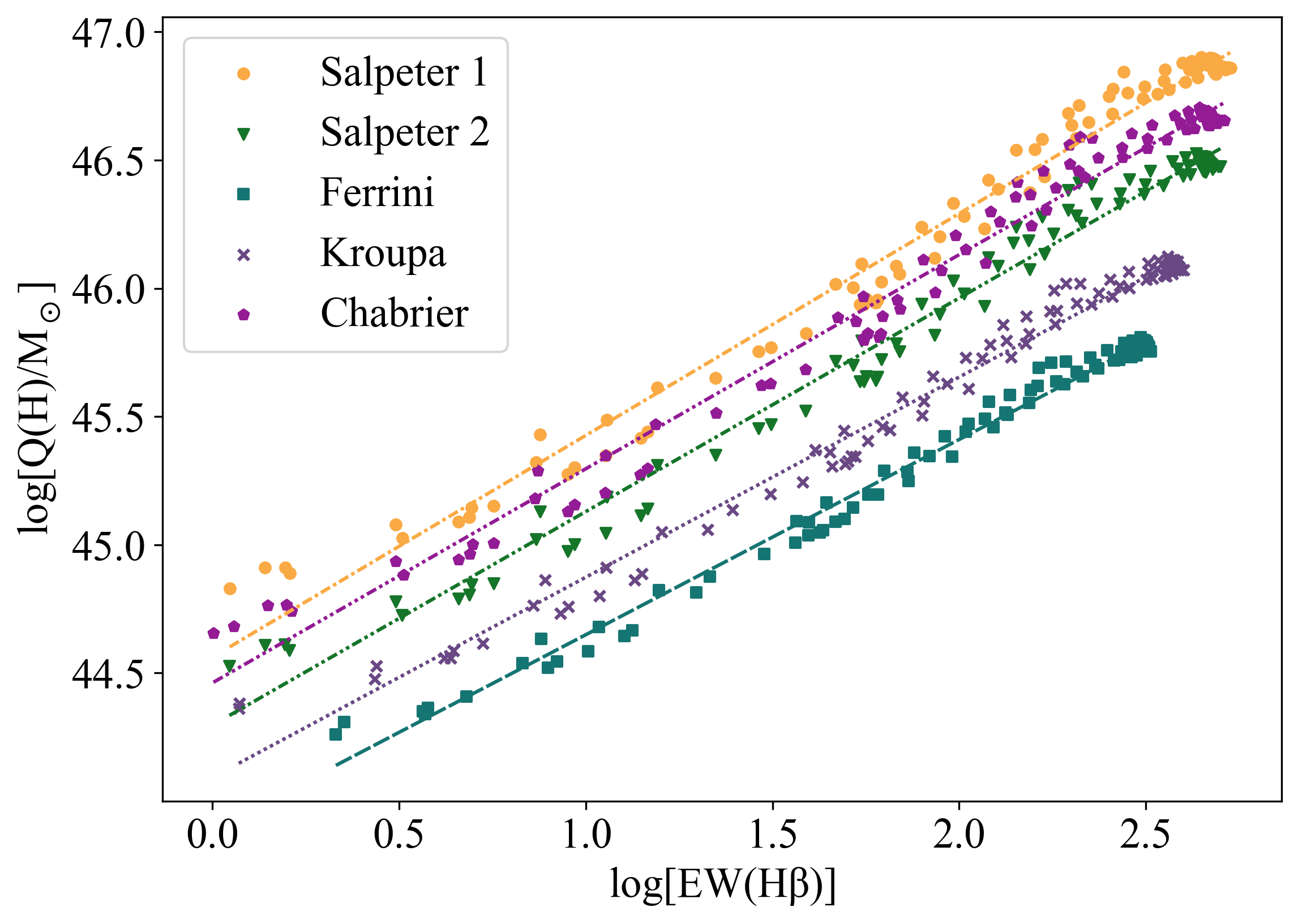}
 \caption{Linear regression between the EW(H$\beta$) and the number of ionising photons for different IMFs according to Tab. \ref{tab:masa_ionizante_salp2} as labelled.}
 \label{fig:masa_cluster}
\end{figure}

\begin{table*}
\centering
\caption{Ionising masses fitting.}
\label{tab:masa_ionizante_salp2}
\begin{tabular}{cccccc}
\hline
IMF & Reference & \begin{tabular}[c]{@{}c@{}}m$_{low}$\\ (M$_\odot$)\end{tabular}&\begin{tabular}[c]{@{}c@{}}m$_{up}$\\ (M$_\odot$)\end{tabular}& a & b \\
\hline
Salpeter1 & \citet{Salpeter1955} & 0.85 & 120 & 44.561$\pm$ 0.025 & 0.865$\pm$ 0.012\\
Salpeter2 & \citet{Salpeter1955} & 0.15 & 100 & 44.296$\pm$ 0.024 &  0.833 $\pm$ 0.012 \\
Ferrini & \citet{Ferrini1990} & 0.15 & 100 &43.887$\pm$ 0.017 &  0.762 $\pm$ 0.009 \\
Kroupa & \citet{Kroupa2002} & 0.15 & 100 & 44.092$\pm$ 0.021&  0.781$\pm$0.010 \\
Chabrier & \citet{Chabrier2003} & 0.15 & 100 & 44.461$\pm$0.024 &  0.835$\pm$ 0.012\\
\hline
\end{tabular}
\end{table*}

The different relations defined in Tab. \ref{tab:masa_ionizante_salp2} by equation \ref{ion-mass} are shown in Fig. \ref{fig:masa_cluster}. The linear regression slopes are very similar among them and also compatible with previous results obtained by \citet{Angeles1998} using stellar populations synthesis models from \citet{1994A&A...284..749C}, \citet{1995A&AS..112...13G} and \citet{1995ApJS...96....9L}. However, we can see that the linear regression intercepts differ by a factor of 5 ($\sim$ 0.7 dex) depending on the chosen IMF.

We have used the Salpeter IMF with $\phi (m) = m^{-\alpha}$, $\alpha = 2.35$, $m_{low}(M_\odot)$ = 0.85 and $m_{up}(M_\odot)$ = 120 that seems the most suitable for our young regions. For regions within the ring we have obtained values between 1.22 $\times$ 10$^4$ (R76) and 5.93 $\times$ 10$^5$ $M_\odot$ (R41). These results are only lower limits to the ionising masses since we are assuming that: (i) there is no dust absorption and reemission at infrared wavelengths and (ii) we are considering there is no photon escape from HII regions \citep[but see][]{photon_scape}. Our derived values are lower than those obtained by \citet{2007MNRAS.382..251D} for CNSFRs and slightly higher than for HII regions  \citep{2000MNRAS.318..462D} both based on lower spatial resolution data (0.4" and 0.7" respectively). 

\begin{figure}
 \includegraphics[width=1\columnwidth]{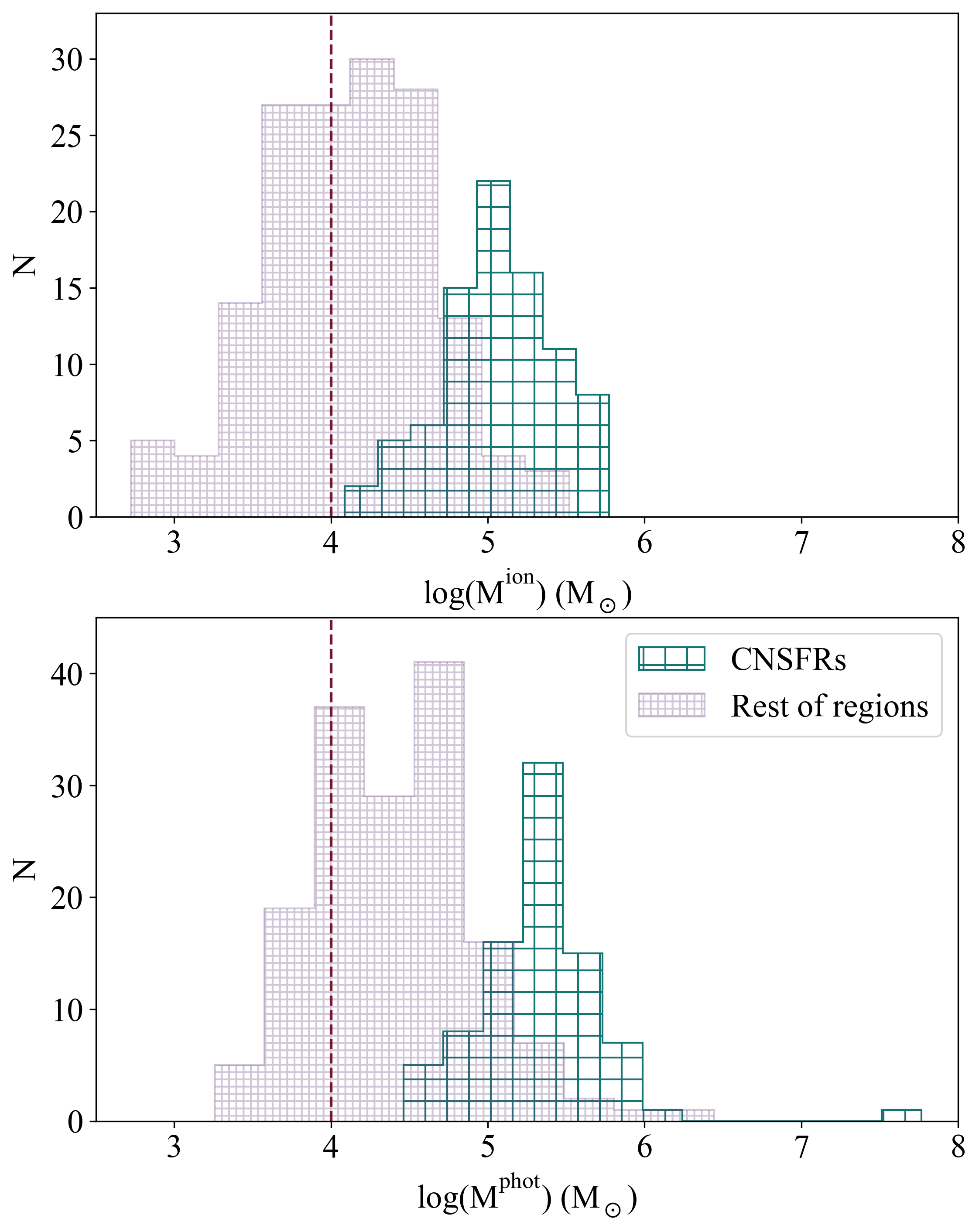}
 \caption{Histograms of the distributions of ionising masses (upper) and the photometric masses (bottom) for the ring HII regions, in green, and outside regions, in purple. The dashed line corresponds to 10$^4$ M$_\odot$ \citep[][see text for details]{1994ApJS...91..553G}.}
 \label{fig:ion_pho_masses}
\end{figure}

The upper panel of Fig. \ref{fig:ion_pho_masses} shows the distribution of ionising masses for the ring HII regions as compared with the ones outside the ring. We can see that masses for the latter ones are an order of magnitude lower with 1.38 $\times$ 10$^4$ and 1.22 $\times$ 10$^5$ as median values respectively. Around $\sim$ 50 \% of regions outside the ring have masses lower than 10$^4$ M$_\odot$, hence the IFM might not be fully sampled \citep{1994ApJS...91..553G,2010A&A...522A..49V}.
For the used stellar population models and the chosen IMF, the ratio of ionising stellar masses and ionised hydrogen masses, M$_{ion}$/M(HII), takes a value of 28. 

Using the same models, we have derived the photometric masses of our CNSFRs from their r-magnitudes. In this case, we cannot establish an analytical equation due to the non linearity between log[EW(H$\beta$)] and M$_r$+2.5$\cdot$log(M$_\odot$). The r magnitude for 1 M$_\odot$ seems to be constant for a chosen IMF although there are variations at ages between 3.5 and 6.5 Ma induced by the presence of WR and red super giants stars (RSG). For the regions within the ring we have obtained values between 2.90 $\times$ 10$^4$ and 1.10 $\times$ 10$^6$ $M_\odot$ corresponding to the two regions mentioned above at the extremes of ionising star masses.  As expected, these latter ones show photometric masses an order of magnitude lower than the former since photometric masses follow ionising star masses in a constant proportion of about 3. An exception to this is the case of R40, that shows the highest value of the photometric mass (7.77 in the log). The physical properties of the ionising cluster are fully compatible with the rest of the ring observed  regions. However, the extracted aperture encloses a non ionising star cluster. 
The bottom panel of Fig. \ref{fig:ion_pho_masses} shows the distribution of the photometric masses for the ring HII regions as compared with the ones outside the ring. 

\begin{table*}
\centering
\caption{Ionising cluster properties. The complete table is available online; here only a part is shown as an example.}
\label{tab:cluster properties}
\begin{tabular}{cccccc}
\hline 

Region ID &log($\eta$) & \begin{tabular}[c]{@{}c@{}}Q(He$_0$) \\ (photons s$^{-1}$)\end{tabular}& \begin{tabular}[c]{@{}c@{}}EW(H$\beta$)  \\ (\AA )\end{tabular} &  \begin{tabular}[c]{@{}c@{}}M$_{ion}$ \\ (M$_\odot$)\end{tabular} &  \begin{tabular}[c]{@{}c@{}}M$_{phot}$ \\ (M$_\odot$)\end{tabular}\\ \hline

R1* & 1.498 $\pm$ 0.043 & (174.0 $\pm$ 6.9) $\times$ 10$^{46}$ & 43.99 $\pm$ 1.60 & (15.9 $\pm$ 1.3) $\times$ 10$^{4}$ & (26.3 $\pm$ 2.2) $\times$ 10$^{4}$\\ 
R2 & 1.383 $\pm$ 0.064 & (132.7 $\pm$ 7.7) $\times$ 10$^{46}$ & 31.80 $\pm$ 1.31 & (15.1 $\pm$ 1.3) $\times$ 10$^{4}$ & (25.9 $\pm$ 2.5) $\times$ 10$^{4}$\\ 
R3 & 0.992 $\pm$ 0.130 & (14.7 $\pm$ 2.1) $\times$ 10$^{47}$ & 19.30 $\pm$ 0.83 & (38.2 $\pm$ 3.3) $\times$ 10$^{4}$ & (66.1 $\pm$ 8.6) $\times$ 10$^{4}$\\ 
R4 & 0.918 $\pm$ 0.100 & (75.4 $\pm$ 8.2) $\times$ 10$^{46}$ & 20.07 $\pm$ 0.76 & (16.8 $\pm$ 1.4) $\times$ 10$^{4}$ & (28.4 $\pm$ 3.6) $\times$ 10$^{4}$\\ 
R5 & 1.072 $\pm$ 0.093 & (71.3 $\pm$ 6.2) $\times$ 10$^{46}$ & 23.64 $\pm$ 0.99 & (12.8 $\pm$ 1.1) $\times$ 10$^{4}$ & (22.8 $\pm$ 2.4) $\times$ 10$^{4}$\\ 
R6 & 0.961 $\pm$ 0.099 & (19.6 $\pm$ 1.6) $\times$ 10$^{46}$ & 28.25 $\pm$ 1.72 & (26.8 $\pm$ 2.6) $\times$ 10$^{3}$ & (43.2 $\pm$ 4.4) $\times$ 10$^{3}$\\ 
R7 & - & (9.0 $\pm$ 1.1) $\times$ 10$^{47}$ & 17.01 $\pm$ 0.63 & (23.4 $\pm$ 1.9) $\times$ 10$^{4}$ & (38.5 $\pm$ 5.5) $\times$ 10$^{4}$\\ 
R8 & - & (12.7 $\pm$ 1.7) $\times$ 10$^{47}$ & 16.04 $\pm$ 0.63 & (33.1 $\pm$ 2.8) $\times$ 10$^{4}$ & (52.4 $\pm$ 8.6) $\times$ 10$^{4}$\\ 
R9 & - & (13.4 $\pm$ 2.0) $\times$ 10$^{47}$ & 15.97 $\pm$ 0.67 & (42.6 $\pm$ 3.7) $\times$ 10$^{4}$ & (6.2 $\pm$ 1.0) $\times$ 10$^{5}$\\ 
R10 & - & (8.7 $\pm$ 1.1) $\times$ 10$^{47}$ & 18.21 $\pm$ 0.69 & (25.9 $\pm$ 2.2) $\times$ 10$^{4}$ & (42.4 $\pm$ 5.8) $\times$ 10$^{4}$\\

\hline
\end{tabular}
\begin{tablenotes}
\centering
\item * Region near SN explosion.
\end{tablenotes}
\end{table*}

Table \ref{tab:cluster properties} shows the ionising cluster properties for each HII region within the ring and lists in columns 1 to 6: (1) the region ID, (2) the logarithmic $\eta$ parameter, (3) the number of helium ionising photons, (4) the measured equivalent width of H$\beta$ line, (5) the ionising mass and (6) the photometric mass.

\subsection{CNSFR evolutionary stage}

\begin{figure}
\includegraphics[width=1\columnwidth]{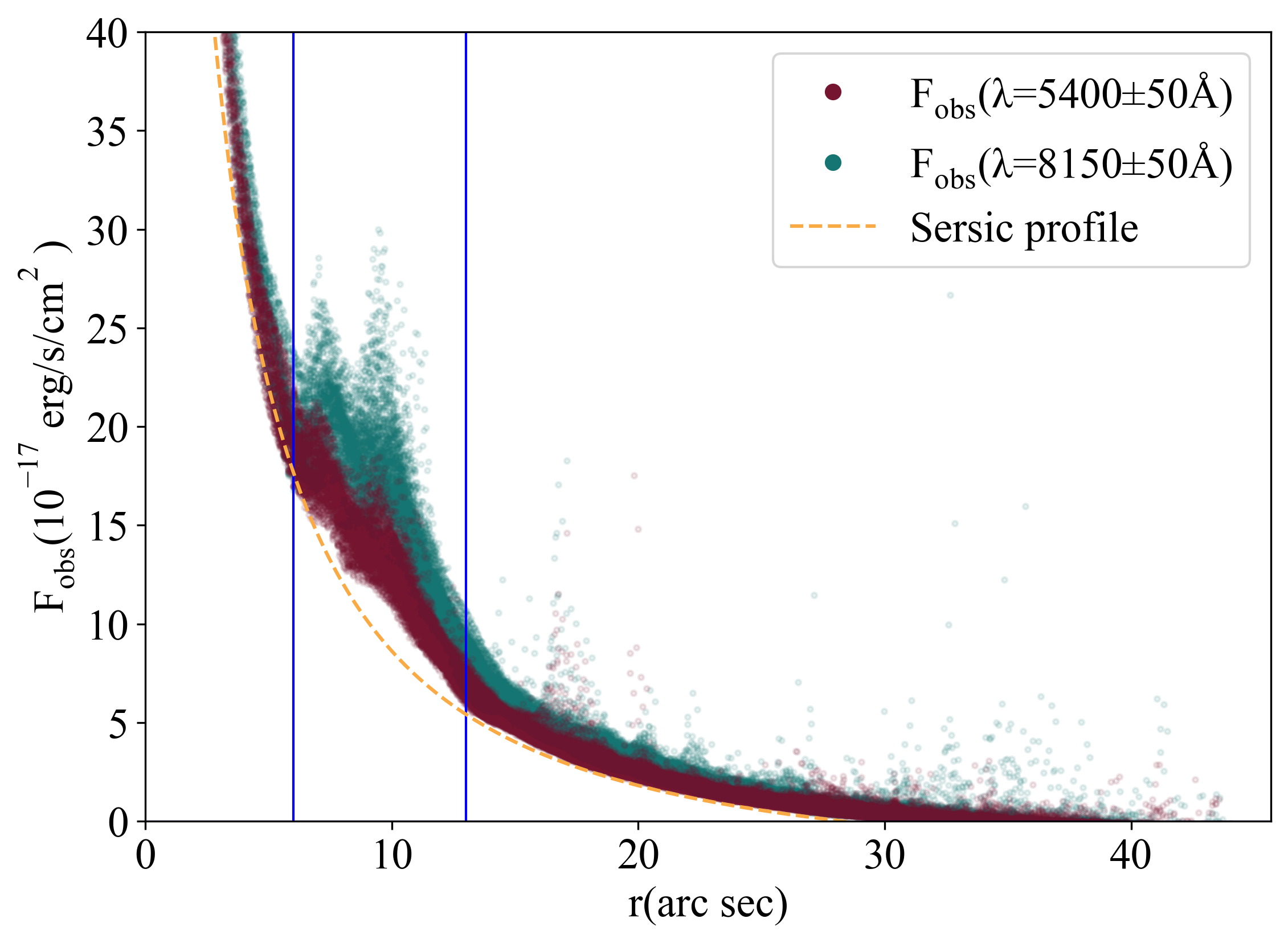}
 \caption{Observed continuum fluxes in individual spaxels as a function of radius in the blue (5400\AA\ ) and red (8150\AA ) spectral ranges. The ring limits are marked with blue vertical lines.}
 \label{fig:cont_profile}
\end{figure}

Up to now, we have assumed in our analysis the presence of a single population. We have checked this hypothesis by looking for the presence of a non-ionising population. In order to do that, we have constructed a pixel-to-pixel intensity profile from the 5400 and 8150 \AA\ continuum maps (see Fig. \ref{fig:Ha_OIII_map}) as can be seen in Fig. \ref{fig:cont_profile}. Two populations can be clearly identified in the ring region enclosed by blue vertical lines. One of them shows up very prominently at the blue continuum wavelength. We identify this flux excess with a young non-ionising population. This is accompanied by a moderate excess of continuum flux at the redder wavelengths that might correspond to the presence of red supergiant stars. To isolate the ionising clusters' contribution, we have corrected the integrated extracted spectra for the presence of the underlying non-ionising cluster population and the galaxy disc underneath that can be fitted by a Sersic light profile. Once this has been accomplished, we have recalculated the H$\beta$ equivalent width and the r and i magnitudes assuming for the stellar population the same extinction as that of the gas.

\begin{figure}
\includegraphics[width=\columnwidth]{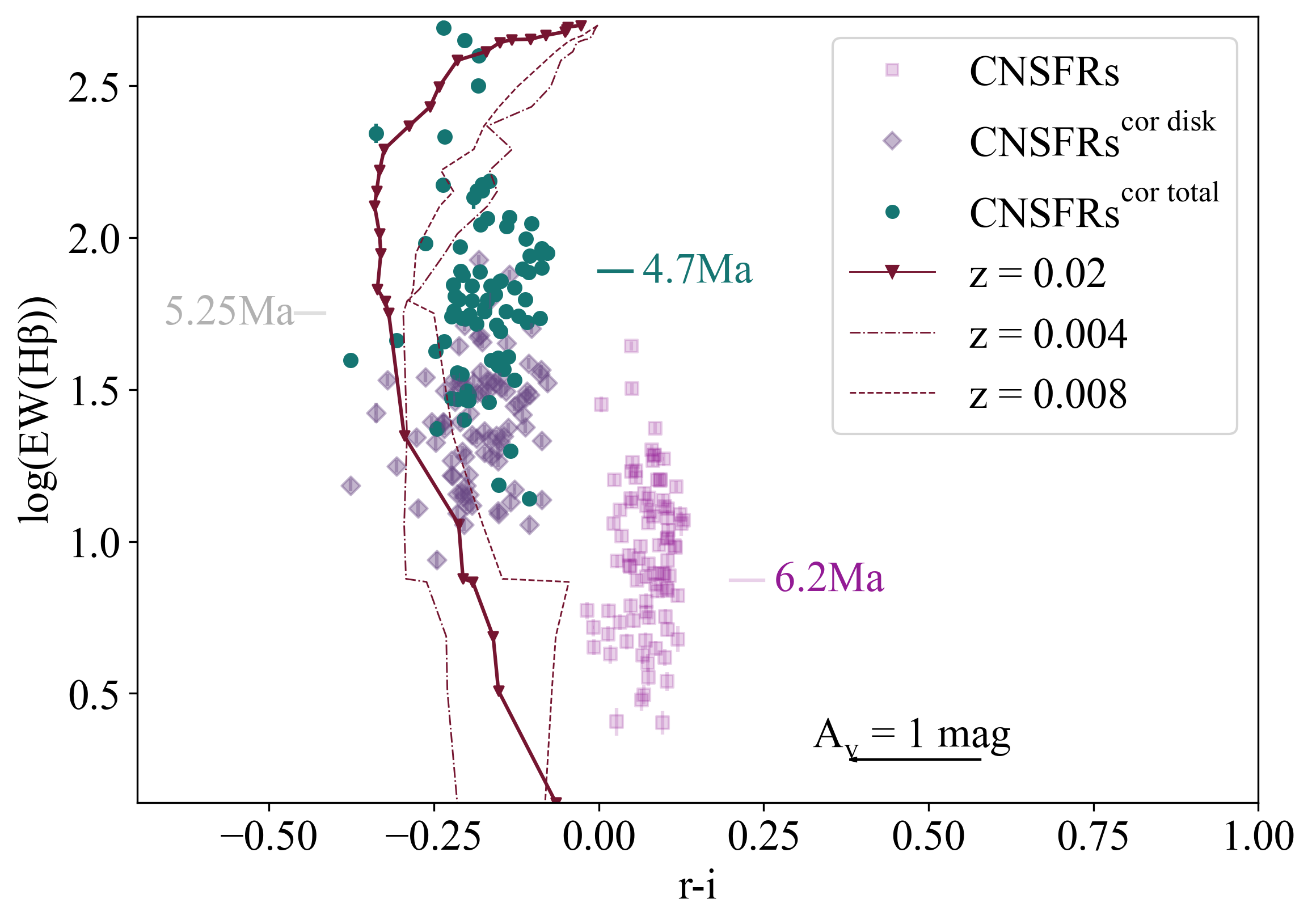}
\caption{The relation between the equivalent width of the H$\beta$ emission line and the r-i colour. The solid line has been calculated with PopStar models \citep{Popstar}. The beginning and end of the line correspond to ages of 0.1 and 8.5 Ma. Observational errors are inside the symbols in the graph.}
\label{fig:EWHb_color}
\end{figure}

Fig. \ref{fig:EWHb_color} shows the relation between the logarithm of the equivalent width of the H$\beta$ line, logEW(H$\beta$), and the r-i colour. Superimposed are single stellar population models from PopStar \citep[][Salpeter's IMF, m$_{low}$ = 0.15 M$_\odot$, m$_{up}$ = 100 M$_\odot$]{Popstar}. EW(H$\beta$) can be related to the time scale of the evolution of ionising star clusters, i.e. up to 10 Ma and, for a single stellar population, decreases with age. On the other hand, the r-i colour samples a longer time scale ( $\geq$ 300 Ma), becoming redder with age \citep[see][]{diaz2000}. For this reason, this graph can be understood as an age balance between the old and young population present in our clusters.

Our observed ring regions, taken at face value (square symbols in the graph), lie to the red from the line defined by single population stellar evolution models and show rather low values of logEW(H$\beta$). However, relaxing the assumption of a single stellar population and correcting for the underlying disc population and young non-ionising population identified above, the data points move up and to the left in the diagram, indicating younger ages for the ionising clusters. Regarding colour correction, we can see that the regions in the galaxy disc (square symbols on the graph) have the reddest colours while both the ring non-ionising population (diamond symbols) and the isolated young ionising clusters (solid circle symbols) show similar r-i values. According to the PopStar models we are using, these colour corresponds to stellar populations of about 300 Ma. The red colours shown by the youngest stellar populations are due to the contribution by a nebular continuum of up to 50 \%.
On the other hand, the EW(H$\beta$) of our ring regions, taken at face value, indicate mean ages of 6.2 Ma while the isolated young ionising population indicates younger ages with a mean value of 4.7 Ma. These latter ages are more consistent with model results than the former ones, since star clusters older than 5.2 Ma do not produce a detectable emission-line spectrum \citep[][]{2010MNRAS.403.2012M}. Composite young stellar populations have also been derived for CNSFRs in selected galaxies using FUV observations and a different methodology to the one described in our work \citep{2022AJ....164..208S}.

\begin{figure}
\centering
\includegraphics[width=0.8\columnwidth]{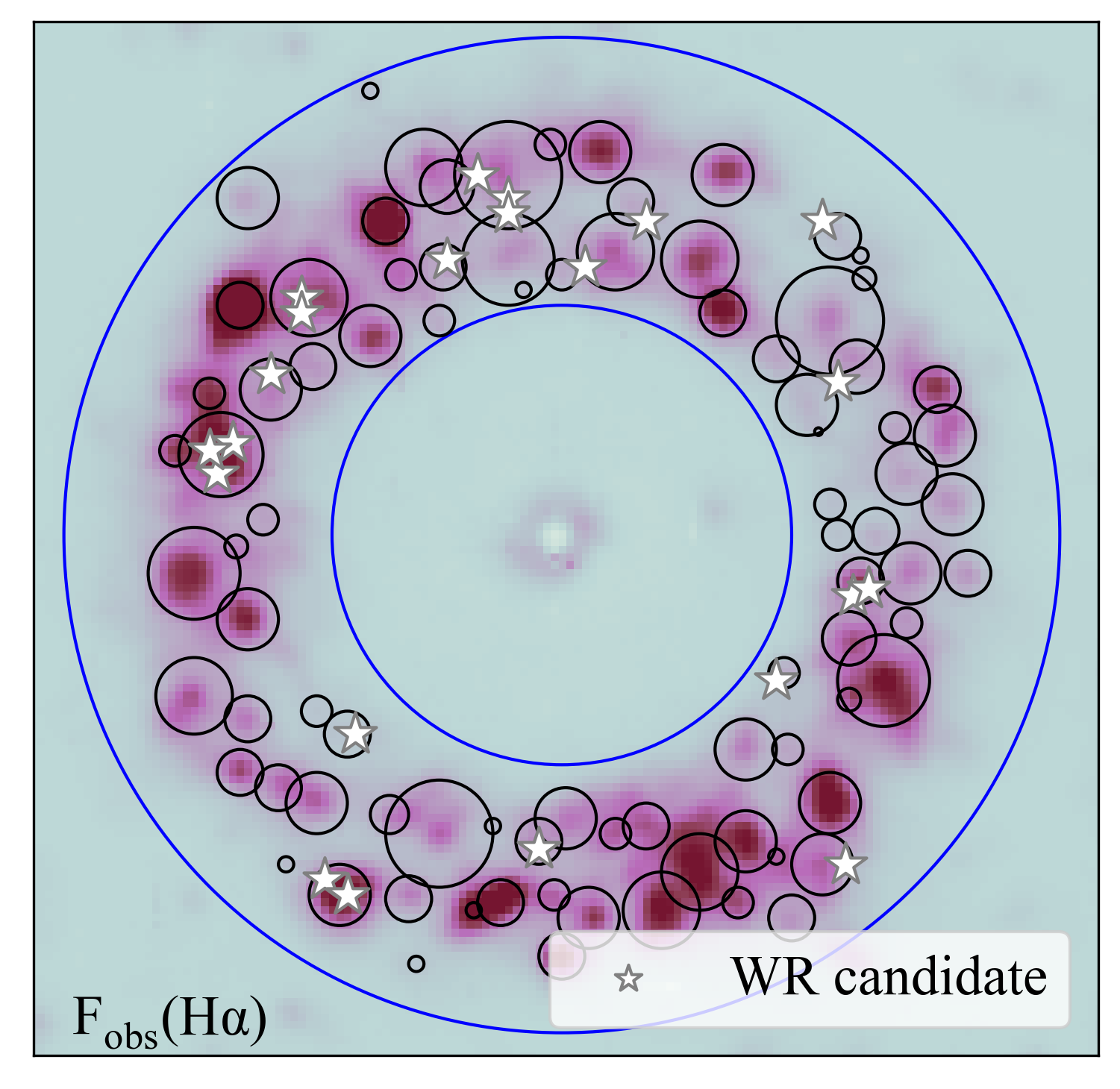}
 \includegraphics[width=\columnwidth]{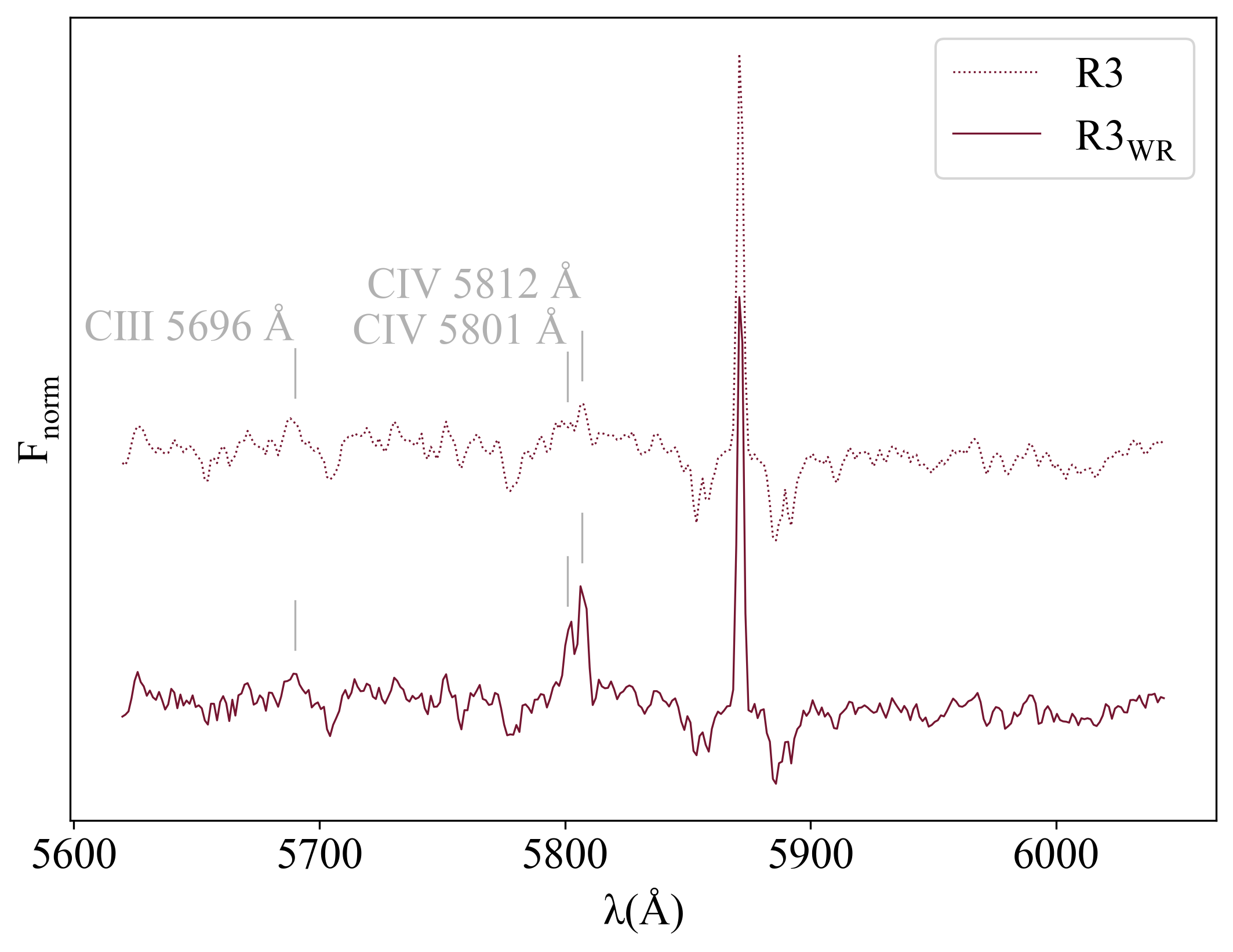}
 \caption{Upper panel: Map of the observed H$\alpha$ flux. HII regions and WR candidates are plotted with black circles and stars respectively. Orientation is North up, East to the left. The limits of the ring are marked by blue circles. Lower panel: Emission lines of WR stars in region R3 \citep[see][]{1992AJ....103.1159M}. Dashed and solid lines correspond to the integrated spectrum and the spectrum of areas showing the WR features respectively.}
 \label{fig:WR}
\end{figure}

Additionally, we have detected carbon Wolf-Rayet (WRC) star features in the spectra of some of the analysed ring HII regions. The upper panel of Fig. \ref{fig:WR} shows the location of these regions, 15 in total, (regions R3, R4, R21, R26, R35, R36, R40, R50, R55, R56, R58, R70, R72, R78, R85; this last one was removed due to its position on the BPT diagram). The lower panel of Fig. \ref{fig:WR} shows an example of the R3 spectrum showing the CIV$\lambda \lambda$ 5801,12 \AA\ and CIII$\lambda$ 5696 \AA\ lines \citep[see][]{1992AJ....103.1159M}. The presence of these features places the age of the regions between 3.2 and 5.25 Ma according to the PopStar models described above.

\begin{figure}
 \includegraphics[width=\columnwidth]{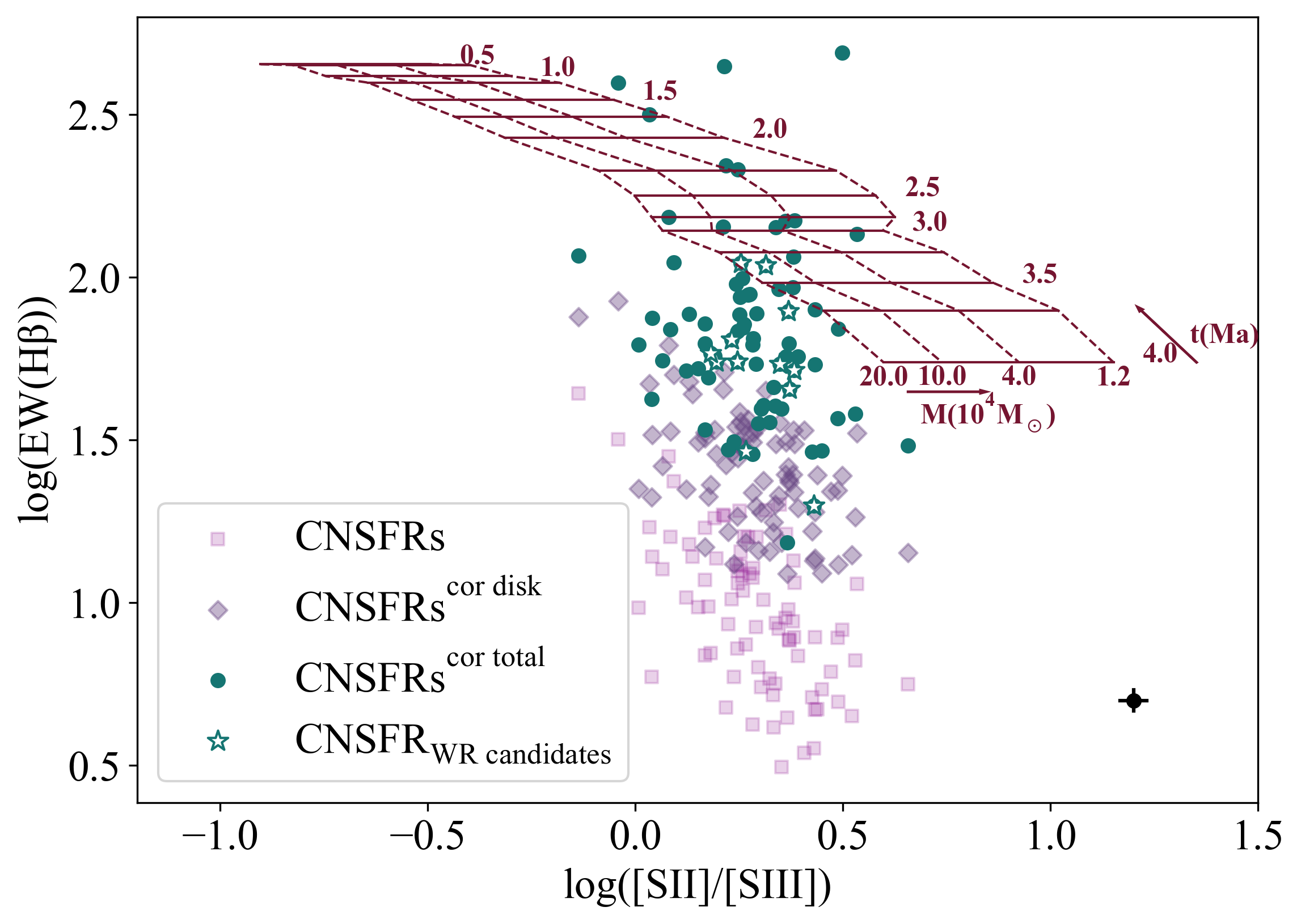}
 \caption{The relation between the equivalent width of the H$\beta$ emission line and the [SII]$\lambda \lambda $ 6717,31 \AA\ / [SIII]$\lambda \lambda $ 9069,9532 \AA\ ratio. Solid and dashed lines are from PopStar models with Cloudy and metallicity z = 0.02 \citep{2010MNRAS.403.2012M,2013MNRAS.432.2746G}. Mean error bars for CNSFRs are shown at the bottom right corner of the panel.}
 \label{fig:EWHb_SII_SIII}
\end{figure}

In order to better constrain the evolutionary stage of our CNSFRs, we have studied the evolution of the [SII]$\lambda \lambda $ 6717,31 \AA\ / [SIII]$\lambda \lambda $ 9069,9532 \AA\ ratio with the age of our ionising regions. The [SII]/[SIII] ratio is a good indicator of ionisation parameter and depends on ionising mass for zero age stellar populations and then decreases as the cluster evolves due to the increasing loss of ionising photons. 

Fig. \ref{fig:EWHb_SII_SIII} shows the relation between the equivalent width of the H$\beta$ emission line and the [SII]/[SIII] ratio. Superimposed are the same solar metallicity PopStar models used before. A trend between the degree of evolution and the degree of ionisation of the nebula seems to exist in the galaxy disc population (magenta squares in the graph). This effect was already noticed by \citet{2006MNRAS.365..454H} being explained by the different contributions of continuum light from underlying populations. The IFU data analysed here allow the subtraction of the disc and the young non-ionising stellar populations. Once this has been done, this trend is lost and the isolated young ionising clusters appear to cover the area occupied by the models. Furthermore, the CNSFRs with WR features are concentrated in a narrow range of ages around 3.5 Ma in agreement with the single stellar population models used. 

This correction also affects our initially derived ionising cluster masses and the photometric masses quoted in Section \ref{clusters}, through the number of ionising photons per unit solar mass which depends on EW(H$\beta$) for the former and the r-magnitude for the latter. The corrected values of the ionising cluster masses have a mean of 3.5 $\times$ 10$^4$ M$_{\odot}$, a factor of about 4.5 smaller than uncorrected ones; in the case of the photometric masses the corrected mean value is 1.8 $\times$ 10$^5$ M$_{\odot}$. This gives a ratio of ionising cluster mass to photometric mass of about 19 \%. 

\section{Conclusions}

In this work we present a study of the physical properties of the CNSFRs in the ring of the face-on spiral NGC~7742 using MUSE observations publicly available and the full spectral region observed, from 4800 to 9300 \AA . The work is centred in the study of the individual ionising clusters that power the HII regions populating the ring of the galaxy. We have used the data cubes from the ESO Science Archive to produce 2D maps in the H$\alpha$ and H$\beta$ emission lines obtaining the spatial distribution of the visual extinction necessary for the nebular analysis. Additionally, two continuum maps at central wavelengths 5400 and 8150 \AA\ and the line maps of [OIII] and [NII] are aso presented. A map of the EW(H$\alpha$) emission shows the circumnuclear regions within the ring, object of this study, having EW(H$\alpha$) > 20 \AA , consistent with the presence of star formation occurred less than 10 Ma ago. We have delimited the ring from the radial distribution of the observed H$\alpha$ flux, as having an inner radius of 6 arcsec (0.75 kpc) and an outer radius of 13 arcsec (1.63 kpc). The observed H$\alpha$ flux map has been used to select the ring ionised regions and also a set of HII regions external to it for comparison purposes. At the end of the procedure, we have obtained a total of 88 HII regions in the ring and 158 regions outside. The emission line ratios of the HII regions within the ring are consistent with the predictions of star forming models. However, maps of the central part of NGC~7742 in the [OIII] and  [NII] emission line ratios  allow the identification of a small circumnuclear ring at about 200 pc from the galaxy nucleus that seems to be dominated by shocks or an AGN non-thermal component of low activity. Three of our segmented ring regions to the South-East, R85, R86 and R87 may be somewhat affected by the radiation from the galaxy nucleus and consequently, they have not be considered in our analysis.

In order to study the properties of the selected CNSFRs, we have measured the most prominent emission lines in their spectra: H$\beta$ and H$\alpha$ Balmer lines; [OIII]$\lambda\lambda$ 4959,5007 \AA , [NII]$\lambda\lambda$ 6548,84 \AA , [SII]$\lambda\lambda$ 6716,31 \AA , [ArIII]$\lambda$ 7136 \AA\ and [SIII]$\lambda$ 9069 \AA\ forbidden lines and also the weaker lines of [SIII]$\lambda$ 6312 \AA , HeI$\lambda$ 6678 \AA and [OII]$\lambda\lambda$ 7320,30 \AA. We have calculated as well integrated fluxes inside the Sloan Digital Sky Survey (SDSS) filters for each selected region by convolving the appropriate filter transmission with their spectral energy distributions. A colour-magnitude diagram r-i vs M$_r$ shows the CNSFRs to have a rather constant value. 

For our observed ring HII regions we have derived: (1) the number of Hydrogen ionising photons per second, Q(H$_0$); (2) the electron density of the emitting gas per cubic centimeter, n$_e$; (3) the ionisation parameter, u; (4) the corresponding angular radius in arcsec; (5) the filling factor; and (6) the mass of ionised hydrogen in solar masses, M(HII). All these values are consistent with those found in other studies of similar regions.  Q(H$_0$) is between 2.57 $\times$ 10$^{49}$ and 1.51 $\times$ 10$^{51}$ which points to these regions being ionised by star clusters; the electron density of the ionised gas is well below the critical one for collisional deexcitation; the ionisation parameter is inside a narrow range centred around log(u) $\simeq$ -3.5; the estimated angular radii are in very good agreement with the measured ones, all of them spatially resolved, and show linear values between 34 and 130 pc; filling factors are low, with a mean value of 0.043, similar to the ones estimated for high metallicity disc HII regions; and, finally, the ionised hydrogen mass has a mean value of 3.07 $\times$ 10$^4$ M$_{\odot}$.

We have used sulphur as a tracer for chemical abundances of the selected HII regions using the methodology developed in \citet{2022MNRAS.511.4377D}, very adequate to the spectral characteristics of the MUSE spectroscopic data:  4800-9300 \AA\ wavelength range and spectral dispersion of 1.25 \AA /pix and the expected abundances of the studied regions, in the high metallicity range. The weak, temperature sensitive  [SIII]$\lambda$ 6312 \AA\ line has been measured with a S/N higher than 1 in $\sim$45 \% (38 out of 88) of the HII regions within the ring. For these regions total sulphur abundances have been derived by the so called "direct method". For the rest of the regions we had to rely on empirical calibrations to derive their sulphur abundances, which has been done through the use of the S$_{23}$ parameter that has little dependence on reddening effects or calibration uncertainties since the lines involved can be measured relative to nearby hydrogen recombination lines. Also, the lines are observable even at over-solar abundances given their lower dependence with electron temperature. Derived sulphur abundances are between 6.53 $\leq$ 12+log(S/H) $\leq$ 7.50, that is between 0.25 and 2.4 times the solar value, with most regions showing values slightly below solar. For a few ring HII regions we derived the total oxygen abundances using the [OII]$\lambda \lambda$ 7220,30 \AA\ to calculate the O$^+$ contribution. For three of the analysed regions oxygen abundances are found to be high, of the order of 12+log(O/H) around 9.0 (2 times solar) with a contribution by O$^+$ to the total abundance as high as 90 \%. These values reflect in very low S/O ratios. Similar values have been found for other high metallicity regions by different authors.

The final part of this work concerns the properties of the CNSFR ionising clusters. For the regions presenting the [OII]$\lambda\lambda$ 7220,30 \AA\ lines we derived the $\eta$ parameter that can be related to the effective temperature of the ionising radiation, finding values close to log($\eta$)  around 1.0, which implies low effective temperatures. An equivalent temperature of the ionised clusters can be estimated from the ratio of helium to hydrogen ionising photons,  Q(He$_0$)/Q(H$_0$). For 63 regions we could derive this ratio using the HeI$\lambda$ 6678 \AA\ line finding a rather constant value of around 10$^{-3}$, corresponding to an equivalent temperature below 40000 K. The masses of ionising clusters, once corrected for the contribution of underlying non-ionising populations, were derived using PopStar models and are found to have a mean value of 3.5 $\times$ 10$^4$ M$_{\odot}$, comparable to the mass of ionised gas and about 19 \% of the corrected photometric mass. The young stellar population of the CNSFRs has contributions of ionising and non-ionising populations in a ratio 0.24 with ages around 5 Ma and 300 Ma respectively.

The homogeneity of abundances and  continuum colours, together with the kinematics and counter rotating nature of the ring fits the minor merger scenario proposed by previous works. This merger would have triggered the star formation in the ring producing massive star clusters showing at present a young stellar population 300 Ma old accompanied by a subsequent young ionising population involving around 20 \% of the integrated cluster masses. Satellite accretion and major or minor mergers have also been suggested as the origin of the galaxy clumps observed at intermediate redshift \citep[see e.g.][]{2005ApJ...627..632E}. However, a recent study of clumps and accreted satellites in 53 star forming galaxies at z $\sim$ 1-3 \citep{2019MNRAS.489.2792Z} shows that, although the more extended clumps are probably formed in merger processes, the identified compact clumps formed \textit{in situ} show physical properties: sizes ($\sim$ 1-2 Kpc), masses ($\sim$ 10$^7$-10$^8$ M$_{\odot}$), ages $\leq$ 10 Ma and metallicities (12+log(O/H) $\simeq$ 8.56), more compatible with the total values found for the ensemble of ionising clusters studied here, as it would be observed at the quoted redshift. Obviously, more data, preferably at low to intermediate redshift, are needed in order to distinguished between these two hypotheses.

%%%%%%%%%%%%%%%%% Acknowledgements %%%%%%%%%%%%%%%%%
\section*{Acknowledgements}
This research has made use of the services of the ESO Science Archive Facility and NASA’s Astrophysics Data System Abstract Service. It is based on observations collected at the European Organisation for Astronomical Research in the Southern Hemisphere under ESO programme 60.A-9301(A) and data products created thereof. Also we have used observations obtained with the NASA/ESA HST and obtained from the Hubble Legacy Archive, which is a collaboration between the Space Telescope Science Institute (STScI/NASA), the Space Telescope European Coordinating Facility (ST-ECF/ESA), and the Canadian Astronomy Data Centre (CADC/NRC/CSA). 

This work has been supported by Spanish grants from the former Ministry of Economy, Industry and Competitiveness through the MINECO-FEDER research grant AYA2016-79724-C4-1-P, the present Ministry of Science and Innovation through research grant PID2019-107408GB-C42 and the National Research Agency through research grant AEI/10.13039/501100011033.

S.Z. acknowledges the support from contract: BES-2017-080509 associated to the first of these grants. 

\section*{Data Availability}
The original data on which this article is based can be found in the ESO Science Archive Facility from ESO telescopes at La Silla Paranal Observatory. 

%%%%%%%%%%%%%%%%% REFERENCES %%%%%%%%%%%%%%%%%%%%%
\bibliographystyle{mnras}
\bibliography{bibliografia.bib} % if your bibtex file is called example.bib

%%%%%%%%%%%%%%%%% APPENDICES %%%%%%%%%%%%%%%%%%%%%

%\appendix
%\section{Tables}
%\label{appendix1}
%\input{Table/TABLAS}

%%%%%%%%%%%%%%%%% Template %%%%%%%%%%%%%%%%%%
\end{document}